\documentclass[]{JHEP3}

\pdfoutput=1

\usepackage{amsmath}
\usepackage{epsfig,multicol,bbm}
\usepackage{graphicx}
\usepackage{bm}
\usepackage{epsfig}

\renewcommand{\emph}[1]{{\it #1}}

\def\nslash{n\!\!\!\slash}
\def\bnslash{\bar n\!\!\!\slash}

\newcommand{\nn}{\nonumber} 
\newcommand{\bn}{{\bar n}}
\newcommand{\mcdot}{\!\cdot\!}

\newcommand{\SCETa}{\mbox{${\rm SCET}_{\rm I}$ }}
\newcommand{\SCETb}{\mbox{${\rm SCET}_{\rm II}$ }}

\newcommand{\vect}[1]{\mathbf{#1}}
\newcommand{\abs}[1]{\left\lvert #1\right\rvert}
\newcommand{\bra}[1]{\left\langle #1\right\rvert}
\newcommand{\ket}[1]{\left\lvert #1\right\rangle}

\newcommand{\Lqcd}{\Lambda_{\text{QCD}}}
\newcommand{\e}{\mathrm{e}}

\newcommand{\eUV}{\epsilon_{\text{UV}}}
\newcommand{\eIR}{\epsilon_{\text{IR}}}
\newcommand{\MeV}{\text{ MeV}}
\newcommand{\GeV}{\text{ GeV}}

\newcommand{\eq}[1]{Eq.~\eqref{#1}}
\newcommand{\eqs}[2]{Eqs.~\eqref{#1} and \eqref{#2}}
\newcommand{\eqss}[3]{Eqs.~\eqref{#1}, \eqref{#2}, and \eqref{#3}}
\renewcommand{\sec}[1]{Sec.~\ref{#1}}
\newcommand{\ssec}[1]{Sec.~\ref{ssec:#1}}
\newcommand{\fig}[1]{Fig.~\ref{#1}}
\newcommand{\tab}[1]{Table~\ref{tab:#1}}

\DeclareMathOperator{\Tr}{Tr}

\DeclareMathOperator{\Disc}{Disc}
\DeclareMathOperator{\sgn}{sgn}

\newcommand{\df}{\mathrm{d}}
\newcommand{\CF}{C_F}

\newcommand{\taus}{\tau_a^s}
\newcommand{\hattaus}{\hat{\tau}_a^s}
\newcommand{\tausp}{\tau_a^{s'}}
\newcommand{\taun}{\tau_a^n}

\newcommand{\taunp}{{\tau_a^{n}}'}

\newcommand{\as}{\alpha_s}

\newcommand{\cV}{\mathcal{V}}
\newcommand{\cO}{\mathcal{O}}
\newcommand{\cR}{\mathcal{R}}
\newcommand{\cS}{\mathcal{S}}
\newcommand{\cSperp}{\mathcal{S}}

\newcommand{\Mae}[3]{\bigl\langle#1\bigl\lvert#2\bigr\rvert#3\bigr\rangle}

\newcommand{\bare}{(0)}
\newcommand{\cusp}{\! \rm cusp}

\newcommand{\delreal}[1]{\delta_{#1}(\taun, q, l^+)}
\newcommand{\delvirt}[1]{\delta_{#1}(\taun, l^+)}

\newcommand{\softreala}{\,  \vphantom{\Big[} \includegraphics[totalheight = .05\textheight,bb=0 50 130 150]{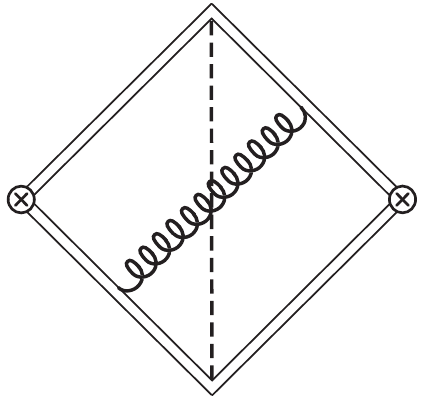}  }

\newcommand{\softvirta}{\,  \vphantom{\Big[} \includegraphics[totalheight = .05\textheight,bb=0 50 130 150]{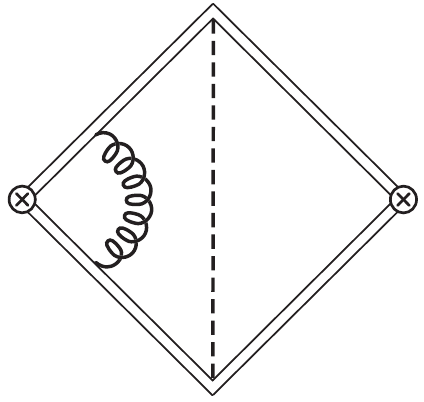}}

\newcommand{\Wilsonb}{\, \includegraphics[totalheight = .03\textheight,bb=0 15 120 65]{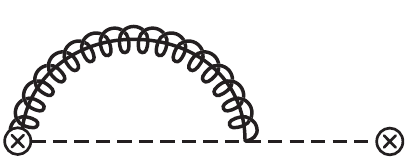}}

\newcommand{\Wilsonbreal}{\, \includegraphics[totalheight = .03\textheight,bb=0 15 120 65]{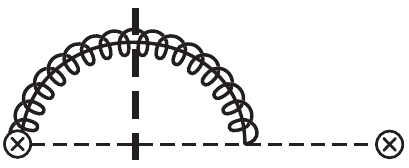}}

\newcommand{\Wilsonbvirt}{\, \includegraphics[totalheight = .03\textheight,bb=0 15 120 65]{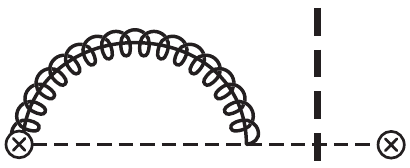}}
\newcommand{\QCDa}{\, \includegraphics[totalheight = .03\textheight,bb=0 15 132 65]{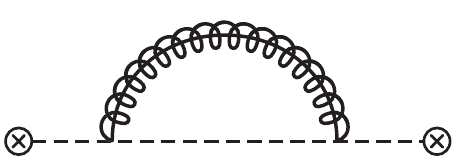}}
\newcommand{\QCDb}{\, \includegraphics[totalheight = .03\textheight,bb=0 15 110 65]{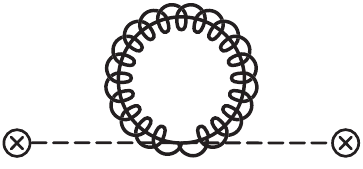}}
\newcommand{\QCDtilde}{ \, P_n \includegraphics[totalheight = .03\textheight,bb=0 15 130 65]{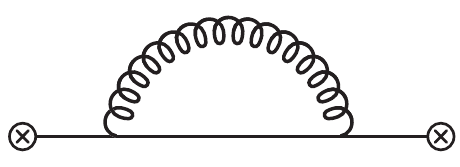} P_{\bar n}}
\newcommand{\QCDtildereal}{ \, P_n \includegraphics[totalheight = .03\textheight,bb=0 15 130 65]{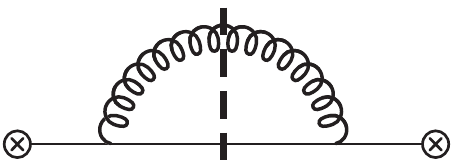} P_{\bar n}}
\newcommand{\QCDtildevirt}{ \,  P_n \includegraphics[totalheight = .03\textheight,bb=0 15 130 65]{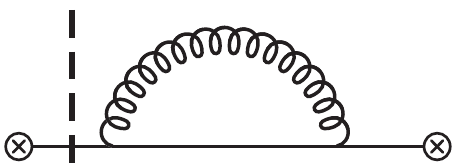} P_{\bar n}}

\newcommand{\Gexp}{K}
\newcommand{\Omeg}{\Omega} 

\newcommand{\fexp}{f^{\rm exp}}
\newcommand{\PT}{\rm PT}
\newcommand{\NLL}{\rm NLL/NLO}

\title{Effective Predictions of Event Shapes: Factorized, Resummed, and Gapped Angularity Distributions}

\author{Andrew Hornig, Christopher Lee, and Grigory Ovanesyan\\ 
Theoretical Physics Group, Lawrence Berkeley National Laboratory, 1 Cyclotron Rd., Berkeley, CA  94720, USA,
  and Center for Theoretical Physics, 366 LeConte Hall \#7300, University of California, Berkeley, CA, 94720, USA 
 \\ E-mail: \email{ahornig@berkeley.edu}, \email{clee@berkeley.edu}, \email{ovanesyan@berkeley.edu} }

\preprint{UCB-PTH-09/03 \\ arXiv:0901.3780}	

\abstract{Using soft-collinear effective theory (SCET), which provides a unified framework for factorization, resummation of logarithms, and incorporation of universal nonperturbative functions in hard-scattering QCD cross-sections, we present a new prediction of angularity distributions in $e^+e^-$ annihilation. Angularities $\tau_a$ are an infinite class of event shapes which vary in their sensitivity to the substructure of jets in the final state, controlled by a continuous parameter $a<2$. We calculate  angularity distributions for all $a<1$ to first order in the strong coupling $\alpha_s$ and resum large logarithms in these distributions to next-to-leading logarithmic (NLL) accuracy. Our expressions for the next-to-leading order (NLO) $\cO(\as)$ partonic jet and soft functions in the factorization theorem for angularity distributions are given for the first time.  We employ a model for the nonperturbative soft function with a  gap parameter which cancels the renormalon ambiguity in the partonic soft function. We explore the relation between the SCET approach to resummation and past approaches in QCD, and discuss the advantages of the effective theory approach.  In addition, we draw from the NLO calculations of the jet and soft functions an intuitive lesson about how factorization breaks down in the effective theory as $a\to 1$. }

\keywords{Event shapes, Factorization, Resummation, Effective Field Theory}

\begin{document}


\section{Introduction}
\label{sec:intro}

Event shapes probe the hadronic final states produced in hard scattering processes for jet-like structure \cite{Dasgupta:2003iq}. Two-jet event shapes $e$ in hadronic $e^+ e^-$ annihilations are constructed so that one of the kinematic endpoints corresponds to the limit of two back-to-back perfectly collimated jets. Different event shapes vary in their sensitivity to particles close to or far away from the jet axis and thus used in tandem probe the substructure of jets \cite{Almeida:2008tp,Almeida:2008yp}. Some examples of two-jet event shapes are the familiar thrust \cite{Brandt:1964sa,Farhi:1977sg}, jet masses \cite{Clavelli:1979md,Chandramohan:1980ry,Clavelli:1981yh}, and jet broadening \cite{Catani:1992jc}, 
and the more recently introduced angularities \cite{Berger:2003iw}. The shape of the distributions in these variables depend on several energy scales, namely, the scale $Q$ of the hard scattering, the scale of the invariant mass or typical transverse momentum of the jet $\mu_J$, and the scale $\Lqcd$ of soft radiation from the jets involved in color recombination occurring during hadronization. Event shapes thus probe the behavior of QCD over a large range of energy scales, and indeed have been the source of some of the most precise extractions of the strong coupling constant $\alpha_s$ \cite{Becher:2008cf,Dissertori:2007xa,Bethke:2006ac}.

Dependence on strong interactions at soft scales  near $\Lqcd$ where QCD is nonperturbative would render predictive calculations impossible, without the use of factorization. Factorization separates an observable into pieces depending on each individual relevant energy scale. Those pieces depending on large scales can be calculated perturbatively, while those depending on soft scales remain nonperturbative. If these soft functions are, however, universal among different observables or physical processes, then  calculations of the factorized observables become predictive. A large number of two-jet event shape distributions in $e^+ e^-$ annihilation can be factorized into hard, jet, and soft functions:
\begin{equation}
\label{fact-theorem}
\frac{1}{\sigma_{\text{tot}}}\frac{\df\sigma}{\df e} = H(Q;\mu)\int \df e_1\,\df e_2\,\df e_s\, J_1(e_1;\mu) J_2(e_2; \mu) S(e_s;\mu)\delta(e - e_1 - e_2 - e_s)\,,
\end{equation}
where $H(Q;\mu)$ is the hard coefficient dependent only on the hard scattering $e^+ e^-\rightarrow q\bar q$ at center-of-mass energy $Q$, $J_{1,2}$ are jet functions describing the perturbative evolution of the initially produced partons $q,\bar q$ into collimated jets of lower-energy partons, and finally $S(e_s;\mu)$ is the soft function describing the color exchange between the two jets leading to the hadronization  of their constituent partons. This description introduces dependence on a factorization scale $\mu$, at which the cross-section is factorized, into each of the individual functions. This dependence must cancel in the whole combination in \eq{fact-theorem}. The full distribution and the individual jet and soft functions contain terms of the form $(1/e)\alpha_s^n\ln^m e$ which become large in the two-jet limit $e\to 0$. The dependence of the hard, jet, and soft functions on the factorization scale $\mu$ can be determined from renormalization group equations, which can be used to resum the large logarithms \cite{Contopanagos:1996nh}.

The formidable achievements of proofs of factorization theorems for hard scattering cross-sections in QCD span a long and monumental history \cite{Collins:1989gx,Sterman:1995fz}. More recently many of these theorems were reformulated in the language of soft-collinear effective theory (SCET)  
\cite{Bauer:2000ew,Bauer:2000yr,Bauer:2001ct,Bauer:2001yt}.  This was done for two-jet event shapes for light quark jets in the series of papers \cite{Bauer:2002ie,Bauer:2003di,Lee:2006nr,Bauer:2008dt} and for top quark jets in \cite{Fleming:2007qr,Fleming:2007xt}. Some of the relations between the full and effective theory formulations of factorization were explored in \cite{Lee:2006nr,Bauer:2008dt}. Equivalent results can be formulated in either language, although our discussion below will be in the context of SCET, which we find advantageous for its intuitive framework for separating physics at hard, collinear, and soft scales and its explicit Lagrangian for interactions between collinear and soft modes. These features facilitate the implementation of factorization and resummation of logarithms of ratios of all the relevant energy scales. At the same time that the effective theory provides us an intuitive framework in which to analyze the behavior of event shape distributions, the properties of the angularities themselves will in turn illuminate properties of the effective theory, and in particular, the conditions under which it is valid for the observables under consideration.

To describe the conditions under which the distribution in a particular event shape factorizes as in \eq{fact-theorem}, it is useful to write  event shapes in a generic form. Many event shapes can be written in the form,
\begin{equation}
e(X) = \frac{1}{Q}\sum_{i\in X} \abs{\vect{p}_\perp^i}f_e(\eta_i)\,,
\label{eventshape}
\end{equation}
where the sum is over all particles  $i$ in the final state $X$, $\vect{p}_\perp^i$ is the transverse momentum of the $i$th particle and $\eta_i$ its rapidity relative to the thrust axis. Each choice of the weight function $f_e$ determines a different event shape. For example, for the thrust and jet broadening, $f_{1-T}(\eta) = \e^{-\abs{\eta}}$ and $f_B (\eta) = 1$. 
A continuous set of event shapes which generalize the thrust and jet broadening are the \emph{angularities} $\tau_a$ \cite{Berger:2003iw}, corresponding to the choice 
\begin{equation}
\label{ftauadef}
f_{\tau_a}(\eta) = \e^{-\abs{\eta}(1-a)}\,,
\end{equation}
where $a$ is any real number $a<2$. For $a\geq 2$, the function in \eq{ftauadef} weights particles collinear to the thrust axis too strongly and makes  the quantity \eq{eventshape} sensitive to collinear splitting, and thus not infrared-safe. The factorization theorem \eq{fact-theorem}, however, is valid only for $a<1$. At $a=1$, the distribution of events in $\tau_1$ is dominated by jets with invariant mass of order $\Lqcd$. Thus, the jet and soft scales coincide, and the distribution cannot be divided into separately infrared-safe jet and soft functions, at least in the traditional form of the factorization theorem. This breakdown can be seen in the uncontrollable growth of a number of nonperturbative power corrections  as $a\rightarrow 1$ \cite{Berger:2003iw,Lee:2006nr}, or in the failure to cancel infrared divergences in the perturbative calculation of the jet or soft functions in the same limit, as we have recently explored in Ref.~\cite{Hornig:2009kv}. We review this breakdown of factorization in the explicit perturbative calculations we perform below. Any choice of weight function $f_e$ that sets a jet scale at or lower than the soft scale will ruin the factorization \eq{fact-theorem}. 

The distributions for which the factorization in \eq{fact-theorem} breaks down might still factorize in a different form, by distinguishing collinear and soft modes not by their invariant mass, but by their rapidity, as proposed in \cite{Manohar:2006nz}. We do not, however, pursue such a strategy here, and focus only on angularities with strictly $a<1$.\footnote{Even though traditional factorization breaks down for $a=1$ (jet broadening), the resummation of jet broadening in QCD was performed in \cite{Catani:1992jc,Dokshitzer:1998kz} and nonperturbative effects were discussed in \cite{Dokshitzer:1998qp}.}

The soft function evaluated at a scale $\mu_s\sim\Lqcd$ is nonperturbative. Evaluated at a higher scale, however, it can be calculated in perturbation theory. An appropriate model for the soft function should interpolate between these two regimes. In our analysis we adopt a model like that proposed for hemisphere jet masses in \cite{Hoang:2007vb} and for $b$-quark distributions in \cite{Ligeti:2008ac}, in which the soft function is a convolution, 
\begin{equation}
\label{softmodel}
S(e_s;\mu) = \int \df e_s' \, S^{\PT}(e_s - e_s';\mu) \fexp(e_s' -\Delta_{e})\,,
\end{equation}
where $S^{\PT}$ is the partonic soft function calculated in perturbation theory, and $\fexp$ is a nonperturbative model function. The gap parameter $\Delta_e$, proposed in Ref.~\cite{Hoang:2007vb}, enters $\fexp$ through a theta function $\theta(e_s - \Delta_e)$ so that the minimum possible value of an event shape $e$ of final states is $\Delta_e$, which is zero in the partonic distribution, but is nonzero due to hadronization in the actual distribution.  The full soft function $S(e_s;\mu)$ inherits its scale dependence from $S^{\PT}(e_s;\mu)$ and thus has a well-defined running with the scale $\mu$.

The partonic soft function $S^{\PT}(e_s; \mu)$ contains a renormalon ambiguity due to the behavior of the perturbative series at high orders. This ambiguity should not be present in the full physical distribution or the soft function, so the ambiguity in $S^{\PT}$ is canceled by a corresponding ambiguity in $\Delta_e$. Shifting from $\Delta_e$ to a renormalon-free gap parameter $\bar{\Delta}_{e}(\mu) = \Delta_e - \delta_e(\mu)$ removes the ambiguity from the entire soft function \eq{softmodel}. This greatly reduces the uncertainty in the predicted distribution due to such renormalon ambiguities. These features were demonstrated in \cite{Hoang:2007vb} for jet mass and thrust distributions. In this paper, we extend the soft function model and demonstrate that a similar cancellation occurs for angularities $\tau_a$. 

Many studies of nonperturbative soft power corrections in event shape distributions have been based on the behavior of the perturbative expansions of the distributions, either the behavior of their renormalon ambiguities \cite{Manohar:1994kq,Beneke:2000kc} or their dependence on a postulated ``infrared'' effective coupling $\alpha_s$ at low scales \cite{Dokshitzer:1995qm,Dokshitzer:1995zt,Dokshitzer:1997ew}. In particular, they led to the proposal of a universal soft power correction to the mean values of event shape distributions in the form   \cite{Dokshitzer:1995zt,Dokshitzer:1997ew}
\begin{equation}
\label{shiftclaim}
\langle e \rangle = \langle e \rangle_{\text{PT}} + \frac{c_e\mathcal{A}}{Q}\,,
\end{equation}
where $\langle e\rangle_{\PT}$ is the mean value of the partonic distribution, and the coefficient of the $1/Q$ power correction is an exactly-calculable number $c_e$ dependent on the choice of event shape multiplied by an unknown nonperturbative parameter $\mathcal{A}$ which is universal for numerous event shape distributions. In \cite{Lee:2006nr} the operator definition of the soft function in the factorization theorem \eq{fact-theorem} was used to prove the relation \eq{shiftclaim} to all orders in $\as$. For angularities, $c_{\tau_a} = 2/(1-a)$. This scaling of the power correction with $a$ was observed in \cite{Berger:2003pk} based on the behavior of the resummed perturbative series for angularity distributions after imposing an IR cutoff on the scale in $\alpha_s(\mu)$ and in \cite{Berger:2004xf} based on analysis of the distributions using dressed gluon exponentiation \cite{Gardi:2001di}. Below we  will review the proof of the scaling in \cite{Lee:2006nr} based on the operator definition of the soft function independently of its perturbative expansion, and later use the scaling rule to constrain the nonperturbative model we adopt for the soft function in angularity distributions.

The history of calculating event shape distributions using  perturbation theory  in QCD goes all the way back to QCD's earliest years. 
The thrust distribution for light quark jets to $\cO(\as)$ was calculated in \cite{DeRujula:1978yh}, to which our fixed-order results for $\df\sigma/\df\tau_a$ reduce at $a=0$. The resummation of the thrust distribution to NLL was performed in QCD in  \cite{Catani:1992ua,Catani:1991kz} and to LL in SCET in  \cite{Bauer:2006mk,Schwartz:2007ib} (and later extended to ${\rm N}^3 {\rm LL}$ in \cite{Becher:2008cf}). Our results are consistent with these SCET results at the appropriate orders for $a=0$. The jet mass distribution for top quark jets was calculated and resummed to the same order in \cite{Fleming:2007xt}, with which we agree on the SCET jet and soft functions for $a=0$ in the limit $m_t = 0$. The jet and soft functions for thrust or jet mass distributions  can be derived easily from the ``ordinary'' SCET jet function $J(k^+)$, and the hemisphere soft function $S(k^+,k^-)$, because the thrust and jet mass depend only on a single light-cone component of the total four-momentum in each hemisphere (cf. \cite{Catani:1992ua}). These standard jet and soft functions were calculated to two-loop order in  \cite{Becher:2005pd,Becher:2006qw,Hoang:2008fs}. Angularities for arbitrary $a$, however, depend on \emph{both} light-cone components $k^\pm$ in \emph{each} hemisphere, thus requiring the new calculations we perform below.

In the original introduction of the angularities $\tau_a$ \cite{Berger:2003iw} the resummation of logarithms was achieved to the same next-to-leading-logarithmic (NLL)  order that we achieve below, but without full inclusion of next-to-leading-order (NLO)  jet and soft functions for the $\tau_a$-distribution, which we calculate explicitly here for the first time. This improves the accuracy of our result for small $\tau_a$. 
Our result is also improved in this region by adopting the soft function model \eq{softmodel} which cures unphysical behavior of the point-by-point distribution $\df\sigma/\df\tau_a$ as $\tau_a\rightarrow 0$ due to renormalon ambiguities.
The results of \cite{Berger:2003iw} converted to the traditional form of an NLL resummed event shape distribution \cite{Catani:1992ua} were subsequently matched to fixed-order QCD at $\cO(\as^2)$ numerically in \cite{Berger:2003pk}, improving the accuracy of the large-$\tau_a$ region. We perform this fixed-order matching only at $\cO(\as^1)$.

Comparing our result to those of \cite{Berger:2003iw,Berger:2003pk} elucidates the relation between SCET and traditional QCD-based approaches to resumming logarithms more generally. 
While the advantages of SCET in achieving factorization or resummation of logarithms through renormalization group evolution  can of course be  formulated without the explicit language of the effective theory (see, e.g., \cite{Berger:2003iw,Contopanagos:1996nh}), the effective theory nevertheless  unifies these concepts and methods in an intuitive framework that, we have found, allows us greater facility in improving the precision and reliability of our predictions of event shape distributions.  
Even though we do not go beyond the existing NLL resummation of logarithms of $\tau_a$  \cite{Berger:2003iw,Berger:2003pk},  the flexibility in the effective theory to vary the scales $\mu_{H,J,S}$, where logarithms in the hard, jet, and soft functions are small and from which we run each function to the factorization scale $\mu$, allows additional improvements. For example, we are able to avoid any spurious Landau pole singularities which the traditional approaches usually encounter. (For previous discussions on how the effective theory avoids spurious Landau poles present in the traditional approach, see Refs.~\cite{Manohar:2003vb,Becher:2006nr,Becher:2006mr}.)

The plan of the paper is as follows. In Sec.~\ref{sec:review}, we review the demonstration of factorization of event shape distributions in the formalism of SCET that was presented in \cite{Bauer:2008dt}, recalling the introduction of the event shape operator $\hat e$ that returns the value of an event shape $e$ of a final state $X$, constructed from the energy-momentum tensor. In Sec.~\ref{sec:NLO}, we calculate the jet and soft functions appearing in the factorization theorem for angularity distributions for $a<1$ to one-loop order in $\as$. We recall the observations of \cite{Hornig:2009kv} about how the breakdown of factorization as $a\rightarrow 1$ is observed in the infrared behavior of these functions in perturbation theory. In Sec.~\ref{sec:NLL} we solve the renormalization group equations obeyed by the hard, jet, and soft functions and resum leading and next-to-leading logarithms of $\tau_a$ in the perturbative expansions of these functions, and explain how we match the resummed distributions onto the fixed-order prediction of QCD at $\cO(\as)$. In Sec.~\ref{sec:model} we construct a model for the soft function in angularity distributions for all $a<1$, based on existing models for hemisphere and thrust soft functions which contain a nonperturbative gap parameter introduced in \cite{Hoang:2007vb}, which  cancels the renormalon ambiguity in the perturbative series for the soft function. In Sec.~\ref{sec:results} we present plots of our final predictions of angularity distributions using all the results of Secs.~\ref{sec:NLO}--\ref{sec:model}.  In Sec.~\ref{sec:relation} we compare and contrast the SCET approach to predicting resummed angularity distributions to those  based on factorization and RG evolution in full QCD \cite{Berger:2003iw} and to the traditional approach to resummation \cite{Berger:2003pk,Catani:1992ua}. In Sec.~\ref{sec:conclusions} we present our conclusions, and in the Appendices, we verify a consistency relation among the hard, jet, and soft anomalous dimensions for arbitrary $a$, provide some technical details necessary for the solution of the RG equations for the jet and soft functions, and explain our procedure to calculate angularity distributions at fixed-order in QCD at $\cO(\as)$, noting the hitherto unnoticed property of the angularities that they fail to separate two- and three-jet-like events for values of $a\lesssim -2$, and so cease to behave exactly as ``two-jet'' event shapes.

\section{Review of Factorization of Event Shape Distributions}
\label{sec:review}

We begin by reviewing the factorization of event shape distributions in the formalism of SCET, presented in \cite{Bauer:2008dt}. 

\subsection{Event shape distributions in full QCD}

The full QCD distribution of events in $e^+ e^- \rightarrow\text{ hadrons}$ in an event shape variable $e$ is given, to leading-order in electroweak couplings, by
\begin{equation}
\frac{\df \sigma}{\df e} = \frac{1}{2Q^2}\sum_X\int \df^4 x\,\e^{iq\cdot x}\sum_{i=V,A}L^i_{\mu\nu}\bra{0} j_i^{\mu\dag}(x)\ket{X}\bra{X}j_i^\nu(0)\ket{0}\delta(e - e(X))\,,
\label{QCDdist}
\end{equation}
where $q = (Q,\vect{0})$ is the total four-momentum in the center-of-mass frame, the sum is over final states $X$, and $e(X)$ is the value of the event shape $e$ of the state $X$. The final state is produced by the vector and axial currents,
\begin{equation}
\label{QCDcurrent}
j_i^\mu = \sum_{f,a}\bar q_f^a \Gamma_i^\mu q_f^a\,,
\end{equation}
where $\Gamma_V^\mu = \gamma^\mu$ and $\Gamma_A^\mu = \gamma^\mu\gamma^5$ and the sum is over quark flavors $f$ and colors $a$. The leptonic tensor, which includes contributions from an intermediate photon and $Z$ boson, is given by
\begin{subequations}
\begin{align}
L^V_{\mu\nu} &= -\frac{e^4}{3Q^2}\left(g_{\mu\nu} - \frac{q_\mu q_\nu}{Q^2}\right)\left[Q_f^2 - \frac{2Q^2 v_e v_f Q_f}{Q^2 - M_Z^2} + \frac{Q^4(v_e^2+a_e^2)v_f^2}{(Q^2-M_Z^2)^2}\right] \\
L^A_{\mu\nu} &= -\frac{e^4}{3Q^2}\left(g_{\mu\nu} - \frac{q_\mu q_\nu}{Q^2}\right)\frac{Q^4(v_e^2+a_e^2)a_f^2}{(Q^2-M_Z^2)^2}\,, 
\end{align}
\end{subequations}
where  $Q_f$ is the electric charge of $f$ in units of $e$, and  $v_f,a_f$ are the vector and axial charges of $f$,
\begin{equation}
v_f = \frac{1}{2\sin\theta_W\cos\theta_W}(T^3_f - 2Q_f\sin^2\theta_W)\,,\quad a_f = \frac{1}{2\sin\theta_W\cos\theta_W}T^3_f.
\end{equation}
As shown in \cite{Bauer:2008dt}, the sum over hadronic final states remaining in \eq{QCDdist} can be performed by introducing an operator $\hat e$ that gives the event shape $e(X)$ of a final state $X$. This operator  can be constructed from a momentum flow operator, which in turn is constructed from the energy-momentum tensor. That is, 
\begin{equation}
\hat e \ket{X} \equiv e(X) \ket{X} = \frac{1}{Q} \int_{-\infty}^\infty \!\!\df\eta \, f_e(\eta) \mathcal{E}_T(\eta;\hat t) \ket{X}\,,
\label{ehatdef}
\end{equation}
where $\hat t$ is the operator yielding the thrust axis of final state $X$, and $\mathcal{E}_T(\eta;\hat t)$ is the transverse momentum flow operator, yielding the total transverse momentum flow in the direction given by rapidity $\eta$, measured with respect to the thrust axis, in a final state $X$,
\begin{equation}
\mathcal{E}_T(\eta;\hat t)\ket{X} \equiv  \frac{1}{\cosh^3\eta}\int_0^{2\pi} \df\phi\lim_{R\rightarrow\infty} R^2\int_0^\infty \df t\,\hat n_i T_{0i}(t,R\hat n)\ket {X} = \sum_{i\in X}\abs{\vect{p}_\perp^i}\delta(\eta-\eta_i)\ket{X} \,,
\label{ETfromT0i}
\end{equation}
which is closely related to the energy flow operator proposed in \cite{Korchemsky:1997sy}.
The thrust axis operator $\hat t$ can be constructed explicitly, as shown in  \cite{Bauer:2008dt}. After matching onto SCET, however, an explicit construction is not necessary, as the thrust axis is simply given by the jet axis $\vect{n}$ labeling the two-jet current. The difference between the two axes introduces power corrections in $\lambda$ which are subleading, as long as $a<1$ \cite{Berger:2003iw, Bauer:2008dt}.  Using the operator $\hat e$, we perform the sum over $X$ in \eq{QCDdist}, leaving
\begin{equation}
\frac{\df\sigma}{\df e} = \frac{1}{2Q^2}\int \df^4 x\,\e^{iq\cdot x}\sum_{i=V,A} L^i_{\mu\nu}\bra{0} j_i^{\mu\dag}(x)\delta(e - \hat e)j_i^\nu(0)\ket{0}\,.
\label{QCD2}
\end{equation}

\subsection{Factorization of event shape distributions in SCET}

To proceed to a factorized form of the distribution \eq{QCD2}, we match the current $j^\mu$ and the operator $\hat e$ onto operators in SCET. To reproduce the endpoint region of the two-jet event shape distribution, we match the QCD currents $j_i^\mu$ onto SCET operators containing fields in just two back-to-back collinear directions,
\begin{eqnarray}
j_i^\mu(x) &=& \sum_{\vect{n}}\sum_{\tilde p_n, \tilde p_\bn}C_{n \bar n}(\tilde p_n,\tilde p_\bn;\mu) {\cal O}_{n \bn}(x; \tilde p_n, \tilde p_\bn)\,,
\label{SCETcurrentO2}
\end{eqnarray}
summing over the direction $\vect{n}$ of the light-cone vectors $n,\bar n = (1,\pm \vect{n})$, and label momenta $\tilde p_n,\tilde p_{\bar n}$.  The two-jet operators \cite{Bauer:2003di,Bauer:2002nz}, after the BPS field redefinition 
\cite{Bauer:2001yt}
with soft Wilson lines, are
\begin{equation}
 {\cal O}_{n \bn}(x; \tilde p_n, \tilde p_\bn)= e^{i(\tilde p_n - \tilde p_\bn)\cdot x}\bar\chi_{n,p_n}(x)Y_{n}(x)\Gamma_i^\mu\overline Y_{\bn}(x)\chi_{\bn,p_\bn}(x)\,,
 \label{O2def}
\end{equation}
where $\Gamma_V^\mu = \gamma_\perp^\mu$ and $\Gamma_A^\mu = \gamma_\perp^\mu\gamma_5$. The soft Wilson lines are the path-ordered exponentials of soft gluons,
\begin{equation}
Y_n(x) = P\exp\left[ig\int_0^\infty n\cdot A_s(ns + x)\right]\,,\qquad \overline Y_\bn(x) =P\exp\left[ig\int_0^\infty \bn\cdot \bar A_s(\bn s + x)\right] \,,
\end{equation}
with $A_s,\bar A_s$ respectively in the fundamental or anti-fundamental representation. 
The jet fields $\chi_{n} = W_n^\dag\xi_n$ and $\chi_{\bn} = W_\bn^\dag\xi_\bn$ are combinations of collinear quark fields made invariant under collinear gauge transformations by Wilson lines of collinear gluons  \cite{Bauer:2000yr,Bauer:2001ct}, where
\begin{equation}
W_n(x) = \sum_{\text{perms}}\exp\left[-g\frac{1}{\bar{\mathcal{P}}}\bn\cdot A_{n,q}(x)\right]\,,
\end{equation}
where $q$ is the label momentum of the collinear gluon field $A_n$, and $\bar{\mathcal{P}}$ is a label momentum operator which acts as $\bar{\mathcal{P}}A_{n,q} = (\bn\cdot q)A_{n,q}$ \cite{Bauer:2001ct}.
 Recall that, in SCET, collinear momenta $p_c^\mu = \tilde p^\mu + k^\mu$ are divided into a large label piece, $\tilde p^\mu = (\bn\cdot \tilde p)n^\mu/2 + \tilde p_\perp^\mu$, and a residual piece, $k^\mu$, where $\bn\cdot\tilde p$ is $\mathcal{O}(Q)$, $\tilde p_\perp$ is $\mathcal{O}(Q\lambda)$, and $k$ is $\mathcal{O}(Q\lambda^2)$. The residual momenta are the same size as soft momenta, $k_s$, of $\mathcal{O}(Q\lambda^2)$. 
 Below, however, we will see how the natural scaling of the collinear modes varies with the choice of observable $\tau_a$. The integral over $x$ in \eq{QCD2} enforces that the label momenta of the jet fields in the two-jet operator satisfy $\bar n\cdot\tilde p_n = -n\cdot\tilde p_\bn = Q$ and $\tilde p_n^\perp = \tilde p_\bn^\perp = 0$.

We must also match the operator $\hat e$ in full QCD onto SCET. To do so we simply replace the QCD energy-momentum tensor $T^{\mu\nu}$ appearing in the definition \eq{ETfromT0i} with the energy-momentum tensor in SCET, and, as noted above, set the thrust axis equal to the jet axis $\vect{n}$ in the two-jet operator $\mathcal{O}_{n\bn}$.  After the BPS field redefinition,  to leading order in $\lambda$ the SCET energy-momentum tensor is a direct sum over contributions from fields in the $n,\bn$ collinear and soft sectors, since the Lagrangian splits into these separate sectors with no interactions between them. (Beyond leading order in $\lambda$, there are power-suppressed terms in the SCET Lagrangian in which interactions between collinear and soft fields do not decouple following the BPS field redefintion \cite{Bauer:2002uv,Chay:2002vy,Pirjol:2002km,Bauer:2003mga}.) Then the event shape operator $\hat e$ splits into separate collinear and soft operators,
\begin{equation}
\label{splite}
\hat e = \hat e_{n} + \hat e_{\bn} + \hat e_s \,,
\end{equation}
where each $\hat e_i$ is constructed only from the energy-momentum tensor of sector $i$ of the effective theory. So, finally, the event shape distribution in SCET factorizes into purely hard, collinear and soft functions,
\begin{equation}
\label{finalformula}
\frac{1}{\sigma_0}\frac{\df\sigma}{\df e} = H(Q;\mu) \int \df e_n\,\df e_{\bar n}\,\df e_s\,\delta(e-e_n-e_{\bar n}-e_s) J_n(e_n;\mu) J_{\bar n}(e_{\bar n};\mu) S(e_s;\mu)\,,
\end{equation}
where the hard coefficient is the squared amplitude of the two-jet matching coefficient, 
\begin{equation}
\label{hardfunction}
H(Q;\mu)  = \abs{C_{n\bn}(Qn/2,-Q\bn/2;\mu)}^2\,,
\end{equation}
and the jet and soft functions are given by the matrix elements of collinear and soft operators,
\begin{equation}
S(e_s;\mu) = \frac{1}{N_C}\Tr\bra{0}\overline Y^\dag_{\bar n}(0)Y_n^\dag(0)\delta(e_s - \hat e_s)Y_n(0)\overline Y_{\bar n}(0)\ket{0} \label{softfunctiondef}
\,,
\end{equation}
and
\begin{equation}
\label{integratedjetfunctions}
J_n(e_n;\mu) = \int\frac{\df l^+}{2\pi}\mathcal{J}_n(e_n,l^+;\mu)\,,\quad J_\bn(e_\bn;\mu) = \int\frac{\df k^-}{2\pi}\mathcal{J}_\bn(e_\bn,k^-;\mu)\,,
\end{equation}
where 
\begin{subequations}
\label{jetfunctions}
\begin{align}
\mathcal{J}_n(e_n,l^+;\mu)\left(\frac{\nslash}{2}\right)_{\!\!\alpha\beta} &=  \frac{1}{N_C}\Tr\int \df^4 x\,\e^{il\cdot x}\bra{0}\chi_{n,Q}(x)_\alpha\delta(e_n  -  \hat e_n)\bar\chi_{n,Q}(0)_\beta\ket{0} \label{njetfunction} \\
\mathcal{J}_\bn(e_\bn,k^-;\mu)\left(\frac{\bnslash}{2}\right)_{\!\!\alpha\beta}  &= \frac{1}{N_C}\Tr\int\df^4 x\,\e^{ik\cdot x} \bra{0}\bar\chi_{\bar n,-Q}(x)_\beta\delta(e_{\bar n} -  \hat e_{\bar n})\chi_{\bar n,-Q}(0)_\alpha\ket{0}\,. \label{nbarjetfunction} 
\end{align}
\end{subequations}
In \eqss{softfunctiondef}{njetfunction}{nbarjetfunction},  the traces are over colors. Also, in \eq{finalformula}, we have divided the distribution by the total Born cross-section for $e^+e^-\rightarrow q\bar q$, 
\begin{equation}
\label{sigmazero}
\sigma_0 = \frac{4\pi\alpha^2 N_C}{3 Q^2}\sum_f \left[Q_f^2 - \frac{2Q^2 v_e v_f Q_f}{Q^2 - M_Z^2} + \frac{Q^4(v_e^2+a_e^2)(v_f^2+a_f^2)}{(Q^2-M_Z^2)^2}\right]\,.
\end{equation}
The $n$-collinear jet function $J_n$ depends only on the $l^+ \equiv n\cdot l$ component of the residual momentum, and $J_\bn$ on $k^- \equiv \bn\cdot k$, as only the $n\cdot \partial$ derivative appears in the $n$-collinear Lagrangian, and $\bn\cdot\partial$ in the $\bn$-collinear Lagrangian, at leading order in $\lambda$ \cite{Bauer:2000yr}. In angularity distributions, the jet functions are independent of the residual transverse momenta $k_\perp,l_\perp$ as long as $a < 1$  \cite{Bauer:2008dt}.
 
In Secs.~\ref{sec:NLO} and \ref{sec:NLL} we calculate the above hard, jet, and soft functions for angularity distributions to next-to-leading order in $\alpha_s$, and solve for their dependence on $\mu$ through the renormalization group equations, which will allow us to sum large logarithms of $\tau_a$.

\subsection{Universal first moment of the soft function}

As shown in \cite{Lee:2006nr}, the behavior of the soft function \eq{softfunctiondef} under Lorentz boosts in the $\vect{n}$ direction implies a universal form for its first moment. The vacuum $\ket{0}$ and the Wilson lines $Y_{n,\bn}(0), \overline Y_{n,\bn}(0)$ are all invariant under such boosts, while the transverse momentum flow operator $\mathcal{E}_T(\eta)$ appearing in the definition of $\hat e_s$ transforms as $\mathcal{E}_T(\eta)\rightarrow \mathcal{E}_T(\eta')$ under a boost by rapidity $\eta' - \eta$.  These properties imply that the first moment of $S(e_s; \mu)$ is given by
\begin{equation}
\int \df e_s\,e_s \, S(e_s; \mu) = \frac {c_e\mathcal{A}(\mu)}{Q}\,,
\label{softmoment1}
\end{equation}
where
\begin{align}
c_e &= \frac{1}{Q}\int_{-\infty}^\infty \df \eta\,f_e(\eta)  \label{cedef} \\
\mathcal{A}(\mu) &=  \frac{1}{N_C}\Tr\bra{0}\overline Y_{\bar n}^\dag(0)Y_n^\dag(0) \mathcal{E}_T(0) Y_n(0)\overline Y_{\bar n}(0)\ket{0}\,.
\end{align}
The coefficient $c_e$ is exactly calculable from the definition of the event shape $e$ in \eq{eventshape} while $\mathcal{A}(\mu)$ is not fully calculable due to the contribution of nonperturbative effects, but is completely independent of the choice of variable $e$. The first moment \eq{softmoment1} is universal for all event shapes of the form \eq{eventshape} in this sense. For angularities, using \eq{ftauadef} and \eq{cedef},
\begin{equation}
c_a = \int_{-\infty}^\infty \df \eta\,\e^{-\abs{\eta}(1-a)} = \frac{2}{1-a}\,.
\label{c-a}
\end{equation}
This scaling of the first moment of the soft function for angularities will constrain the parameterization of the nonperturbative model for the soft function that we introduce in Sec.~\ref{sec:model}.

\section{Fixed-order Perturbative Calculations of Hard, Jet, and Soft Functions}
\label{sec:NLO}

In this section we calculate at next-to-leading  order, that is, $\mathcal{O}(\as)$, in perturbation theory the hard, jet, and soft functions, $H(Q;\mu),J_a^{n,\bn}(\tau_a^{n,\bn};\mu)$, and $S_a(\tau_a^s;\mu)$, in the factorization theorem for angularity distributions, which is given by \eq{finalformula} with $e = \tau_a$.\footnote{Note that here and below a superscript $n$ on a quantity is not a power but denotes ``$n$-collinear'' just as $\bn$ denotes ``$\bn$-collinear'' and $s$ denotes ``soft''.}

\subsection{Hard function at NLO}
\label{ssec:hard}

The hard function $H(Q;\mu)$, given by \eq{hardfunction}, is the squared amplitude of the two-jet matching coefficient $C_{n\bn}(Q,-Q;\mu)$. This matching coefficient was calculated, for example, in \cite{Manohar:2003vb} in the context of DIS and in \cite{Bauer:2003di} for $e^+e^-$ annihilation, to NLO.  It is found by calculating a matrix element of the QCD current \eq{QCDcurrent} and SCET current \eq{SCETcurrentO2} (for example, $\bra{q(p_q)\bar q(p_{\bar q})}j_i^\mu\ket{0}$), and requiring that the two match. Since the matching of the currents is independent of the observable being calculated, we do not need to repeat the matching calculation here, and simply quote the result. The matching coefficient $C_{n\bn}(\tilde p_n,\tilde p_\bn;\mu)$ in the SCET current \eq{SCETcurrentO2} is given by
\begin{equation}
\label{C2NLO}
C_{n\bn}(\tilde p_n,\tilde p_\bn ; \mu) = 1 - \frac{\alpha_s C_F}{4\pi}\left[ 8 - \frac{\pi^2}{6} + \ln^2\left(\frac{\mu^2}{2\tilde p_n\cdot\tilde p_\bn}\right) + 3\ln\left(\frac{\mu^2}{2\tilde p_n\cdot\tilde p_\bn}\right)\right]\,.
\end{equation}
Here and in the remainder of this section, $\as \equiv \as(\mu)$. The hard function $H(Q;\mu)$ in \eq{hardfunction} is thus
\begin{equation}
\label{hardNLO}
H(Q;\mu) = 1 - \frac{\as C_F}{2\pi}\left(8 - \frac{7\pi^2}{6} + \ln^2\frac{\mu^2}{Q^2} + 3\ln \frac{\mu^2}{Q^2}\right)\,.
\end{equation}
The additional contribution to the coefficient of $\pi^2$ in going from \eq{C2NLO} to \eq{hardNLO}  is due to the sign of $2\tilde p_n\cdot\tilde p_\bn = -Q^2$, following the conventions of \cite{Bauer:2001ct}.

The bare SCET two-jet operators in \eq{O2def} are renormalized by the relation
\begin{equation}
\label{O2renorm}
\mathcal{O}_{n\bn}^{\bare}(x;\tilde p_n,\tilde p_\bn) = Z_{\mathcal{O}}(\tilde p_n,\tilde p_\bn;\mu)\mathcal{O}_{n\bn}(x;\tilde p_n,\tilde p_\bn)\,,
\end{equation}
where the renormalization constant, calculated using dimensional regularization to regulate the UV divergences in $d= 4 - 2 \epsilon$ dimensions, is given by
\begin{equation}
\label{ZO}
Z_{\mathcal{O}}(\tilde p_n,\tilde p_\bn;\mu) = 1 + \frac{\as C_F}{4\pi}\left[ \frac{2}{\epsilon^2}  + \frac{2}{\epsilon}\ln\left(\frac{\mu^2}{2\tilde p_n\cdot\tilde p_\bn}\right) + \frac{3}{\epsilon}\right]\,.
\end{equation}

Matching the QCD current \eq{QCDcurrent} onto only two-jet operators in SCET is sufficient to describe accurately the two-jet region near $\tau_a = 0$ of angularity distributions. To calculate accurately also the tail region to $\mathcal{O}(\as)$, where the jets broaden and an additional jet begins to form, we would need to include a basis of three-jet operators in \eq{SCETcurrentO2} as well
\cite{Bauer:2006mk,Marcantonini:2008qn}. 
But since we are mainly interested in obtaining the correct shape of the two-jet region, we do not pursue this approach here. We will simply calculate the whole distribution in SCET with only two-jet operators, and then match the tail region numerically onto the fixed-order prediction of full QCD. This will be described more precisely in \ssec{match}. 

\subsection{Cutting rules for weighted matrix elements}
\label{ssec:taudelta}
The jet and soft functions that typically appear in factorizations of hard cross-sections in SCET are defined in terms of matrix elements of the products of collinear and soft fields, which are related to the imaginary part of the matrix element of a time-ordered product of the fields according to the optical theorem,
\begin{align}
\label{top}
\int \df^4 x\, \e^{iq\cdot x}\Mae{0}{\phi(x) \phi^\dagger(0)}{0} = \Disc \left[ \int \df^4 x \, \e^{iq\cdot x} \Mae{0}{{\rm T} \, \phi(x) \phi^\dagger(0)}{0}\right]
\,.\end{align}
The right-hand side is then related to the sum of all cuts of the relevant Feynman diagrams using the standard Cutkosky cutting rules. 

However, for more generic jet observables such as angularities for $a \neq 0$, the jet and soft functions that appear in factorization proofs contain matrix elements in which additional operators are inserted between the collinear and soft fields in the definition of the traditional jet and soft functions \cite{Bauer:2008jx}. For the matrix elements involving the extra insertion of such operators, we need to generalize the cutting rules for calculating these matrix elements from Feynman diagrams.

For the case of angularities, the jet and soft functions given in \eqss{softfunctiondef}{njetfunction}{nbarjetfunction} differ from the traditional jet and soft functions by the insertion of the delta function operator $\delta(\tau_a - \hat\tau_a)$. We denote the appropriate generalized prescription for calculating the new matrix element from the Feynman diagrams of  time-ordered perturbation theory as the ``$\tau_a$-discontinuity,''
\begin{align}
\label{general-taudisc}
\int \df^4 x\, \e^{iq\cdot x}\Mae{0}{\phi(x) \, \delta (\tau_a - \hat{\tau}_a) \, \phi^\dagger(0)}{0} \equiv \Disc_{\tau_a}\left[\int \df^4 x\, \e^{iq\cdot x} \Mae{0}{{\rm T} \phi(x) \phi^\dagger(0)}{0}\right]
\,.\end{align}
The $\Disc_{\tau_a}$ prescription is to cut the diagrams contributing to the matrix element of time-ordered operators just as for the usual matrix elements in \eq{top} but to insert an additional factor of $\delta(\tau_a - \tau_a(X))$ for each cut, where $X$ is the final state created by the cut.\footnote{The operator-based method that was developed in \cite{Ore:1979ry} for calculating weighted cross-sections can be used to relate matrix elements such as in the left-hand side of \eq{general-taudisc} directly to the ordinary discontinuity of matrix elements of time-ordered products of fields. However, for the scope of this paper, we choose simply to apply the prescription \eq{general-taudisc}.} This prescription corresponds to reinserting a sum over a complete set of final states between the delta function operator and $\phi^\dag(0)$ in \eq{general-taudisc}, and is precisely how we would calculate the full differential cross-section as written in \eq{QCDdist}. In the next two subsections we illustrate extensively the use of the $\Disc_{\tau_a}$ prescription.

\subsection{Calculation of the soft function to NLO}
\label{ssec:soft}

\FIGURE[t]{
\centerline{\mbox{ \hbox{\epsfysize=3.75truecm\epsfbox{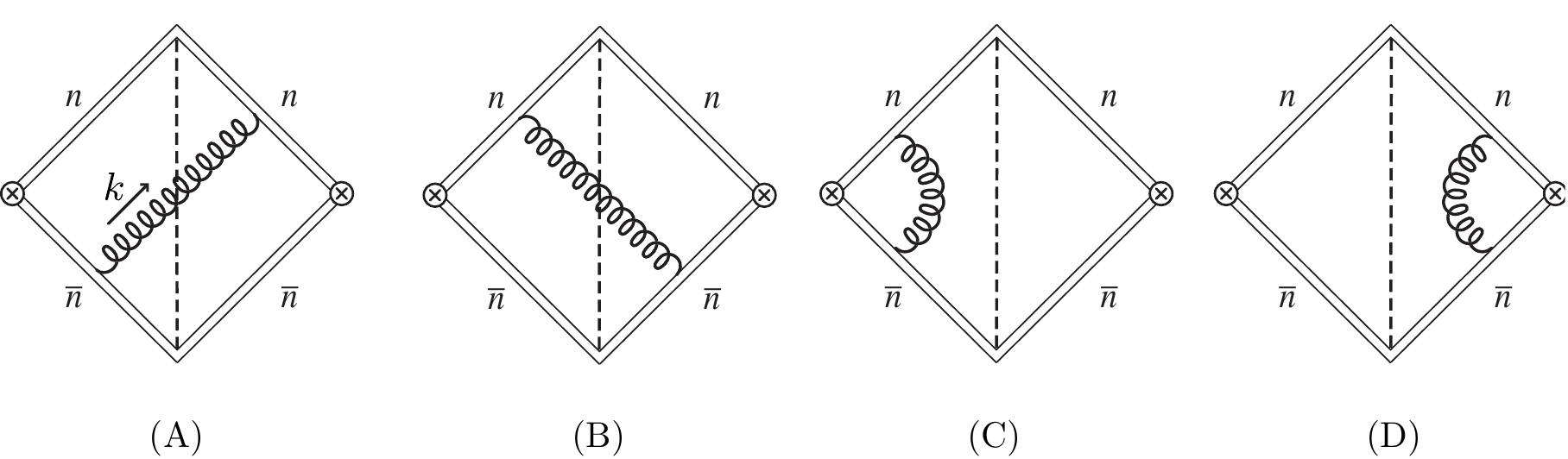}}  }}
\vspace{-.5cm}
{\caption[1]{The (A), (B) real and (C), (D) virtual contributions to the soft function. The gluons all have momentum $k$.}
\label{softfunction} }
}

The diagrams that contribute to the soft function are shown in \fig{softfunction}. From \eqss{ftauadef}{ehatdef}{ETfromT0i}, the contribution to the angularity from an on-shell soft gluon with momentum $k$ is
\begin{align}
\taus = \frac{\abs{\vect{k}_\perp}}{Q} e^{-\frac{1-a}{2} \abs{\ln \frac{k^+}{k^-}} } =
\begin{cases}
& \frac{1}{Q} \abs{k^+}^{1-\frac{a}{2}} \abs{k^-}^{\frac{a}{2}} \quad {\rm for} \quad k^-\geq k^+ \\
&\frac{1}{Q} \abs{k^-}^{1-\frac{a}{2}} \abs{k^+}^{\frac{a}{2}} \quad {\rm for} \quad k^+ \geq k^-
\end{cases}
\label{taucalc}
\,.\end{align}
Since cutting a gluon puts it on shell, the operator $\hattaus$ returns these values when acting on a cut soft gluon. When no gluon is in the final state cut, the operator $\hattaus$ simply returns zero. The real and virtual diagrams then contain delta functions, which we denote $\delta_R$ and $\delta_V$, respectively, 
\begin{subequations}
\begin{align}
\delta_R &\equiv \delta_R(\taus, k) = \theta(k^- - k^+) \, \delta\Bigl(\taus - \frac{1}{Q} \abs{k^+}^{1-\frac{a}{2}} \abs{k^-}^{\frac{a}{2}} \Bigr)  \\
&\qquad \qquad \qquad \; + \theta(k^+ - k^-) \, \delta\Bigl(\taus - \frac{1}{Q} \abs{k^-}^{1-\frac{a}{2}} \abs{k^+}^{\frac{a}{2}} \Bigr) \,, \nn\\
\delta_V &\equiv \delta_V(\taus) = \delta(\taus)
\,.\end{align}
\end{subequations}
In terms of these delta functions, the (bare) perturbative soft function can be written
\begin{align}
S^{\PT\bare}_{a}(\taus; \mu) =\delta(\taus) + 2 \softreala \, \delta_R + 2 \softvirta \delta_V 
\,,\end{align}
where we used that the tree-level contribution is just $\delta(\taus)$ and that the two real and the two virtual diagrams in Fig.~\ref{softfunction} give identical contributions.

In pure dimensional regularization, the virtual contributions are scaleless and hence vanish so we only need to evaluate the real diagrams. They add to
\begin{align}
\label{realsoft}
 2 \softreala \delta_R  &= 2 g^2 \mu^{2 \epsilon } \CF \, n \mcdot \bn \int\!\frac{\df^d k}{(2 \pi)^d} \frac{1}{k^-} \frac{1}{k^+} 2\pi \delta(k^- k^+ - \abs{\vect{k}_\perp}^2) \, \theta(k^-) 
 \delta_R(\taus, k) 
\,. \end{align}
Performing the $k$ integrals gives
\begin{align}
S^{\PT\bare}_{a}(\taus; \mu) &= \delta(\taus) + \theta(\taus) \frac{\as \CF \, n \mcdot \bar n}{\pi (1-a)} \left ( \frac{4\pi \mu^2}{Q^2} \right )^\epsilon \frac{1}{\Gamma(1-\epsilon) } \frac{1}{\epsilon}\left (\frac{1}{\taus} \right)^{1+ 2 \epsilon} 
 \label{softplus}
\,.\end{align}
Nonzero values of $\taus$ regulate the IR divergences, and so here the $1/\epsilon$ pole is of UV origin, $\epsilon = \eUV$.

Applying the distribution relation (valid for $\epsilon<0$)
\begin{align}
\frac{\theta(x)}{x^{1+2 \epsilon}} = - \frac{\delta(x)}{2 \epsilon} + \left[ \frac{\theta(x)}{x}\right]_+  - 2 \epsilon \left[ \frac{\theta(x) \ln{x}}{x} \right]_+ + \mathcal{O}(\epsilon^2)
\label{omegaexpansion}
\,,\end{align}
where
\begin{equation}
     \left[\frac{\theta(x)\ln^n(x)}{x}\right]_+ \equiv \, \lim_{\beta\to 0} \left[\frac{\theta(x-\beta) \ln^n(x)}{x}+\frac{\ln^{n+1}\beta}{n+1} \delta(x-\beta)\right]
     \label{lognplusdef}
\,,\end{equation}
to \eq{softplus} we obtain the final result for the (bare) angularity soft function,
\begin{align}
S^{\PT\bare}_{a}(\taus; \mu) 
&= \int\!\df\tausp\, Z_S(\taus - \tausp ; \mu) S_a (\tausp ; \mu)
\label{softNLObare}
\,,\end{align}
where to NLO the renormalized soft function, $S_a^{\PT}$, is given by 
\begin{align}
S_a^{\PT}(\taus; \mu)  &= \delta(\taus) \left [1- \frac{\as \CF}{ \pi (1-a)}\left ( \frac{1}{2}\ln^2{\frac{\mu^2}{Q^2}} -\frac{\pi^2}{12} \right ) \right ] +\frac{2 \as \CF}{ \pi (1-a)} \left [ \frac{\theta(\taus)}{\taus} \ln{\frac{\mu^2}{(Q \taus)^2}}\right ]_+  
\label{softNLO}
\,,\end{align}
and the renormalization factor, $Z_S$, is given by
\begin{align}
Z_S(\taus; \mu) &= \delta(\taus ) \left[ 1- \frac{\as \CF}{\pi(1-a)} \left( \frac{1}{\epsilon^2} + \frac{1}{\epsilon} \ln{\frac{\mu^2}{Q^2}}\right) \right]  + \frac{1}{\epsilon} \frac{2 \as \CF}{\pi (1-a)}\left[ \frac{\theta(\taus )}{\taus}\right]_+
\label{Zsoft}
\,.\end{align}

\subsection{IR structure of the soft function}
\label{ssec:softIR}

While the mathematical identity in \eq{omegaexpansion} allowed us to arrive at our final result, \eq{softNLO}, the origin of the $1/\epsilon$ poles became obscured through its use. In fact, the use of \eq{omegaexpansion} is only valid for $\epsilon<0$ which suggests that the $1/\epsilon$ pole on the right-hand side of \eq{omegaexpansion} is of IR origin. The virtual diagrams, while formally zero in pure dimensional regularization, play the role of converting this IR divergence into a UV divergence by adding a quantity proportional to $(1/\epsilon_{\rm UV} - 1/\epsilon_{\rm IR})$ to the coefficient of $\delta(\taus)$, if the final result is in fact free of IR divergences. Na\"{\i}vely it seems that this conversion cannot possibly occur for arbitrary $a$, because the $1/\epsilon$ poles in the real diagrams have $a$-dependent coefficients (see \eq{Zsoft}), while the virtual diagrams contain no apparent $a$ dependence.   Nevertheless, by carefully examining the contribution of both the real and virtual diagrams, we will show that, for $a<1$, the virtual diagrams play precisely this role and convert each IR divergence in the real graphs into UV, but that for $a\geq 1$, this cancellation is incomplete.  This is accomplished through an analysis of integration regions in the loop momentum integrals that avoids the use of explicit IR regulators. Our presentation here complements our discussion of these issues in \cite{Hornig:2009kv}.

Using that 
$
\int_0^1\! \df x \left[\ln^n(x)/x\right]_+ =0 ,
$
the contribution to the coefficient of $\delta(\taus)$ can be isolated by integrating the diagrams over $\taus$ from 0 to 1. We find that the contribution from the real diagrams can be written as 
\begin{align}
\label{realregion}
\int_0^1 \df \taus \Bigg[ 2 \softreala \delta_R \Bigg] = \!\frac{\as \CF \, n \mcdot \bar n}{2 \pi}  \frac{\left ( 4\pi \mu^2 \right )^\epsilon}{\Gamma(1-\epsilon) } \int_{\cR}\! \df k^+ \df k^- (k^+ k^-)^{-1-\epsilon}
\,, \end{align}
where $\cR$ is given by the region of positive $k^{+}$ and $k^-$ such that
\begin{align}
 & \qquad (k^-)^\frac{a}{2}(k^+)^{1-\frac{a}{2}} < Q \quad \text{for} \quad k^-\geq k^+ \nn\\
& \qquad (k^+)^\frac{a}{2}(k^-)^{1-\frac{a}{2}} < Q \quad \text{for} \quad  k^-\leq k^+
\,.\end{align}
This region is plotted in \fig{kpluskminus}A for various values of $a$.

\FIGURE[t]{
\includegraphics[totalheight=.5\textheight]{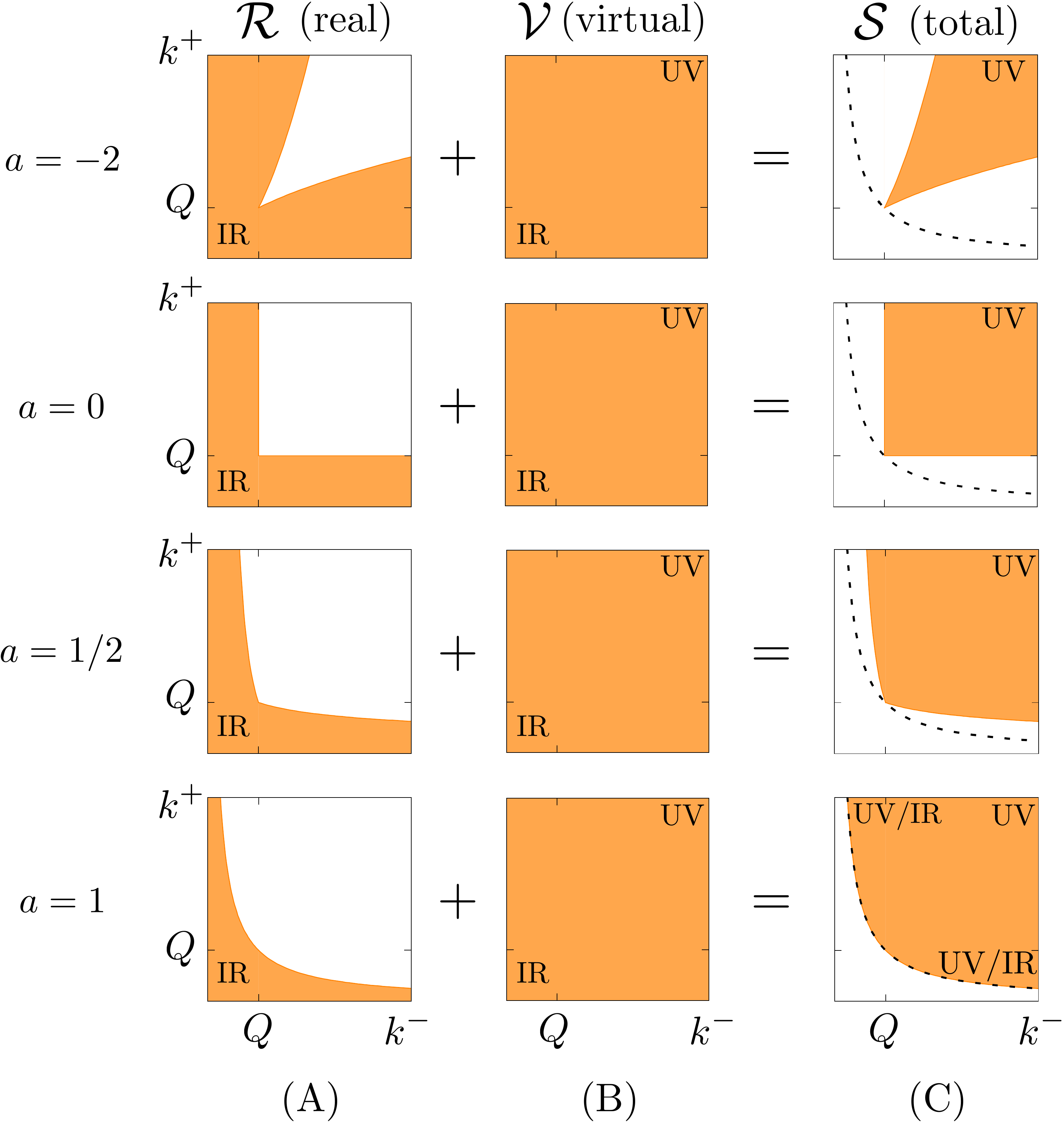}
\hfill
\includegraphics[totalheight=.5\textheight]{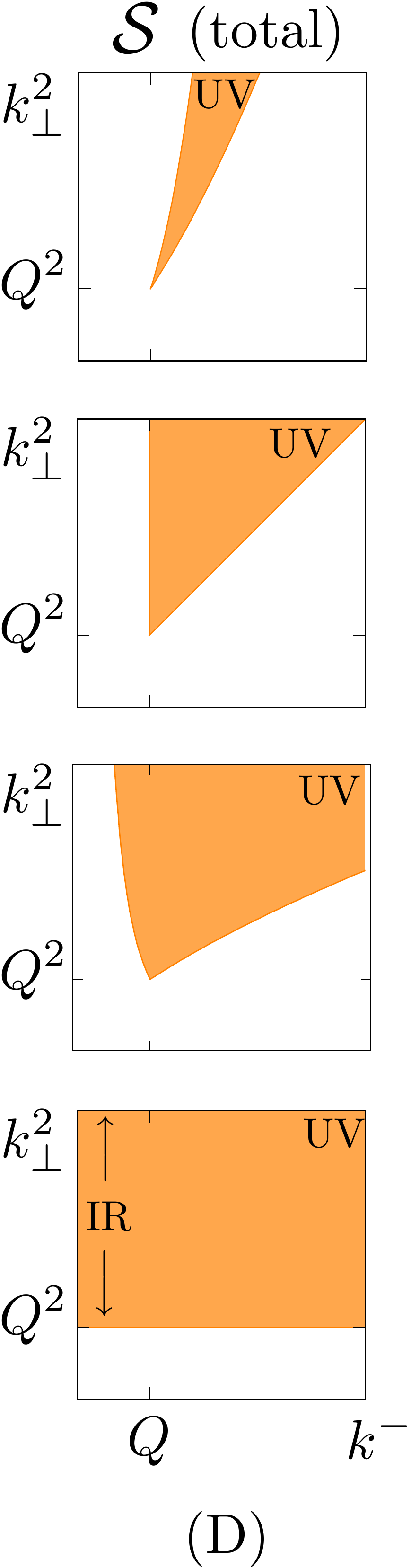}
{\caption[1]{The regions of integration for the coefficent of $\delta(\taus)$ in $S^{\bare}_a(\taus)$ in the (A), (B), (C) $k^-\!,k^+$ and (D) $k^-\!, \vect{k}_\perp^2 $ planes. The regions of integration for both (A) the real contribution $\cR$ and (B) the virtual contribution $\cV$ contain both UV and IR divergences. Since the integrands for the two contributions differ only by an overall minus sign, (C) the region resulting in their sum $\cS$, is the complement of $\cR$ and contains only UV divergences for $a<1$. The dashed line in (C) represents the line of constant $k^+ k^-= Q^2$.}
\label{kpluskminus} }
}

The contribution of the virtual diagrams to the coefficient of $\delta(\taus)$ sums to
\begin{align}
\int_0^1 \df \taus \Bigg[ 2 \softvirta \delta_V \Bigg] &= 2 g^2 \mu^{2 \epsilon } \CF \, n \mcdot \bn \int\!\frac{\df^d k}{(2 \pi)^d} \frac{1}{k^- - i 0^+} \frac{1}{k^+ + i 0^+} \frac{i}{k^+ k^- -\abs{\vect{k}_\perp}^2 - i 0^+} \nn\\
&=-\frac{\as \CF \, n \mcdot \bar n}{2 \pi} \frac{ \left ( 4\pi \mu^2 \right )^\epsilon}{\Gamma(1-\epsilon) } \int_\cV \! \df k^+ \df k^- (k^+ k^-)^{-1-\epsilon}
\label{virtregion}
\,,\end{align}
where $\cV$ is the entire positive $k^{+}, k^{-}$ quadrant, plotted in \fig{kpluskminus}B.

The two contributions to $\delta(\taus)$, \eqs{realregion}{virtregion}, are each both UV and IR divergent, but as we will show, their sum is convergent for $\epsilon >0$ and so is only UV divergent. Since the form of the integrand is the same and the virtual contribution differs only by an overall minus sign, it converts the region of integration of the real contribution, $\cR$, into the complementary part of the positive $k^+, k^-$ quadrant (see \fig{kpluskminus}) which does not include the IR divergent regions $k^{\pm} \to 0$. Note that as $a \rightarrow 1$, 
the boundary of the region of integration $\cR$
approaches the curve of constant $k^+ k^-= Q^2$. With this boundary, the integral over the region $\cS$ does not converge for either positive or negative $\epsilon$, implying that both IR and UV divergences are present. 

That the region $\cS$ has only UV divergence for $a<1$ and has both UV and IR divergence for $a=1$ is perhaps more clearly seen in the $k^-\!, \vect{k}_\perp^2 $ plane. The integral of the soft diagrams over $\taus$ in terms of these variables is given by
\begin{align}
 \int_0^1\! \df \taus \, \Bigg[ 2 \softreala \delta_R+ 2 \softvirta \delta_V \Bigg] = - \frac{\as \CF \, n \mcdot \bar n}{2 \pi}  \frac{\left ( 4\pi \mu^2 \right )^\epsilon}{\Gamma(1-\epsilon) } \int_{\cS}  \frac{\df k^- \df \vect{k}_\perp^2}{k^- ( \vect{k}_\perp^2)^{1+\epsilon} } \nn
\,,\end{align}
and the resulting region $\cSperp$ in terms of $k^-$ and $\vect{k}_\perp^2$ for $a \le 1$ is
\begin{align}
\left( \frac{\vect{k}_\perp^2}{Q^2} \right)^{-\frac{a}{2(1-a)}}< \left( \frac{k^-}{Q} \right) <\left( \frac{\vect{k}_\perp^2}{Q^2} \right)^{\frac{2-a}{2(1-a)}} \quad {\rm with} \quad \vect{k}_\perp^2>Q^2
\,.\end{align}
The region $\cSperp$ is plotted for several values of $a$ in \fig{kpluskminus}D. The limiting case $a=1$ clearly includes the IR divergent region $k^- \rightarrow 0$ for all $\vect{k}_\perp^2 > Q^2$.

Performing the integral over $\cS$ we obtain
\begin{align}
\int_0^1\! \df \taus \, S^{ \bare}_a(\taus; \mu) = 1-\frac{\as \CF \, n \mcdot \bar n}{2 \pi(1-a)} \left ( \frac{ 4\pi \mu^2 }{Q^2} \right )^\epsilon \frac{1}{\epsilon^2 \, \Gamma(1-\epsilon) }
\label{totalsoftdelta}
\,.\end{align}
After expanding \eq{totalsoftdelta} in $\epsilon$, we find that the coefficient of $\delta(\taus)$ in \eq{softNLO} is unchanged, except that for $a<1$ all the $1/\epsilon$ poles are unambiguously of UV origin. 

A lesson from this analysis is that in pure dimensional regularization, the coefficient of $(1/\eUV-1/\eIR)$ in a virtual diagram cannot be determined from the virtual diagram alone, but only together with the real diagram whose IR divergence it is supposed to cancel. The reason that the virtual subtraction can depend on $a$ even though by itself it is independent of $a$ is that the area of overlap between the integration regions of real and virtual diagrams depends on $a$. 

\subsection{Calculation of the jet functions to NLO}
\label{ssec:jet}

Now we proceed to calculate the jet functions given by \eqs{integratedjetfunctions}{jetfunctions}. The diagrams that contribute to $J_a^n$ are shown in \fig{jet-function}, and the Feynman rules necessary to calculate these diagrams are found in \cite{Bauer:2000yr}. The total momentum flowing through each diagram is $Qn/2 + l$, with the label component $Qn/2$ specified by the labels on the jet fields in the matrix elements in \eq{njetfunction}, and $l$ the residual momentum. The total momentum of the gluon in each loop is $q$, which has both label and residual components.
All results for the anti-quark jet function $J_a^{\bar n}$ can be found from those for the quark jet function $J_a^n$ with the replacement $n \leftrightarrow \bar n$ and so we calculate explicitly only $J_a^n$. 

Cutting the diagrams in Fig.~\ref{jet-function} in all possible places, we can cut through the gluon loops or through one of the individual quark propagators connected to a current. We naturally call these classes of cut diagrams ``real'' and ``virtual'' respectively. The real and virtual diagrams contain the delta functions,
\begin{align}
\delta_R & \equiv \delreal{R} \equiv \delta \Big( \taun - \frac{1}{Q} \left[(q^-)^{\frac{a}{2}}(q^+)^{1-\frac{a}{2}} +(Q-q^-)^{\frac{a}{2}}(l^+-q^+)^{1-\frac{a}{2}} \right]  \Big) \,,\nn\\
\delta_V & \equiv \delvirt{V} \equiv \delta \Big(\taun - \big( l^+ /Q \big)^{1-\frac{a}{2}} \Big)\,,
\label{deltaRVsoft}
\end{align}
which are obtained using \eq{taucalc}.  
In this case we simply consider the contribution to $\tau_a$ from a final state with a single on-shell collinear quark of momentum $l$ for $\delta_V$ and from a final state consisting of an on-shell collinear gluon of momentum $q$ together with an on-shell collinear quark of momentum $l - q$ for $\delta_R$, and use that the `$-$' component of momentum is always larger than the `$+$' component for on-shell collinear particles. The momentum $l$ flowing through the diagrams in Fig.~\ref{jet-function} has a label component which is fixed to be $Qn/2$ by the labels on the collinear fields in the matrix element in \eq{njetfunction}.

\FIGURE[t]{
\centerline{\mbox{ \hbox{\epsfysize=1.75truecm\epsfbox{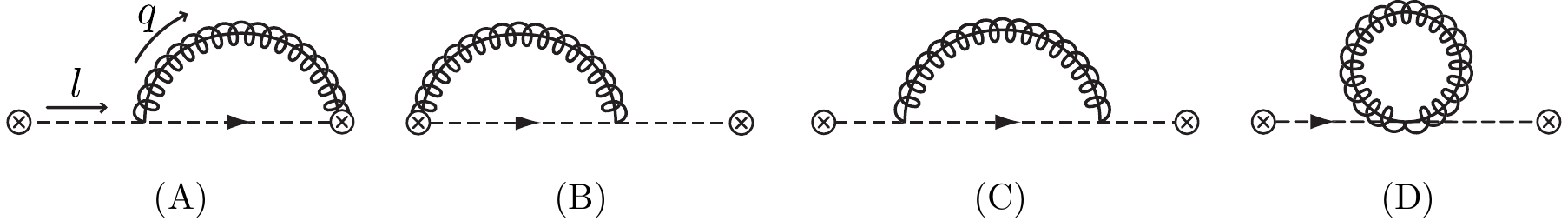}}  }}
\vspace{-.5cm}
{ \caption[1]{Diagrams contributing to the angularity jet
function $\mathcal{J}_a^n(\taun, l^+)$ with incoming momentum $l = \frac{n}{2} Q + \frac{\bar n}{2} l^+$ and gluon momentum $q$: (A) Wilson line emission diagram and (B) its mirror; (C) sunset and (D) tadpole QCD-like diagrams. The contributions to the jet function $J_a^n(\taun)$ are given by the integrals of these diagrams over the $+$ component of the incoming momentum, $ \int\!\df l^+ \mathcal{J}_a^n(\taun, l^+) = 2 \pi J_a^n(\taun) $.}
\label{jet-function} }
}

Before turning to evaluate the diagrams in \fig{jet-function}, we first perform a few simplifications to facilitate the computation. First, we note that the Wilson line emission diagram, \fig{jet-function}A, and its mirror, \fig{jet-function}B, give identical contributions. Second, we employ the
fact that the number and complexity of jet function diagrams needed in loop calculations is reduced by noticing that the QCD-like diagrams can be computed using ordinary QCD Feynman rules with appropriate insertions of the projection operators $P_n = \nslash\bnslash/4$ and $P_{\bar n} = \bnslash\nslash/4$ \cite{Becher:2006qw, Bauer:2008qu}. In particular, for our one-loop example we use that the sum of \fig{jet-function}C and \fig{jet-function}D reduces to
\begin{align}
\QCDa + \QCDb = \QCDtilde 
\label{QCDtilde}
\,.\end{align}
Next, we relate the $\taun$-discontinuity to the ordinary discontinuity,
\begin{align}
\label{taudisc}
&  {\rm Disc}_{\, \taun} \!\Big[ 2 \Wilsonb + \QCDtilde \Big]  \\
& \equiv  \!\Big[ 2 \Wilsonbreal + \QCDtildereal \Big]  \delta_{R}  + \Big[ 2 \Wilsonbvirt + 2\QCDtildevirt \Big] \delta_{V} \nn\\
& =    {\rm Disc} \Big[ 2 \Wilsonb + \QCDtilde \Big]  \delta_V +\Big[  2 \Wilsonbreal + \QCDtildereal \Big]  \Big( \delta_{R}- \delta_{V} \Big) \nn
\,,\end{align}
where in the third line the we used that the real diagrams induced by taking the discontinuity in the first term cancel the coefficient of $\delta_V$ in the second term. 

Now, since $\delvirt{V}$ has no dependence on the loop momentum $q$, it factors out of the $\df^d q$ integrand. This implies that, after adding the tree-level contribution to the one-loop $\taun$-discontinuity in \eq{taudisc}, we can write the NLO jet function as
\begin{align}
\label{jscript}
\mathcal{J}^{n \bare}_a(\taun, l^+; \mu) \frac{\nslash}{2} &= 2 \pi \delta(l^+) \delta(\taun) \frac{\nslash }{2} + {\rm Disc}_{\, \taun} \!\Big[2 \Wilsonb + \QCDtilde \Big] \\
&= J^{n\bare}(l^+; \mu) \frac{\nslash }{2} \delta_V+\Big[  2 \Wilsonbreal + \QCDtildereal \Big]  \Big( \delta_{R}- \delta_{V} \Big) \nn
\,,\end{align}
where $J^{\bare}_n(l^+; \mu)$ is the standard jet function \cite{Bauer:2001yt},
\begin{align}
J^{n\bare}(l^+; \mu) \frac{\nslash }{2} &  \equiv \frac{1}{N_C} {\rm Disc} \left[ \int\!\df^4 x \, e^{i l \cdot x} \,{\rm Tr}  \, \Mae{0}{ {\rm T} \, \chi_{n, Q} (  x  ) \, \bar{\chi}_{n, Q} (0) }{0} \right] \nn\\
& = 2 \pi \delta(l^+) \frac{\nslash }{2} +  {\rm Disc} \Big[ 2 \Wilsonb + \QCDtilde \Big] +\mathcal{O}(\as^2)\,,
\label{oldjetfunction}
\end{align}
containing no additional operator insertions. Each term on the second line of \eq{jscript} is then well-defined\footnote{By this we mean that had we evaluated the individual cut virtual QCD-like diagrams contained in the first line of \eq{jscript} directly, we would have encountered the  complication of cutting one lone quark propagator and thus putting  the second lone, uncut quark propagator on shell also.}  and straightforwardly calculable.  In fact, $J^n(l^+; \mu)$ has been calculated to two loops~\cite{Becher:2006qw}, and we expect that the techniques we employed above are the most practical way to extend our results to two loops. The additional term on the second line of \eq{jscript} is a sum of real emission diagrams containing a difference of the delta functions $\delta_R$ and $\delta_V$.  Note that for the special case $a=0$, $\delvirt{V} = \delreal{R}$ and this additional term vanishes, so $J^n = J^n_{a=0}$. This is why only the standard jet function is needed when $a=0$. 

To find the angularity jet function $J_a^n(\taun; \mu)$, we must integrate \eq{jscript} over $l^+$ as in \eq{integratedjetfunctions}, 
\begin{align}
J^{n \bare}_a(\taun; \mu) = \int\!\frac{\df l^+}{2 \pi} \mathcal{J}^{n \bare}_a(\taun, l^+; \mu)
\,.\end{align}
By integrating the known one-loop expression for $J^{\bare}_n(l^+; \mu)$ (see, e.g., \cite{Bauer:2003pi,Bosch:2004th}), we find that the contribution of the first term in \eq{jscript} is
\begin{align}
\label{aeq0-contribution}
\int\!\frac{\df l^+}{2 \pi} J^{n \bare}(l^+; \mu) \, \delta_{V} &=  \delta(\taun) \bigg \{ 1+ \frac{\as \CF}{4 \pi} \bigg[ \frac{4}{ \epsilon^2} + \frac{3}{\epsilon} +  \frac{4}{\epsilon} \ln{\frac{\mu^2}{Q^2}}+ 2 \ln^2{\frac{\mu^2}{Q^2}} \nn\\
& \qquad \qquad \qquad \qquad + 3 \ln{\frac{\mu^2}{Q^2}} + 7 - \pi^2 \bigg] \bigg \} \nn\\
& \qquad - \frac{1}{1-a/2} \bigg[ \bigg( \frac{4}{\epsilon}+ 3 + 8 \ln{\frac{\mu}{Q (\taun)^{1/(2-a)}}} \bigg) \bigg( \frac{\theta(\taun)}{\taun} \bigg) \bigg]_+
\,.\end{align}
It is well known that all $1/\epsilon$ poles in this expression are of UV origin.

We find that the term involving the real QCD-like diagram in \eq{jscript} is
\begin{align}
\label{qcdtildereal}
 &\int\!\frac{\df l^+}{2 \pi} \Big[  \QCDtildereal \Big] \Big( \delta_{R} - \delta_{V} \Big) \\
 & \quad = -g^2 \mu^{2 \epsilon} \CF (d-2) \frac{\nslash}{2} \int\!\frac{\df l^+}{2\pi} \left( \frac{1}{l^+} \right)^2 \int\!\frac{\df^d q}{(2\pi)^d} (l^+-q^+) \nn\\
 & \quad \qquad \times \left ( (-2\pi i) \delta(q^+ q^- - \abs{\vect{q}_\perp}^2) \theta(q^-) \right )  \frac{1}{q^-} \Big( \delreal{R} - \delvirt{V} \Big)  \nn\\
& \quad \qquad \times \left ( (-2\pi i) \delta \big((Q-q^-)(l^+-q^+) - \abs{\vect{q}_\perp}^2 \big) \theta(Q-q^-)  \right )\nn\\
& \quad = \frac{\as \CF }{2 \pi (2-a)} \frac{\nslash}{2} \left ( \frac{4 \pi \mu^2}{Q^2}\right )^\epsilon \frac{2(1-\epsilon)}{\Gamma(1 - \epsilon)} \left( \frac{1}{\taun} \right) ^{1+ \frac{\epsilon}{1-a/2}} \nn\\
& \qquad \quad \times \int_0^1\!\df x \, x \left[ \left( x^{a-1} + (1-x)^{a-1} \right)^{\frac{\epsilon}{1-a/2}}  -  \left(x(1-x) \right)^{-\epsilon} \right]  \nn
\,,\end{align}
where we defined $x \equiv q^-/Q$. This expression is finite as $\epsilon \to 0$.

For the term involving the real Wilson line diagram, we find
\begin{align}
\label{wilsonreal}
 &\int\!\frac{\df l^+}{2 \pi} \Big[ \Wilsonbreal \Big] \Big( \delta_{R} - \delta_{V} \Big) \\
 & \quad = -g^2 \mu^{2 \epsilon} \CF \, n \mcdot \bar n  \frac{\nslash}{2} \int\!\frac{\df l^+}{2\pi} \frac{1}{l^+} \int\!\frac{\df^d q}{(2\pi)^d} \frac{1}{q^-} \left ( (-2\pi i) \delta(q^+ q^- - \abs{\vect{q}_\perp}^2) \theta(q^-) \right )\nn\\
& \quad \qquad \times \bigg[ (Q-q^-) \Big ( (-2\pi i) \delta \big((Q-q^-)(l^+-q^+) - \abs{\vect{q}_\perp}^2 \big) \theta(Q-q^-) \Big)  \nn\\
& \qquad \qquad \quad - Q \Big ( (-2\pi i)\delta \big(Q(l^+-q^+) \big) \Big)\bigg] \Big( \delreal{R} - \delvirt{V} \Big)  \nn
\,.\end{align}
The piece with $\delta_R$ can be written as
\begin{align}
\label{wilsonrealR}
\int\!\frac{\df l^+}{2 \pi} \Big[ \Wilsonbreal \Big] \delta_{R} &= \theta(\taus)\frac{\as \CF n \mcdot \bar n}{2 \pi (2-a)} \frac{\nslash}{2} \left ( \frac{4 \pi \mu^2}{Q^2}\right )^\epsilon \frac{1}{\Gamma(1 - \epsilon)} \left( \frac{1}{\taun} \right) ^{1+ \frac{\epsilon}{1-a/2}} \\
& \quad\times  \left [  \int_0^1\!\frac{\df x}{x} \, (1-x)  \left( x^{a-1} + (1-x)^{a-1} \right)^{\frac{\epsilon}{1-a/2}}  - \int_0^{\infty} \!\frac{\df x}{x} x^{-\epsilon\frac{1-a}{1-a/2}} \right] \nn
\,,\end{align}
and the piece with $\delta_V$ is
\begin{align}
\label{wilsonrealV}
\int\!\frac{\df l^+}{2 \pi} \Big[ \Wilsonbreal \Big] \delta_{V} &= \theta(\taus)\frac{\as \CF n \mcdot \bar n}{2 \pi (2-a)} \frac{\nslash}{2} \left ( \frac{4 \pi \mu^2}{Q^2}\right )^\epsilon \frac{1}{\Gamma(1 - \epsilon)} \left( \frac{1}{\taun} \right) ^{1+ \frac{\epsilon}{1-a/2}} \nn\\
& \quad \times  \left [  \int_0^1\!\frac{\df x}{x} \, (1-x) \left( x(1-x) \right)^{-\epsilon}  - \int_0^{\infty} \!\frac{\df x}{x} x^{-\epsilon} \right] 
\,.\end{align}
The second term in brackets in each of \eqss{wilsonreal}{wilsonrealR}{wilsonrealV} corresponds to the zero-bin subraction \cite{Manohar:2006nz} needed to avoid the double counting of soft modes \cite{Lee:2006nr,Idilbi:2007ff,Idilbi:2007yi}. Note that from the expressions in both \eqs{wilsonrealR}{wilsonrealV}, the zero-bin contributions are scaleless and hence formally zero. Their role is to convert the IR divergence ($q^- \to 0$) in each integrand into a UV divergence ($q^- \to \infty$) for $a<1$. After this subtraction, both of the integrals over $x$ in brackets are convergent for $\epsilon>0$.

Subtracting \eq{wilsonrealV} from \eq{wilsonrealR} and performing the integral over $x$ we find that 
\begin{align}
\label{R-Vcounterterms}
&\int\!\frac{\df l^+}{2\pi}{\rm Disc} \Big[  2 \Wilsonbreal + \QCDtildereal \Big]  \Big( \delta_{R}- \delta_{V} \Big) \\
& \quad = - \frac{\as \CF }{2 \pi (2-a)} \frac{\nslash}{2} \left ( \frac{4 \pi \mu^2}{Q^2}\right )^\epsilon \frac{1}{\Gamma(1 - \epsilon)}\frac{1}{\epsilon} \left( \frac{1}{\taun} \right)^{1 + \frac{\epsilon}{1-a/2}} \bigg[ \frac{2a}{1-a} + \epsilon^2 \frac{2a(\pi^2-9)}{3(2-a)}  \nn\\
& \qquad \qquad \qquad \quad- \epsilon^2 \frac{4}{1-a/2} \int_0^1\!\df x \frac{1-x +x^2/2}{x} \ln[(1-x)^{1-a}+x^{1-a}]  + \mathcal{O}(\epsilon^3) \bigg] \nn
\,,\end{align} 
where the overall $1/\epsilon$ pole is of UV origin from the discussion above.

Applying the relation \eq{omegaexpansion} to \eq{R-Vcounterterms} and adding the result to \eq{aeq0-contribution}, we arrive at our final expression for the (bare) NLO angularity jet function,
\begin{align}
\label{jetNLObare}
J_a^{n\bare} (\taun; \mu) 
&= \int\!\df\taunp\, Z_J(\taun - \taunp; \mu) J_a^n(\taunp; \mu)
\,,\end{align}
where the renormalized jet function, $J_a^n$, is
\begin{align}
\label{jetNLO}
J_a^n (\taun; \mu) &=  \delta(\taun) \bigg \{ 1+ \frac{\as \CF}{\pi} \bigg [  \frac{1-a/2}{2(1-a)} \ln^2{\frac{\mu^2}{Q^2}} + \frac{3}{4} \ln{\frac{\mu^2}{Q^2}} + f(a) \bigg] \bigg \}\nn\\
&  \qquad - \frac{\as \CF}{\pi} \left [ \bigg( \frac{3}{4} \frac{1}{1-a/2}+\frac{2}{1-a}\ln{\frac{\mu}{Q (\taun)^{1/(2-a)}}}\bigg )\bigg ( \frac{\theta(\taun)}{\taun}\bigg )\right ]_+ 
\,,\end{align}
where we defined
\begin{align}
\label{fa}
f(a) \equiv & \,\frac{1}{1-a/2}\bigg ( \frac{7-13a/2}{4} - \frac{\pi^2}{12}\frac{3 - 5a + 9a^2/4}{1-a} \nn\\
& \qquad \qquad -\int_0^1\!\df x \frac{1-x +x^2/2}{x} \ln[(1-x)^{1-a}+x^{1-a}] \bigg )
\,,\end{align}
and the $Z$-factor is given by
\begin{align}
\label{Zjet}
Z_{J}(\taun; \mu) &= \delta(\taun) \left[ 1+ \frac{\as \CF}{\pi}\left( \frac{1-a/2}{1-a} \left( \frac{1}{\epsilon^2} + \frac{1}{\epsilon} \ln{\frac{\mu^2}{Q^2}}\right) + \frac{3}{4 \epsilon}\right)\right]  - \frac{1}{\epsilon} \frac{\as \CF}{\pi (1-a)}\left[ \frac{\theta(\taun)}{\taun}\right]_+
\,.\end{align}

\subsection{IR structure of the jet functions}
\label{ssec:jetIR}

As we showed in \sec{ssec:jet}, the $1/\epsilon$ pole in front of the plus-distribution corresponds to a UV divergence. However, as we discussed for the case of the soft function in \sec{ssec:softIR}, the use of \eq{omegaexpansion} means that we can not immediately make the same claim for the poles in the coefficient of  $\delta(\taun)$. We now perform an analysis similar to that in \sec{ssec:softIR} by integrating over $0<\taun<1$ to isolate this coefficient and study its divergent structure in the resulting $q^-$, $\vect{q}_\perp^2$ integration regions, complementing our discussion in \cite{Hornig:2009kv}.

The diagrams (C) and (D) in Fig.~\ref{jet-function}, being equivalent to diagrams in full QCD as noted above, are manifestly infrared-finite and do not need to be analyzed in further detail. The Wilson line graphs (A) and (B) potentially contain infrared divergences that we must identify more carefully.

If the jet function is infrared-safe, infrared divergences in virtual and real diagrams, with proper zero-bin subtractions taken, should cancel and leave purely UV divergent integrals. The contribution of the sum of the real and virtual Wilson line diagrams to the coefficient of $\delta(\tau_a^n)$ in the jet function $J_a^{n\bare}(\tau_a^n)$ is
\begin{align}
2 \int_0^1 &\df \tau_a\int \frac{\df l^+}{2\pi} \Big[\Wilsonbvirt \delta_V + \Wilsonbreal \delta_R\Big] \\
&= - \frac{\as C_F}{\pi}\frac{(4\pi\mu^2)^\epsilon}{\Gamma(1-\epsilon)} \Biggl[\int_{\tilde{\mathcal{J}}} \df q^- \df \vect{q}_\perp^2 \frac{1}{(\vect{q}_\perp^2)^{1+\epsilon}}\left(\frac{1}{q^-} - \frac{1}{Q}\right)
- \int_{\mathcal{J}_0} \df q^- \df \vect{q}_\perp^2 \frac{1}{(\vect{q}_\perp^2)^{1+\epsilon}}\frac{1}{q^-} \Biggr] \,,\nn
\end{align}
where the last integral  is the zero-bin subtraction of the na\"{\i}ve collinear integral in the first term. The na\"{\i}ve integration region $\tilde{\mathcal{J}}$ is shown in Fig.~\ref{jet-regions} and is given by $0<q^-<Q$ and
\begin{equation}
\vect{q}_\perp^2 > \left\{Q \left[\frac{1}{(Q-q^-)^{1-a}} + \frac{1}{(q^-)^{1-a}}\right]^{-1}\right\}^{\frac{1}{1-a/2}}\,.
\end{equation}
The zero-bin region $\mathcal{J}_0$ is given by $q^->0$ and
\begin{equation}
\vect{q}_\perp^2 > \left[ Q(q^-)^{1-a}\right]^{\frac{1}{1-a/2}}\,.
\end{equation}
The resulting integral for the total contribution of the zero-bin-subtracted Wilson line diagrams to the coefficient of $\delta(\tau_a^n)$ in the jet function is
\begin{align}
\label{taunUV}
2 \int_0^1 \df \tau_a\int \frac{\df l^+}{2\pi} \bigg[\Wilsonbvirt \delta_V &+ \Wilsonbreal \delta_R\bigg] \\
= -\frac{\as C_F}{\pi}\frac{(4\pi\mu^2)^\epsilon}{\Gamma(1-\epsilon)} \Biggl[&\int_{{\mathcal{J}}} \df q^- \df \vect{q}_\perp^2 \frac{1}{(\vect{q}_\perp^2)^{1+\epsilon}}\frac{\sgn(q^- - Q)}{q^-} 
- \int_{\tilde{\mathcal{J}}} \df q^- \df \vect{q}_\perp^2 \frac{1}{(\vect{q}_\perp^2)^{1+\epsilon}}\frac{1}{Q} \Biggr]\,, \nn
\end{align}
where the region $\mathcal{J}$ resulting from combining $\tilde{\mathcal{J}}$ and $\mathcal{J}_0$, with a relative minus sign in the integrands,  is also shown in Fig.~\ref{jet-regions}. 
\FIGURE[t]{
\includegraphics[width=1\textwidth]{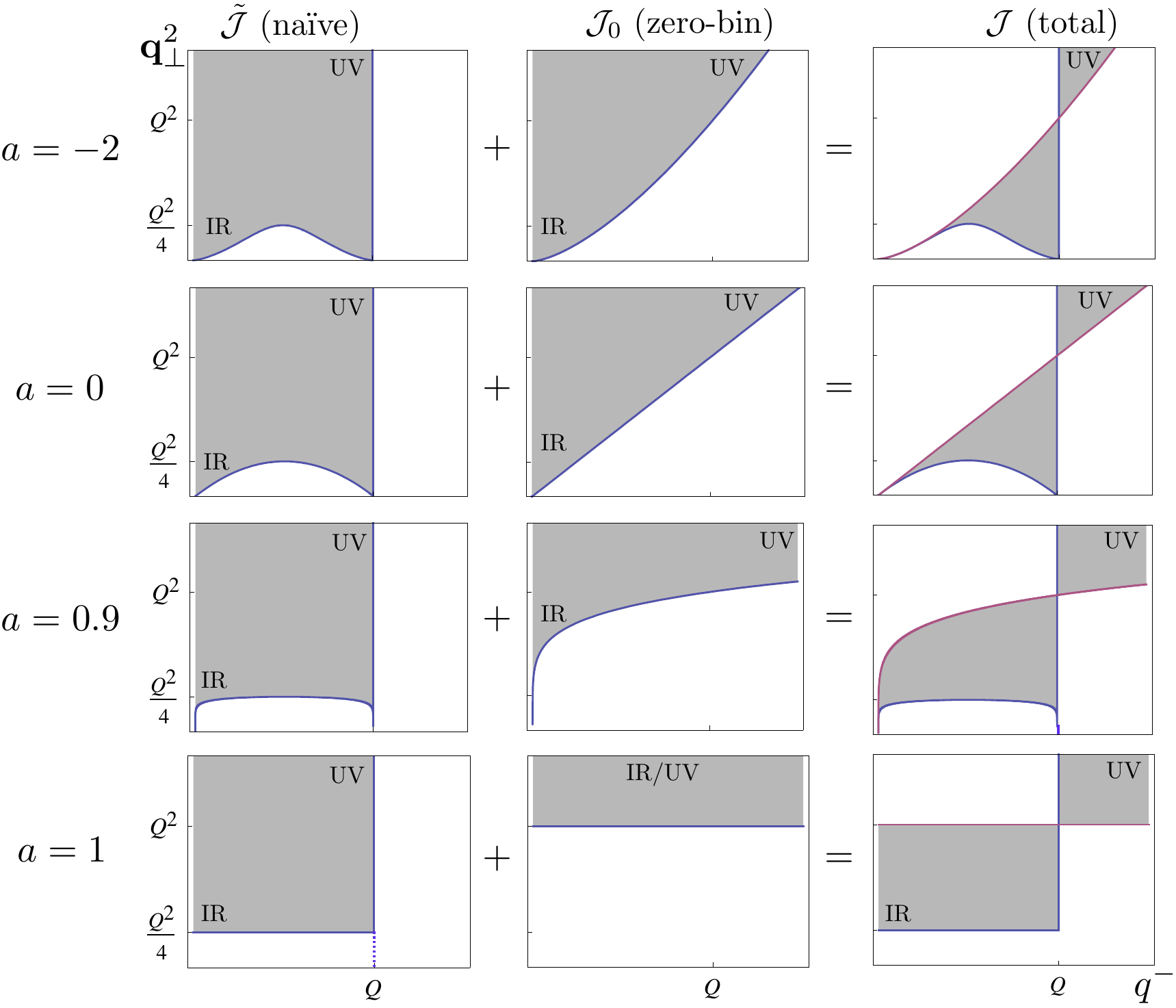} 
{ \caption[1]{Regions of integration for the coefficient of $\delta(\tau_a^n)$ in the jet function $J_a^{n\bare}(\tau_a^n)$. The sum of na\"{\i}ve real and virtual Wilson line diagrams are integrated over the region $\tilde{\mathcal{J}}$ in the $q^-,\vect{q}_\perp^2$ plane. The sum of real and virtual zero-bin subtractions are integrated over $\mathcal{J}_0$, and the resulting sum of na\"{\i}ve diagrams and zero-bin subtractions over the region $\mathcal{J}$. Integrals over $\mathcal{J}$ have only UV divergences as long as $a<1$. For $a=1$, an IR divergent region remains.
\label{jet-regions} }}
}

The shape of the final integration region $\mathcal{J}$ in \fig{jet-regions} demonstrates that the scaleless virtual and zero-bin integrals succeed in converting IR divergences in the real diagram contributions into UV divergences for all $a<1$. The integral over $\mathcal{J}$ in \eq{taunUV} converges for $\epsilon>0$ if and only if $a<1$.  The result of performing this integration, after including the contributions of the QCD-like diagrams in Fig.~\ref{jet-function}C and D, 
agrees with the coefficient of $\delta(\taun)$ that is obtained by (na\"{\i}vely) using the relation \eq{omegaexpansion} in \eq{R-Vcounterterms}.

\subsection{Infrared safety, factorizability, and the effective theory}
\label{ssec:IRsafety}

In the one-loop calculations of soft and jet functions above, we observed that infrared safety of these functions, and, thus, factorizability of the angularity distributions, required $a<1$.  By analyzing explicitly the regions of integration over loop momenta in real and virtual graphs, we were able to identify when the loop integrals contained infrared or ultraviolet divergences. Cancellations of regions in real gluon diagrams sensitive to IR divergences relied crucially not only on the addition of virtual diagrams but also on zero-bin subtractions from collinear diagrams (see also examples in \cite{Manohar:2006nz,Idilbi:2007ff,Idilbi:2007yi,Chiu:2009yx}). 

The shape of the momentum regions contributing to the one-loop soft function in Fig.~\ref{kpluskminus} suggest a simple physical interpretation of the breakdown of factorization as $a\rightarrow 1$. In the $k^\pm$ plane, the region of integration in the sum of real and virtual graphs for $a=1$ is the region above the line $k^+ k^- = Q^2$.  For angularity soft functions with $a<1$, as $k^\pm\rightarrow\infty$, the loop integral goes over a region with $k^+ k^-$ strictly greater than $Q^2$, and in fact, $k^+k^-\rightarrow\infty$, while for $a>1$, the loop integral enters the region with $k^+ k^- < Q^2$, and in fact, $k^+ k^-\to 0$. But this latter region, $k^+\to \infty$ while $k^-\to 0$ or vice versa, is the region where collinear modes live, illustrated in Fig.~\ref{SCETfig}.  This means that collinear modes still contribute to the soft function even after the attempted factorization. 

\FIGURE[t]{
\includegraphics[width=1\textwidth]{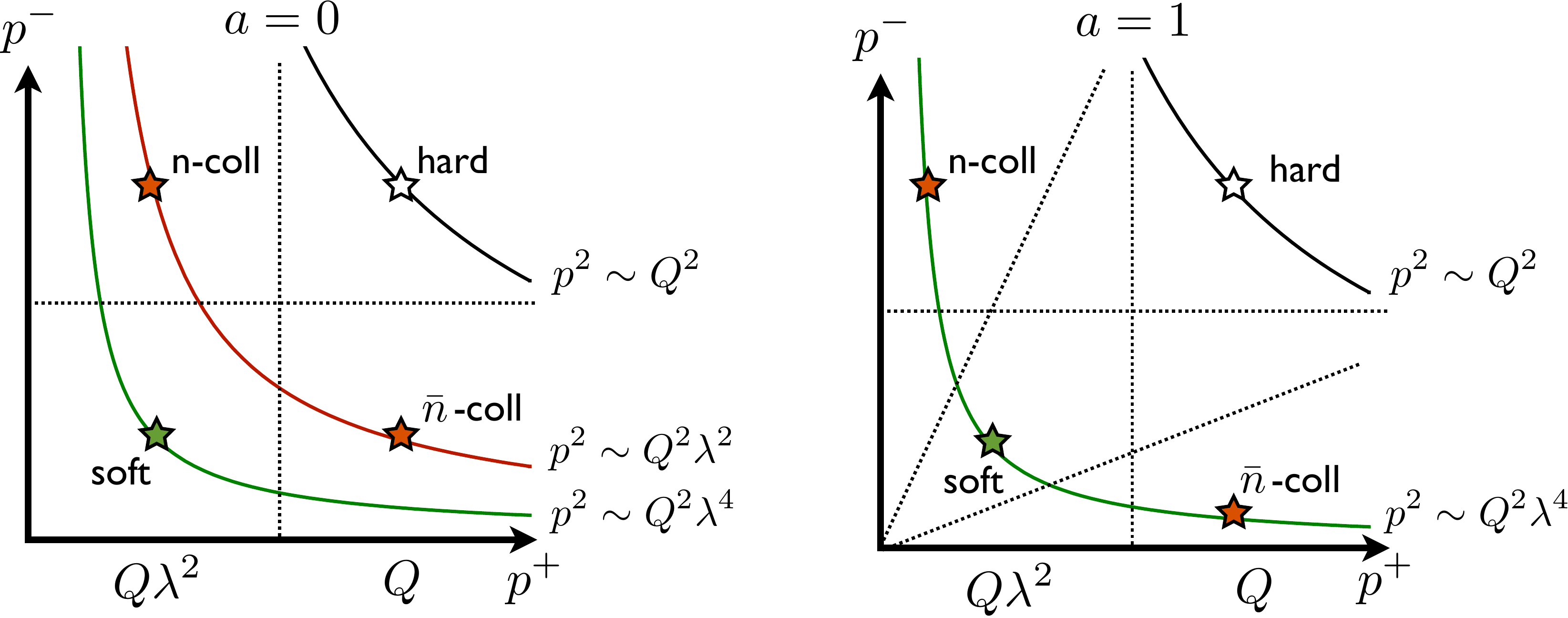} 
{ \caption[1]{Scaling of SCET modes appropriate for angularities $\tau_a$, $a=0,1$. For $a=0$, the collinear modes dominating the $\tau_a$ distribution have virtualities $p^2\sim (Q \lambda)^2$, parametrically separated from the soft scale $p^2\sim (Q \lambda^2)^2$. These scalings correspond to the effective theory known as \SCETa. For $a=1$, the collinear modes in the distribution have typical $p^2\sim(Q \lambda^2)^2$, coinciding with the soft scale. The collinear and soft modes are no longer separated by virtuality but instead by rapidity. These scalings correspond to \SCETb\!. Collinear modes dominating angularity distributions for other values of $a$ between 0 and 1 live at scales intermediate between these limits.
\label{SCETfig} }}
}

This suggests that for $a\geq 1$, the contributions of \SCETa soft and collinear modes to the angularity distribution have not actually been separated. In \SCETa, soft, collinear, and hard modes can be distingushed by their well-separated virtualities, namely, $p_S^2\sim (Q\lambda^2)^2$, $p_J^2 \sim (Q\lambda^{1/(1-a/2)})^2$, and $p_H^2 \sim Q^2$. At $a=1$, the virtualities of soft and collinear modes contributing to the $\tau_a$ distribution coincide, and \SCETa must be matched onto \SCETb where collinear and soft modes both have virtualities $p^2\sim (Q\lambda^2)^2$. In this case, the modes are no longer distinguished by their virtuality, but instead by their rapidity, as illustrated in Fig.~\ref{SCETfig}.  Ref.~\cite{Manohar:2006nz} suggested a modified version of the factorization theorem \eq{fact-theorem} in which soft and jet functions are defined either with cutoffs on rapidity or in dimensional regularization with the scale $\mu$ separated into two light-cone scales $\mu^\pm$, which must satisfy $\mu^+\mu^- = \mu^2$, with each of the two jet functions depending on one of these scales, and the soft function on both. However, in the present paper we do not pursue such a strategy and limit our analysis to angularities with strictly $a<1$.  For arbitrary values of $a$, the virtuality of collinear modes $p_J^2\sim (Q\lambda^{1/(1-a/2)})^2$ suggests an interpretation as the modes of an effective theory ``$\text{SCET}_{1+a}$.''\footnote{We would like to thank M. Strassler for suggesting this terminology to CL.} Since our analysis and calculations utilize the framework of \SCETa, we may expect non-negligible corrections to our results to arise for values of $a$ less than but approaching 1, and for reasonable criteria for when corrections are negligible, our analysis is reliable for values of $a\lesssim 1/2$ \cite{Lee:2006nr}. 

\section{NLL Resummation of Logarithms and Fixed-order Matching to QCD}
\label{sec:NLL}

The fixed-order NLO cross-section, obtained by using the fixed-order expressions for the hard, jet, and soft functions in \eqss{hardNLO}{softNLO}{jetNLO} in the factorization formula \eq{finalformula}, contain logarithms of $\mu$ divided by the scales $Q$, $Q\tau_a$, and the intermediate scale $Q \tau_a^{1/(2-a)}$. This means that there is no single choice for the scale $\mu$ that will simultaneously set all of the logarithms in the NLO cross-section to zero. For small $\tau_a$, these scales become widely separated and the logarithms of ratios of these scales become large, which causes the perturbative series to break down. In \ssec{hardNLL} and \ssec{jet-softNLL}, we take advantage of the effective theory framework separating the hard, jet, and soft contributions by evolving each of them separately through renormalization-group (RG) evolution which resums these logarithms. We then combine these RG-evolved functions into the full cross-section accurate to NLO at fixed order in $\alpha_s$ and resummed to NLL accuracy in \ssec{fullNLL}.

Since our final result for the NLL/NLO resummed distribution is derived using an effective theory which is valid only in the small-$\tau_a$ limit, it does not get the larger-$\tau_a$ region as accurately as QCD at $\cO(\as)$. To arrive at a result that retains NLL/NLO accuracy in the small-$\tau_a$ region while retaining the accuracy of QCD at $\cO(\as)$ in the larger-$\tau_a$ region, we need to match our distribution onto QCD. This matching is constructed such that if we turn off the resummation, the distributions should agree with full QCD to $\mathcal{O}(\alpha_s)$. We perform this matching in \ssec{match}.

\subsection{Hard function at NLL}
\label{ssec:hardNLL}

The anomalous dimension of the hard function in \eq{hardNLO} can be found by requiring that matrix elements of the bare two-jet operator in \eq{O2renorm} are independent of the scale $\mu$, and is given by
\begin{equation}
\gamma_H(\mu) = - \gamma_{\mathcal{O}}(Qn/2,-Q\bn/2;\mu) - \gamma_{\mathcal{O}}^*(Qn/2,-Q\bn/2;\mu) \,,
\end{equation}
where
\begin{align}
\label{anomO}
\gamma_{\mathcal{O}}(\tilde p_n,\tilde p_\bn;\mu) = - Z_\mathcal{O}^{-1}(\tilde p_n,\tilde p_\bn;\mu) \mu \frac{\df}{\df \mu} Z_{\mathcal{O}}(\tilde p_n,\tilde p_\bn;\mu) =  \frac{\as \CF}{ 2\pi} \left( 2 \ln{ \frac{\mu^2}{2\tilde p_n\cdot\tilde p_\bn}} +3 \right) 
\,,
\end{align}
so that
\begin{align}
\label{anomhard}
\gamma_H(\mu) = - \frac{\as \CF}{ \pi} \left( 2 \ln{ \frac{\mu^2}{Q^2}} +3 \right) 
\,,
\end{align}
which is the first term in the expansion of the anomalous dimension to all orders in $\alpha_s$,
\begin{equation}
\label{anomhardallorders}
\gamma_H(\mu) = \Gamma_H[\alpha_s]\ln\frac{\mu^2}{Q^2} + \gamma_H[\alpha_s]\,.
\end{equation}
Solving the RG equation,
\begin{align}
\mu \frac{\df }{\df \mu} H(Q; \mu) = \gamma_H(\mu)  H(Q; \mu) 
\,,\end{align}
for $H(Q; \mu)$ gives
\begin{align}
\label{hardresum}
H(Q; \mu) = H(Q; \mu_0) e^{K_H}  \left(\frac{\mu_0}{Q} \right)^{\omega_H}
\,,\end{align}
where $\omega_H$ and $K_H$ are defined as
\begin{subequations}
\label{omega-Khard}
\begin{align}
\label{omegahard}
\omega_H &\equiv \omega_H(\mu, \mu_0) \equiv  \frac{8 \CF}{\beta_0}  \left[\ln{r}+\left(\frac{\Gamma_{\cusp}^1}{\Gamma_{\cusp}^0}-\frac{\beta_1}{\beta_0}\right)\frac{\as(\mu_0)}{4\pi}(r-1)\right] \\
\label{Khard}
K_H & \equiv K_H(\mu, \mu_0) \equiv \frac{6 \CF}{ \beta_0} \ln{r} + \frac{16\pi\CF}{(\beta_0)^2}\bigg[\frac{r-1-r\ln{r}}{\as(\mu)} \\
&\qquad \qquad \qquad\quad +\left(\frac{\Gamma^1_{\cusp}}{\Gamma^0_{\cusp}}-\frac{\beta_1}{\beta_0}\right)\frac{1-r+\ln{r}}{4\pi}+\frac{\beta_1}{8\pi\beta_0}\ln^2{r}\bigg] \nn
\,.\end{align}
\end{subequations}
Here $r=\frac{\as(\mu)}{\as(\mu_0)}$, and $\beta_0, \beta_1$  are the one-loop and two-loop  coefficients of the beta function,
\begin{equation}
\beta[\alpha_s] = \mu\frac{\df\alpha_s}{\df\mu} = -2\alpha_s\left[\beta_0\left(\frac{\alpha_s}{4\pi}\right) + \beta_1\left(\frac{\alpha_s}{4\pi}\right) ^2 +\cdots\right]\,,
\end{equation}
where
\begin{align}
   \beta_0=\frac{11C_A}{3} - \frac{2n_f}{3} \qquad {\rm and } \qquad \beta_1= \frac{34C_A^2}{3}-\frac{10C_A n_f}{3} -2\CF n_f
\,.\end{align}
The two-loop running coupling $\as(\mu)$ at any scale is given by
\begin{equation}
\label{2loopalphas}
\frac{1}{\as(\mu)} = \frac{1}{\as(M_Z)} + \frac{\beta_0}{2\pi}\ln\left(\frac{\mu}{M_Z}\right) + \frac{\beta_1}{4\pi\beta_0}\ln\left[1+\frac{\beta_0}{2\pi}\as(M_Z)\ln\left(\frac{\mu}{M_Z}\right)\right]\,.
\end{equation}
In \eq{hardresum}, we have used the fact that to all orders in perturbation theory, $\Gamma_H[\alpha_s]$ is proportional to $\Gamma_{\rm cusp}[\alpha_s]$, where
 \begin{align}
 \label{gamma-gammacusp}
 & \Gamma_{\cusp}[\as]=\left(\frac{\as}{4\pi}\right)\Gamma^0_{\cusp}+\left(\frac{\as}{4\pi}\right)^2 \Gamma^1_{\cusp}+\cdots
 \,.\end{align}
The ratio of the one-loop and two-loop coefficients of $\Gamma_{\rm cusp}$ is \cite{Korchemsky:1987wg}
 \begin{align}
 &\frac{\Gamma_{\cusp}^1}{\Gamma^0_{\cusp}}=\left(\frac{67}{9}-\frac{\pi^2}{3}\right)C_A-\frac{10n_f}{9}
\,.\end{align}
$\Gamma_{\cusp}^1$ and $\beta_1$ are needed in the expressions of $\omega_H$ and $K_H$ for complete NLL resummation since we formally take $\as^2 \ln\tau_a \sim \mathcal{O}(\as)$. 

\subsection{Jet and soft functions at NLL}
\label{ssec:jet-softNLL}

The jet and soft functions obey the RG equation
\begin{align}
   \label{RGEtau}
  & \mu \frac{d}{d\mu} F(\tau; \mu)= \int_{-\infty}^{+\infty}\!\!\df \tau' \, \gamma_F (\tau-\tau'; \mu) F(\tau'; \mu)
\,,\end{align}
where $F=J,S$.
The anomalous dimensions $\gamma_{F}$ can be found from the $Z$-factors (given in \eqs{Zsoft}{Zjet}) via the relation
\begin{align}
\gamma_F(\tau - \tau'; \mu) = - \int\!\df\tau''\, Z_F^{-1} (\tau - \tau''; \mu) \, \mu \frac{\df}{\df \mu} Z_F(\tau'' - \tau'; \mu)
\,.\end{align}
We find that
\begin{align}
\label{anomjet}
\gamma_J(\tau - \tau'; \mu) &= \frac{2 \as \CF}{\pi} \left \{ \delta(\tau - \tau') \left( \frac{1-a/2}{1-a} \ln{\frac{\mu^2}{Q^2}} + \frac{3}{4} \right) - \frac{1}{1-a} \left [\frac{\theta(\tau - \tau')}{\tau - \tau'}\right]_+ \right \}
\,,\end{align}
and
\begin{align}
\label{anomsoft}
\gamma_S(\tau - \tau'; \mu) &= \frac{2 \as \CF}{\pi (1-a)} \left \{ - \delta(\tau - \tau') \ln{\frac{\mu^2}{Q^2}}  +2 \left [\frac{\theta(\tau - \tau')}{\tau - \tau'}\right]_+ \right \}
\,.\end{align}
Both anomalous dimensions are the first terms in the perturbative expansion of the general form to all orders in $\alpha_s$ \cite{Fleming:2007xt,Grozin:1994ni},
\begin{align}
\label{anomgeneral}
\gamma_F (\tau - \tau'; \mu) = -  \Gamma_F [\as] \left( \frac{2}{j_F} \left[ \frac{\theta(\tau - \tau')}{(\tau - \tau')}\right]_+ -\ln \frac{\mu^2}{Q^2} \,\delta(\tau-\tau')\right)  + \gamma_F[\as] \delta(\tau - \tau')
\,,\end{align}
\TABLE[r]{
\begin{tabular}{ | c || c | c |}
\hline
& $F = S$ & $F = J$ \\
\hline
$j_F$ & $1$ & $2-a$\\
\hline
$\Gamma_{\! F}^0$ & \rule{-10pt}{3ex} \rule[-1.5ex]{0pt}{0pt} $-8 \CF \frac{1}{1-a} \, $ & $ \, 8 \CF \frac{1-a/2}{1-a} $ \\
\hline
$\gamma_F^0$ & $0$& $6 \CF$ \\
\hline
\end{tabular}
{ \caption[1]{$\Gamma_{\!F}^0$, $\gamma_F$ and $j_F$ for the jet and soft functions.}
\label{tab:jvalue} }}
\noindent where the coefficients $\Gamma_F[\as],\gamma_F[\as]$ have the expansions
\begin{align}
\label{Gammaexpansion}
\Gamma_{\!F}[\as] = \left( \frac{\as}{4 \pi}\right) \Gamma_{\!F}^0 + \left( \frac{\as}{4 \pi}\right)^2 \Gamma_{\!F}^1 + \cdots 
\end{align}
and
\begin{align}
\label{gammaexpansion}
\gamma_F[\as] = \left( \frac{\as}{4 \pi}\right) \gamma_{F}^0 + \left( \frac{\as}{4 \pi}\right)^2 \gamma_{F}^1 + \cdots 
\,.\end{align}
We summarize the coefficients $\Gamma_{\!F}^0$ and $\gamma_F^0$ and the $j_F$-values for the jet and soft functions in \tab{jvalue}.

The solution of the RG equation \eq{RGEtau} with the anomalous dimension $\gamma_F$ of the form given in \eq{anomgeneral} with particular values of $j_F$ was developed in the series of papers \cite{Becher:2006mr,Korchemsky:1993uz,Balzereit:1998yf,Neubert:2005nt}.  Later, it was solved for arbitrary $j_F$ in \cite{Fleming:2007xt} using a convolution variable $t=Q^j \tau$ with mass dimension $j = j_F$. The resulting evolution equation for $F$ is
\begin{align}
F(\tau; \mu) = \int\!\df \tau' \, U_F(\tau-\tau'; \mu, \mu_0) F(\tau'; \mu_0)
\label{Fconvol}
\,,\end{align}
where the evolution kernel $U_F$ is given to all orders in $\alpha_s$  by the expression
\begin{align}
 U_F(\tau-\tau'; \mu, \mu_0)= \frac{e^{ \tilde{K}_F + \gamma_E\tilde{\omega}_F}}{\Gamma(-\tilde{\omega}_F)} \left(\frac{\mu_0}{Q}\right)^{j_F\tilde\omega_F} \left[\frac{\theta(\tau-\tau')}{(\tau-\tau')^{1+\tilde{\omega}_F}}\right]_+
 \label{kernelF}
\,,\end{align}
where $\gamma_E$ is the Euler constant and where $\tilde{\omega}_F$ and $\tilde{K}_F$ are defined as
\begin{subequations}
   \label{kernelparams}
\begin{align}
\label{omegaF}
   \tilde{\omega}_F(\mu,\mu_0) & \equiv \frac{2}{j_F}\int_{\as(\mu_0)}^{\as(\mu)}\frac{d\alpha}{\beta[\alpha]} \Gamma_{\!F}[\alpha] \,, \\
   \label{KF}
    \tilde{K}_F(\mu,\mu_0)& \equiv \int_{\as(\mu_0)}^{\as(\mu)}\frac{d\alpha}{\beta[\alpha]} \gamma_F[\alpha]+2\int_{\as(\mu_0)}^{\as(\mu)}\frac{d\alpha}{\beta[\alpha]} \Gamma_{\!F}[\alpha]\int_{\as(\mu_0)}^{\alpha}\frac{d \alpha'}{\beta[{\alpha'}]}
\,.\end{align}
\end{subequations}
The plus function in \eq{kernelF} for all $\omega<1$ and $\omega \neq 0$ is defined as\footnote{Note that from the definition in \eq{omegaplusdef}, for $\omega<0$ the `$+$' label can be dropped and so \eq{omegaplusdef} is consistent with the distribution relation \eq{omegaexpansion}.}
\begin{align}
   \left[\frac{\theta(x)}{x^{1+\omega}}\right]_{+} &\equiv \lim_{\beta\to 0} \left[\frac{\theta(x-\beta)}{x^{1+\omega}}-\frac{\beta^{-\omega}}{\omega} \delta(x-\beta)\right] \nn\\
  & = - \frac{\delta(x)}{\omega} + \sum_{n=0}^\infty  (-\omega)^n \left[ \frac{\theta(x) \ln^n{x}}{x}\right]_+ 
   \label{omegaplusdef}
\,,\end{align}
with the latter plus functions defined in \eq{lognplusdef}.

For the NLL parameters of the evolution kernel $U_F$,  \eq{kernelparams} gives
\begin{subequations}
  \label{kernelparamsNLL}
\begin{align}
\label{omegaFNLL}
  \omega_F(\mu, \mu_0) &=-\frac{\Gamma_{\!F}^0}{j_F\, \beta_0} \left[\ln{r}+\left(\frac{\Gamma_{\cusp}^1}{\Gamma_{\cusp}^0}-\frac{\beta_1}{\beta_0}\right)\frac{\as(\mu_0)}{4\pi}(r-1)\right] \,,\\
  \label{KFNLL}
  K_F(\mu_,\mu_0) &=-\frac{\gamma_{F}^0}{2\beta_0}\ln {r}+ \frac{-2\pi\Gamma_{\!F}^0}{(\beta_0)^2}\bigg[\frac{r-1-r\ln{r}}{\as(\mu)} \nn\\
  & \qquad \qquad \quad +\left(\frac{\Gamma^1_{\cusp}}{\Gamma^0_{\cusp}}-\frac{\beta_1}{\beta_0}\right)\frac{1-r+\ln{r}}{4\pi}+\frac{\beta_1}{8\pi\beta_0}\ln^2{r}\bigg] 
\,,\end{align}
\end{subequations}
where we have used the fact that $\Gamma_{\!F} \propto \Gamma_{ \cusp}$. This proportionality is well known for the $a=0$ jet and soft functions. In Appendix~\ref{app:cusp} we verify that it remains true for all $a<1$. 

From \eq{Fconvol} we can write explicit formulas for the resummed jet and soft functions at any scale $\mu$. Details of evaluating the integral over the convolution variable $\tau'$ are given in Appendix~\ref{app:RGE}. For the soft function, we plug the fixed-order NLO result \eq{softNLO} at the scale $\mu_0$ into \eq{Fconvol}, and obtain at the scale $\mu$,
\begin{equation}
\label{SoftNLL}
\begin{split}
S_a(\tau_a;\mu) &= \frac{\e^{K_S+\gamma_E\omega_S}}{\Gamma(-\omega_S)} \left(\frac{\mu_0}{Q}\right)^{j_S\omega_S}\\
&\quad \times \biggl[ \biggl\{1 - \frac{\as(\mu_0) C_F}{2\pi}\frac{1}{1-a}\biggl(\ln^2\frac{\mu_0^2}{(Q\tau_a)^2} + 4H(-1-\omega_S)\ln\frac{\mu_0^2}{(Q\tau_a)^2} \\
&\quad\qquad\quad + \frac{\pi^2}{2} + 4\Bigl[[H(-1-\omega_S)]^2 - \psi^{(1)}(-\omega_S)\Bigr]\biggr)\biggr\}  \biggl(\frac{\theta(\tau_a)}{\tau_a^{1+\omega_S}}\biggr)  \biggr]_+\,,
\end{split}
\end{equation}
and for the jet function, plug in the fixed-order NLO result \eq{jetNLO} at $\mu_0$ into \eq{Fconvol}, and obtain at $\mu$,
\begin{align}
\label{JetNLL}
J_a^n(\tau_a;\mu) &= \frac{\e^{K_J + \gamma_E\omega_J}}{\Gamma(-\omega_J)} \left(\frac{\mu_0}{Q}\right)^{j_J\omega_J}\\
&\quad \times \biggl[ \biggl\{ 1 + \frac{\as(\mu_0) C_F}{4\pi}\biggl(\frac{2-a}{1-a}\ln^2\frac{\mu_0^2}{Q^2\tau_a^{\frac{2}{2-a}}} + \Bigl(3 + \frac{4H(-1-\omega_J)}{1-a}\Bigr)\ln\frac{\mu_0^2}{Q^2\tau_a^{\frac{2}{2-a}}} \nn \\
&\quad + 4 f(a) + \frac{4}{(1-a)(2-a)}\Bigl[\frac{\pi^2}{6} + [H(-1-\omega_J)]^2 - \psi^{(1)}(-\omega_J)\Bigr]\biggr)\biggr\} \biggl(\frac{\theta(\tau_a)}{\tau_a^{1+\omega_J}}\biggr)  \biggr]_+\,, \nn
\end{align}
where in the above two equations $K_F \equiv K_F(\mu,\mu_0)$, $\omega_F \equiv \omega_F(\mu,\mu_0)$, $H(z)$ is the harmonic number function, and $\psi^{(\nu)}(z)$ is the polygamma function.

\subsection{Full distribution at NLL}
\label{ssec:fullNLL}

By running the hard, jet, and soft functions from the scales $\mu_0 = \mu_H$, $\mu_J$, and $\mu_S$, respectively, to the common factorization scale $\mu$ and performing the convolution in \eq{finalformula} (see Appendix~\ref{app:RGE} for details), we find for the final resummed expression for the two-jet angularity distribution with NLL/NLO perturbative accuracy
\begin{equation}
\frac{1}{\sigma_0} \frac{\df \sigma_2}{\df {\tau_a}}^{\!\PT} \bigg \vert_{\NLL}  = \left [  \bigg ( 1+ f_H +  2 f_J + f_S \bigg )  U_a^\sigma(\tau_a; \mu, \mu_H,\mu_J,\mu_S) \right]_+
\label{sigmaNLL}
\,,\end{equation}
where we defined 
\begin{align}
U_a^\sigma(\tau_a; \mu, \mu_H,\mu_J,\mu_S) &\equiv \frac{e^{\Gexp + \gamma_E\Omega}}{\Gamma(- \Omeg)}\left(\!\frac{\mu_H}{Q}\!\right)^{\!\omega_H}\!\left(\!\frac{\mu_J}{Q}\!\right)^{\!2j_J\omega_J}\!\left(\!\frac{\mu_S}{Q}\!\right)^{\!j_S \omega_S} \bigg( \frac{\theta({\tau_a})}{{\tau_a}^{1+\Omeg}} \bigg) \,, \label{Usigma}
\end{align}
where
\begin{align}
\label{omegG}
\Omeg& \equiv 2 \,\omega_J(\mu, \mu_J)+\omega_S(\mu, \mu_S) \\
\Gexp&\equiv K_H(\mu, \mu_H)+2K_J(\mu, \mu_J)+K_S(\mu, \mu_S)\,,
\end{align}
with $\omega_H, K_H$ given by \eq{omega-Khard} and $\omega_{J,S}$ and $K_{J,S}$ given by \eq{kernelparamsNLL} and
\begin{subequations}
\begin{align}
f_H &= \frac{\as(\mu_H) \CF}{\pi}  \left (-4 + \frac{7 \pi^2}{12}  - 2 \ln^2{\frac{\mu_H}{Q}} - 3\ln{\frac{\mu_H}{Q}} \right) \\
f_J &= \frac{\as (\mu_J) \CF}{\pi} \bigg[ f(a) + \frac{3/4}{1-a/2} H(-1-\Omeg) + \frac{\frac{\pi^2}{6}+ H(-1-\Omeg)^2 - \psi^{(1)}(-\Omeg) }{2(1-a)(1-a/2)} \\
&\qquad\qquad \qquad + \frac{2-a}{1-a} \ln^2{\frac{\mu_J}{Q {\tau_a}^{1/(2-a)}}} + \left( \frac{3}{2} + \frac{2}{1-a} H(-1-\Omeg) \right)  \ln{\frac{\mu_J}{Q {\tau_a}^{1/(2-a)}}}\bigg] \nn \\
f_S &= \frac{\as (\mu_S) \CF}{\pi} \bigg[\frac{1}{1-a} \left( -\frac{\pi^2}{4}  - 2 H(-1-\Omeg)^2 + 2 \psi^{(1)}(-\Omeg) \right) \\
& \qquad\qquad\qquad - 2 \ln^2{\frac{\mu_S}{Q {\tau_a}}} - 4 H(-1-\Omeg) \ln{\frac{\mu_S}{Q {\tau_a}}} \bigg] \nn
\,,\end{align}
\end{subequations}
and $f(a)$ was defined in \eq{fa}. 

From these expressions, it is clear that the logarithms are minimized by choosing $\mu_H $, $\mu_J$, and $\mu_S$ of order $Q$, $Q {\tau_a}^{1/(2-a)}$, and $Q{\tau_a}$, respectively. We will describe in more detail precisely which values we choose for these scales when we plot the full distributions in Sec.~\ref{sec:results}. 

\subsection{Matching to QCD}
\label{ssec:match}

One way to achieve matching onto QCD is to include three-jet operators in the matching of the QCD current onto the SCET operators in \eq{SCETcurrentO2} \cite{Bauer:2006mk,Marcantonini:2008qn}. For the scope of this paper, however, we simply adopt the matching  procedure described by \cite{Catani:1992ua}, as implemented in \cite{Becher:2008cf}.

To $\cO(\as)$ the full QCD distribution will take the form
\begin{equation}
\label{QCDdistribution}
\frac{1}{\sigma_0}\frac{\df\sigma}{\df\tau_a} = \delta(\tau_a) + \left( \frac{\as}{2\pi} \right) A_a(\tau_a) + \cO(\as^2) \,.
\end{equation}
In Appendix~\ref{app:QCD} we describe how to calculate $A_a(\tau_a)$ numerically.
Meanwhile, the fixed-order two-jet angularity distribution in SCET at $\cO(\as)$ is given by the convolution \eq{finalformula} of the fixed-order hard, jet, and soft functions \eqss{hardNLO}{softNLO}{jetNLO}. The result is independent of $\mu$ (except through $\as \equiv \as(\mu)$), and is given by 
\begin{equation}
\label{singulardist}
\frac{1}{\sigma_0}\frac{\df\sigma_2}{\df\tau_a} = \delta(\tau_a)D^\delta_a + \frac{\as }{2\pi} [D_a(\tau_a)]_+\,,
\end{equation}
where
\begin{align}
D^\delta_a &= 1 - \frac{\as C_F}{2\pi}\frac{1}{2-a}\bigg\{2+5a- \frac{\pi^2}{3}(2+a) \nn\\
& \qquad \qquad \qquad \qquad \qquad + 4\int_0^1\df x\frac{x^2-2x+2}{x}\ln[x^{1-a} + (1-x)^{1-a}]\bigg\} \label{Ddelta} \\
D_a(\tau_a) &= -\frac{2 \CF}{2-a}\frac{\theta(\tau_a)(3+4\ln\tau_a)}{\tau_a}\,. 
\label{DA}
\end{align}

The two-jet fixed-order SCET distribution \eq{singulardist} reproduces the most singular parts of the full QCD distribution\footnote{Technically, we mean that the difference of the two distributions integrated from $0$ to $\epsilon$ vanishes as $\epsilon \to 0$.} \eq{QCDdistribution}, that is, the coefficient of the $\delta(\tau_a)$, $1/\tau_a$ and $(1/\tau_a)\ln{\tau_a}$ pieces. The expression for $D_a(\tau_a)$ in \eq{DA} makes explicit that the angularities are not infrared-safe for $a = 2$.

The difference of the two fixed-order distributions \eq{QCDdistribution} and \eq{singulardist} away from $\tau_a = 0$ is a purely integrable function,
\begin{align}
r_a(\tau_a) \equiv \frac{1}{\sigma_0}\left(\frac{\df\sigma}{\df\tau_a}  - \frac{\df\sigma_2}{\df\tau_a}\right) 
&= \left( \frac{\as}{2\pi} \right)[A_a(\tau_a) - D_a(\tau_a)] 
\,.\label{rtau}
\end{align} 
By adding this remainder function to the NLL resummed SCET distribution, we obtain a result which both agrees with QCD to $\cO(\as)$ and resums large logarithmic terms in the entire perturbative series with NLL/NLO accuracy. The matched distributions are thus defined as
\begin{equation}
\label{QCDmatch}
\frac{1}{\sigma_0}\frac{\df\sigma}{\df\tau_a}^{\!\PT} \bigg \vert_{\NLL} \!\!\!= \frac{1}{\sigma_0}\frac{\df\sigma_2}{\df\tau_a}^{\!\!\PT} \bigg \vert_{\NLL}  \!\!\!+ r_a(\tau_a)\,.
\end{equation}
To find $r_a(\tau_a)$, we numerically obtain $A_a(\tau_a)$ from an analysis of the full QCD distributions away from $\tau_a=0$ using the procedure described in Appendix~\ref{app:QCD}, and then subtract out the expression for $D_a(\tau_a)$ given in \eq{DA}. 

For the case $a=0$ (thrust), the analytic form of $\df\sigma^{\PT}/\df \tau_0$ is known \cite{DeRujula:1978yh}, with which our formula \eq{thrustdist} for $A_0(\tau_0)$ agrees. Using \eqs{thrustdist}{DA}, we obtain the remainder function
\begin{equation}
r_0(\tau_0) = \frac{\as C_F}{2\pi}\left[\frac{2(2-3\tau_0+3\tau_0^2)}{1-\tau_0}\frac{\ln(1-2\tau_0)}{\tau_0} - \frac{2(1-3\tau_0)}{1-\tau_0}\ln\tau_0 + 6 + 9\tau_0\right]\,,
\end{equation}
which we see is integrable down to $\tau_0  = 0$. 

As a consistency check of this matching technique, we calculated the total integral\footnote{The upper limit on $\tau_a$ in \eq{totalsigma}, $\tau_a^{\rm max}$, is that of the maximally symmetric three-jet  configuration, $\tau_{\rm sym}(a) = 1/3^{1-a/2}$ \cite{Berger:2003pk}, but only for $a \gtrsim -2.6$ (see Appendix~\ref{app:QCD}).}  of our fixed-order result,
\begin{align}
\label{totalsigma}
\sigma_{\rm total}= \int_0^{\tau_a^{\rm max}}\!\df\tau_a \left( \frac{1}{\sigma_0}\frac{\df\sigma_2}{\df\tau_a}^{\!\!\PT}+ r_a(\tau_a) \right) \,,
\end{align}
 and compared with the total inclusive cross-section, $\sigma(e^+ e^ - \to X) =  \sigma_0 (1 + \as/\pi )$. We found that our results agreed to any arbitrary precision which could be achieved by our numerical computation. 
 
\section{Nonperturbative Model for the Soft Function}
\label{sec:model}

In this section we adapt the model for the soft function used in jet mass and thrust distributions as constructed in \cite{Hoang:2007vb} to work for all angularities with $a<1$. This model is designed to describe  the small-${\tau_a}$ region where perturbation theory breaks down, while leaving the perturbatively reliable large- and intermediate-${\tau_a}$ regions unaffected. The gap parameter in this model is designed to turn off the soft function at energies below a minimum hadronic threshold. Such a parameter is known to have renormalon ambiguities \cite{Hoang:2007vb}, which must cancel those in the perturbative soft function (which we denote in this section as $S^{\PT}$) to yield a renormalon-free total soft function $S$. To ensure perturbative stability, a scheme is needed to explicitly enforce this cancellation order-by-order in perturbation theory. Recently, the position-mass scheme developed in Ref.~\cite{Jain:2008gb} was used to define a renormalon-free gap parameter for hemisphere jet masses in Ref.~\cite{Hoang:2008fs}. This gap parameter obeys transitive RG evolution and has a well-behaved perturbative expansion. We implement this scheme generalized to arbitrary angularity.

\subsection{Review of hemisphere and thrust soft function models}

To motivate the functional form of the model function that we will use for all angularity distributions, we begin with the model hemisphere soft function constructed in \cite{Korchemsky:2000kp}. This model is a function of two variables which can be chosen to be $l^+$ and $l^-$, defined as the $+$ and $-$ components of the momentum in the $n$ and $\bar n$ hemispheres, respectively. It takes the form
\begin{align}
\fexp(l^+, l^-) =\theta(l^+)\theta(l^-)\frac{{\cal{N}}(A,B)}{\Lambda^2}\left(\frac{l^+ l^-}{\Lambda^2}\right)^{A-1} \exp\left(\frac{-(l^+)^2-(l^-)^2-2B l^+ l^-}{\Lambda^2}\right)
\label{fexphemi}
\,.\end{align}
The parameter $A$ controls how steeply the soft function falls as $l^\pm \to 0$, and $B$ contains information about the cross-correlation of the soft particles in the two hemispheres. $\fexp$ is normalizable for $A>0$ and $B>-1$. $\Lambda$ is an $\mathcal{O}(\Lqcd)$ parameter that describes the range that hadronic effects can smear the soft function around a given $l^+, l^-$. Finally, ${\cal{N}}(A,B)$ is chosen such that $\fexp$ is normalized to unity, $\int_{-\infty}^{+\infty} \df l^+ \df l^- \fexp(l^+, l^-) = 1$. 

In Ref.~\cite{Fleming:2007xt}, this model was used to relate the total hemisphere soft function $S_{\rm hemi}(l^+, l^-)$ to the perturbative hemisphere soft function  $S_{\rm hemi}^{\PT}(l^+, l^-) $ via the convolution
\begin{align}
S_{\rm hemi}(l^+, l^-; \mu) = \int_{-\infty}^{+\infty} \df \tilde{l}^+ \df \tilde{l}^- S^{\PT}_{\rm hemi}(l^+ - \tilde{l}^+, l^- - \tilde{l}^-; \mu)  \fexp(\tilde{l}^+ - \Delta, \tilde{l}^- - \Delta)
\label{stotalhemi}
\,.\end{align}
where $\Delta$ is the gap parameter. This method of implementing the model function ensures a smooth continuation between the nonperturbative, model-dominated and the perturbative regions of the cross-section.

To use this expression in our formalism, we first relate the $a=0$ soft function, $S_0(\tau_0, \mu)$, and the hemisphere soft function, $S_{\rm hemi}(l^+, l^-, \mu)$. Using that $\tau_0 = (l^++l^-)/Q$, we find
\begin{align}
S_0(\tau_0; \mu) &= \int\!\df l^+ \df l^- S_{\rm hemi}(l^+, l^-; \mu) \, \delta \left(\tau_0 - \frac{l^++ l^-}{Q} \right) \nn\\
&=Q\int\!\df l \,S_{\rm hemi} (l, Q \tau_0 - l; \mu)
\,.\end{align}
This gives the model function convolution for $S_0(\tau_0; \mu)$ as
\begin{align}
S_0(\tau_0; \mu) &= Q\int\!\df l \int\!\df l^+ \df l^- \, S_{\rm hemi}^{\PT} (l - l^+, Q \tau_0 - l - l^-; \mu) \fexp (l^+ - \Delta, l^- - \Delta)\nn\\
&= \int\! \df \tau_0' \, S_0^{\PT} (\tau_0 - \tau_0'; \mu) \fexp \bigg(\tau_0' -  \frac{2 \Delta}{Q} \bigg)
\label{SptFexpThrust}
\,,\end{align}
where (absorbing $A$ and $B$ dependent constants into the normalization $\cal N$)
\begin{align}
\fexp(\tau) & \equiv Q^2 \int\!\df \tau' \, \fexp(Q\tau - Q\tau', Q\tau') \nn\\
& =  \theta(\tau) \, {\cal N}(A,B) \frac{Q}{\Lambda} \left(\frac{Q \tau}{\Lambda}\right)^{2A-1}  {_1}F_1\left(\frac{1}{2},\frac{1}{2}+A, (B-1)\frac{(Q\tau)^2}{2\Lambda^2}\right) e^{-(B+1)\frac{ (Q\tau)^2}{2\Lambda^2} } 
\label{fexpdef}
\,.\end{align}
$\fexp(\tau)$ inherits its normalization from $\fexp(l^+, l^-)$, $\int_{-\infty}^\infty\! \df \tau \fexp(\tau) = 1$.

\subsection{Adaptation to all angularities}

For nonzero $a$, we still want to use a convolution of the form
\begin{align}
S_a(\tau_a; \mu) = \int\! \df \tau_a' \, S_a^{\PT} (\tau_a - \tau_a'; \mu) \, \fexp_{a} \bigg(\tau_a' -  \frac{2 \Delta_a}{Q} \bigg)
\label{SptFexp}
\,.\end{align}
Moreover, we would like to retain the functional form of $\fexp$ since it has had relatively good success in describing different event shapes with the same values of $A$ and $B$ \cite{Korchemsky:2000kp}. However, we must at a minimum modify $\fexp$ so that the first moment of $S_a(\tau_a; \mu)$ satisfies the scaling relation given in \eqs{softmoment1}{c-a}. In terms of the first moment of $S^{\PT}_a(\tau_a; \mu)$ and $\fexp_a$, the first moment of $S_a(\tau_a; \mu)$ is
\begin{align}
\int\!\df \tau_a \, \tau_a \, S_a(\tau_a; \mu) &= \int\!\df \tau_a \,\tau_a\int\! \df \tau_a' \, S_a^{\PT} (\tau_a - \tau_a'; \mu) \, \fexp_{a} \bigg(\tau_a' -  \frac{2 \Delta_a}{Q} \bigg) \nn\\
&= S^{ \PT[1]}_{a}(\mu) + \bigg[ \int\!\df \tau_a \, S_a^{\PT}(\tau_a; \mu) \bigg] \bigg( \frac{2 \Delta_a}{Q} + f^{{\rm exp}[1]}_{a} \bigg)  \nn\\
&= S^{\PT [1]}_{a}(\mu) +  \frac{2 \Delta_{a}}{Q} +f^{{\rm exp}[1]}_{a}
\label{Smoment}
\,,\end{align}
where here $S^{\PT[1]}_{a}(\mu)$ and $f^{{\rm exp}[1]}_{a}$ are the first moments of $S_a^{\PT}(\tau_a; \mu)$ and $\fexp_{a}(\tau_a)$, respectively, and in the third line we 
dropped $\as$ corrections to the $\mathcal{O}(\Lqcd/Q)$ power corrections $\Delta_a/Q$ and $f^{{\rm exp}[1]}_{a}$.

Since the first moment of the perturbative soft function, $S_a^{\PT[1]}$, already obeys the proper scaling (cf. \eq{softNLO}) we simply rescale the gap parameter,
\begin{align}
\Delta_a = \frac{\Delta}{1-a}
\label{deltascaling}
\,,\end{align}
and require that the parameters of $\fexp_a$ vary from those in $\fexp$ such that
\begin{align}
f^{{\rm exp}[1]}_{a}\equiv \int\!\df \tau_a \, \tau_a \, \fexp_a(\tau_a ) = \frac{1}{1-a} \int\!\df \tau \, \tau \, \fexp\left(\tau\right) = \frac{1}{1-a} f^{{\rm exp}[1]}
\label{fexpscaling}
\,.\end{align}
This latter condition is most easily satisfied by fixing $A$ and $B$ to their value at $a=0$ and allowing $\Lambda \to \Lambda_a$ to vary accordingly. 
Note from the definition of $\fexp$, \eq{fexpdef}, $\Lambda \fexp(\Lambda \tau/Q) $ is independent of $\Lambda$ and hence $\Lambda_a \fexp_a(\Lambda_a \tau/Q)  =  \Lambda \fexp(\Lambda \tau/Q)$ when $A$ and $B$ are fixed. This implies that
\begin{align}
f^{{\rm exp}[1]}_{a} &= \left ( \frac{\Lambda_a}{Q} \right)^2 \int\!\df \tau_a \, \tau_a \, \fexp_a \left( \frac{\Lambda_a}{Q} \tau_a \right)  = \left ( \frac{\Lambda_a \Lambda}{Q^2} \right) \int\!\df \tau \, \tau \, \fexp \left( \frac{\Lambda}{Q} \tau \right) = \left ( \frac{\Lambda_a}{\Lambda} \right) f^{{\rm exp}[1]}
\,,\end{align}
and so to satisfy \eq{fexpscaling} we take $\fexp_a$ to be defined as in \eq{fexpdef} but with $\Lambda$ replaced with $\Lambda_a$ where
\begin{align}
\label{lambdascaling}
\Lambda_a = \frac{\Lambda}{1-a} 
\,.\end{align}

\subsection{Renormalon cancellation}

We want to ensure that the $1/Q$ renormalon ambiguity in $S^{\PT}(\tau_a; \mu)$ is cancelled order-by-order in perturbation theory. 
To implement the position-mass renormalon cancellation scheme defined in Ref.~\cite{Jain:2008gb} for jet-masses and applied to the $a=0$ gap parameter in Ref.~\cite{Hoang:2008fs}, we first take the Fourier transform of $S_a(\tau_a; \mu)$ with respect to $Q \tau_a$,
\begin{align}
\label{S-pspace}
S_a(x_a; \mu) &\equiv \int\!\df \tau_a \, e^{-i Q \tau_a x_a} S_a(\tau_a; \mu) \nn\\
&= \int\!\df \tau_a \, e^{-i Q\tau_a x_a} \int\!\df \tau'_a \, S_a^{\PT}(\tau_a - \tau'_a; \mu) \, \fexp_a\Bigl(\tau'_a - \frac{2 \Delta_a}{Q}\Bigr) \nn\\ 
&=  S_a^{\PT}(x_a; \mu) \, \fexp(x_a) e^{-2 i \Delta_a x_a}  \nn\\
&= \Big[ S_a^{\PT}(x_a; \mu) e^{-2 i \delta_a(\mu) x_a} \Big] \Big[  \fexp_a(x_a) e^{-2 i \bar{\Delta}_a(\mu) x_a} \Big]
\,,\end{align}
where in the second line we used  \eq{SptFexp} and in fourth line we split $\Delta_a$ into two $\mu$ dependent pieces, $\Delta_a = \bar{\Delta}_a(\mu) + \delta_a(\mu)$. Note that since $\Delta_a$ is $\mu$-independent, $S^{\PT}_a$ and $S_a$ obey the same RG equation.

Next, we demand that for some value $R$, the term in the first pair of brackets in the last line of \eq{S-pspace} satisfies
\begin{align}
\frac{\df}{\df ( ix_a)} \ln[S_a^{\PT}(x_a; \mu) e^{-2 i \delta_a(\mu)x_a}]_{i x_a  = (R e^{\gamma_E})^{-1}}  
= 0
\,,\end{align}
a condition which guarantees no ambiguity in $S_a^{\PT}$ at order $1/Q$.
This gives $\delta_a(\mu)$ to all orders in terms of $S_a^{\PT}(\tau_a; \mu)$ as
\begin{align}
\delta_a(\mu) = - \frac{Q}{2}\frac{\int\!\df \tau_a \, \tau_a \, e^{-Q \tau_a/(Re^{\gamma_E})} S_a^{\PT}(\tau_a; \mu)}{\int\!\df \tau_a \, e^{-Q\tau_a/(Re^{\gamma_E})} S_a^{\PT}(\tau_a; \mu)}
\,,\end{align}
which to leading order is given by the expression
\begin{align}
\delta_a^1(\mu) = - R e^{\gamma_E} \frac{8 \CF}{1-a} \left( \frac{\as(\mu)}{4 \pi} \right) \ln \frac{\mu}{R}
\label{deltamu}
\,.\end{align}

Since $\Delta_a = \bar{\Delta}_a(\mu) + \delta_a(\mu)$ is $\mu$-independent we find that to $\mathcal{O}(\as)$,
\begin{align}
\mu \frac{\df}{\df \mu} \bar{\Delta}_a(\mu) = - \mu \frac{\df}{\df \mu} \delta_a(\mu) = R e^{\gamma_E} \bigg[ \frac{8 \CF}{1-a}\bigg( \frac{\as(\mu)}{4\pi}\bigg) \bigg] 
\equiv - R e^{\gamma_E} \bigg[\Gamma_S^0 \bigg( \frac{\as(\mu)}{4\pi}\bigg) \bigg]
\,.\end{align}
Using that $\Gamma_{\! \bar \Delta} [\as] \propto\Gamma_{\!S}[\as]$ (cf. Refs.~\cite{Hoang:2008fs,Jain:2008gb})
to all orders, we find that the NLL expression for $\mu \,\df \bar{\Delta}_a/\df\mu$ is
and that, for arbitrary $a$, $\Gamma_{\!S}[\as]\propto \Gamma_{\cusp}[\as]$ (cf. App.~\ref{app:cusp})
\begin{align}
\mu \frac{\df}{\df \mu} \bar{\Delta}_a(\mu) =  -R e^{\gamma_E} \bigg[\Gamma_S^0 \bigg( \frac{\as(\mu)}{4\pi}\bigg) \bigg( 1+ \frac{\Gamma^1_{\cusp}}{\Gamma^0_{\cusp}} \frac{\as(\mu)}{4 \pi} \bigg) \bigg] 
\,,\end{align}
which has the solution
\begin{align}
\bar{\Delta}_a(\mu) = \bar{\Delta}_a(\mu_0) -  \frac{R e^{\gamma_E}}{2} \omega_S(\mu, \mu_0)
\label{Deltamu}
\,,\end{align}
where $\omega_S(\mu, \mu_0)$ is given in \eq{kernelparamsNLL}.
Note that since $\delta_a^1(\mu)$ and $\bar{\Delta}_a(\mu) - \bar{\Delta}_a(\mu_0)$ are proportional to $1/(1-a)$, \eq{deltascaling} suggests that we should choose $\bar{\Delta}_a(\mu_0)$ to be $\bar{\Delta}(\mu_0)/(1-a)$, where $\bar{\Delta}(\mu_0)$ is the best choice for $a=0$.

Expanding \eq{SptFexp} in powers of $\as$ to $\mathcal{O}(\as)$ gives
\begin{align}
S_a(\tau_a; \mu) &= \int\!\df \tau'_a \, \bigg[ S_a^{\PT} (\tau_a - \tau'_a; \mu)  + \frac{2 \delta_a^1(\mu)}{Q} \frac{\df}{\df \tau'_a} S_a^{\PT}(\tau_a - \tau'_a; \mu) \bigg] \fexp \bigg(\tau'_a - \frac{2\bar{\Delta}_a(\mu)}{Q} \bigg)
\label{SNLO}
\,,\end{align}
where $S_a^{\PT}$ at NLO in the first term in brackets and at LO in the second term should be used since $\delta_a^1$ is $\mathcal{O}(\as)$. Using the fixed-order expression $S_a^{\PT}(\tau_a; \mu) = \delta(\tau_a) + \mathcal{O}(\as)$ in the second term and integrating this term by parts gives
\begin{align}
S_a(\tau_a; \mu) &= \int\!\df \tau_a'  \bigg[ S_a^{\PT}(\tau_a - \tau_a'; \mu) \, \fexp_a \bigg(\tau_a' - \frac{2 \bar{\Delta}_a(\mu)}{Q} \bigg)\bigg] - \frac{2 \delta_a^1(\mu)}{Q} \frac{\df}{\df \tau_a} \fexp_a \bigg(\tau_a - \frac{2 \bar{\Delta}_a(\mu)}{Q} \bigg) 
\label{SNLO2}
\,.\end{align}
Evolving $S_a(\tau_a; \mu_S)$ to the scale $\mu$ with $U_S(\tau_a - \tau_a'; \mu, \mu_S)$ as in \eq{Fconvol} gives
\begin{align}
S_a(\tau_a; \mu) &= \int\!\df \tau_a'  \bigg[ S_a^{\PT}(\tau_a - \tau_a'; \mu) \, \fexp_a \bigg(\tau_a' - \frac{2 \bar{\Delta}_a(\mu_S)}{Q} \bigg) \nn\\
& \qquad \qquad - \frac{2 \delta_a^1(\mu_S)}{Q} \, U_S(\tau_a - \tau_a'; \mu, \mu_S)\, \frac{\df}{\df \tau_a'} \fexp_a \bigg(\tau_a' - \frac{2 \bar{\Delta}_a(\mu_S)}{Q} \bigg) \bigg]
\label{SNLL}
\,.\end{align}
Here we keep $\bar{\Delta}_a$ and $\delta_a$ at the scale $\mu_S$ which is needed to achieve the $1/Q$ renormalon cancellation \cite{Hoang:2005zw}.

Finally, \eq{SNLL} implies that the total resummed distribution at NLL convoluted with the model function $\fexp_a$ is 
\begin{equation}
\label{totaldist}
\begin{split}
\frac{1}{\sigma_0}\frac{\df \sigma}{\df \tau_a} \bigg \vert_{\NLL} &= \int \!\df \tau_a' \bigg\{ \frac{1}{\sigma_0} \frac{\df \sigma}{\df {\tau_a}}^{\!\PT} \!\!\!\!\!(\tau_a - \tau_a'; \mu)\bigg\vert_{\NLL} \, \fexp_a \bigg(\tau_a' - \frac{2 \bar{\Delta}_a(\mu_S)}{Q} \bigg) \\
& \qquad - \frac{2 \delta_a^1(\mu_S)}{Q} \bigg[ U_a^\sigma(\tau_a - \tau_a'; \mu, \mu_H,\mu_J,\mu_S) \bigg]_+ \frac{\df}{\df \tau_a'} \fexp_a \bigg(\tau_a' - \frac{2 \bar{\Delta}_a(\mu_S)}{Q} \bigg)\bigg\}  \,,
\end{split}
\end{equation}
where the resummed two-jet distribution matched to QCD, $\df \sigma^{\PT}/\df \tau_a  \vert_{\NLL} $,  is given in \eq{QCDmatch} and $U_a^\sigma$ is given in \eq{Usigma}.

\subsection{Numerical results for the soft function}

By plugging the partonic soft function \eq{softNLO} into the model \eq{SNLO2}, we obtain for the full convoluted model soft function to $\mathcal{O}(\alpha_s)$,
\begin{equation}
\begin{split}
S_a(\tau_a;\mu) &= \left\{1 - \frac{\as C_F}{2\pi}\frac{1}{1-a}\left[\ln^2\left(\frac{\mu^2}{Q^2(\tau_a^\Delta)^2}\right) - \frac{\pi^2}{6}\right]\right\}\fexp_a\left(\tau_a^\Delta\right)- \frac{2\delta_1^a(\mu)}{Q}   \frac{\df}{\df \tau_a}\fexp (\tau_a^\Delta) \\ 
&\quad + \frac{2\as C_F}{\pi}\frac{1}{1-a}\int_0^{\tau_a^\Delta} \df\tau' \frac{1}{\tau'}\ln\left(\frac{\mu^2}{Q^2 {\tau'}^2}\right)\left[\fexp_a\left(\tau_a^\Delta - \tau' \right) - \fexp_a\left(\tau_a^\Delta\right)\right]  \,,
\end{split}
\end{equation}
where $\tau_a^\Delta \equiv \tau_a - 2\bar\Delta_a(\mu)/Q$. To integrate against the plus distributions in \eq{softNLO}, we used the prescription
\begin{subequations}
\begin{align}
\int_0^a \df x \left[\frac{\theta(x)}{x}\right]_+ f(x) &= \int_0^a \df x \frac{\theta(x)}{x} [f(x) - f(0)] + f(0)\ln a \\
\int_0^a \df x \left[\frac{\theta(x)\ln x}{x}\right]_+ f(x) &= \int_0^a \df x \frac{\theta(x)\ln x}{x} [f(x) - f(0)] + \frac{1}{2}f(0)\ln^2 a\,,
\end{align}
\end{subequations}
which correspond to the definition of plus-functions given in \eq{lognplusdef}.
To minimize the logarithms in the peak region of the soft function while also avoiding the Landau pole in $\as$, it is natural to choose the scale to be of order $\mu\gtrsim\Lqcd$. To minimize the logarithms for larger values of $\tau_a$, it is natural to choose $\mu\sim Q\tau_a$. A scale choice that interpolates between these two regions is
\begin{equation}
\label{musoft}
\mu = \sqrt{\theta(Q\tau_a-\mu_S^{\rm min})(Q\tau_a-\mu_S^{\rm min})^2 + (\mu_S^{\rm min})^2}\,, 
\end{equation}
where the minimum scale is $\mu_S^{\rm min}\gtrsim \Lqcd$. 


\FIGURE{
\centerline{\resizebox{15cm}{!}{\includegraphics{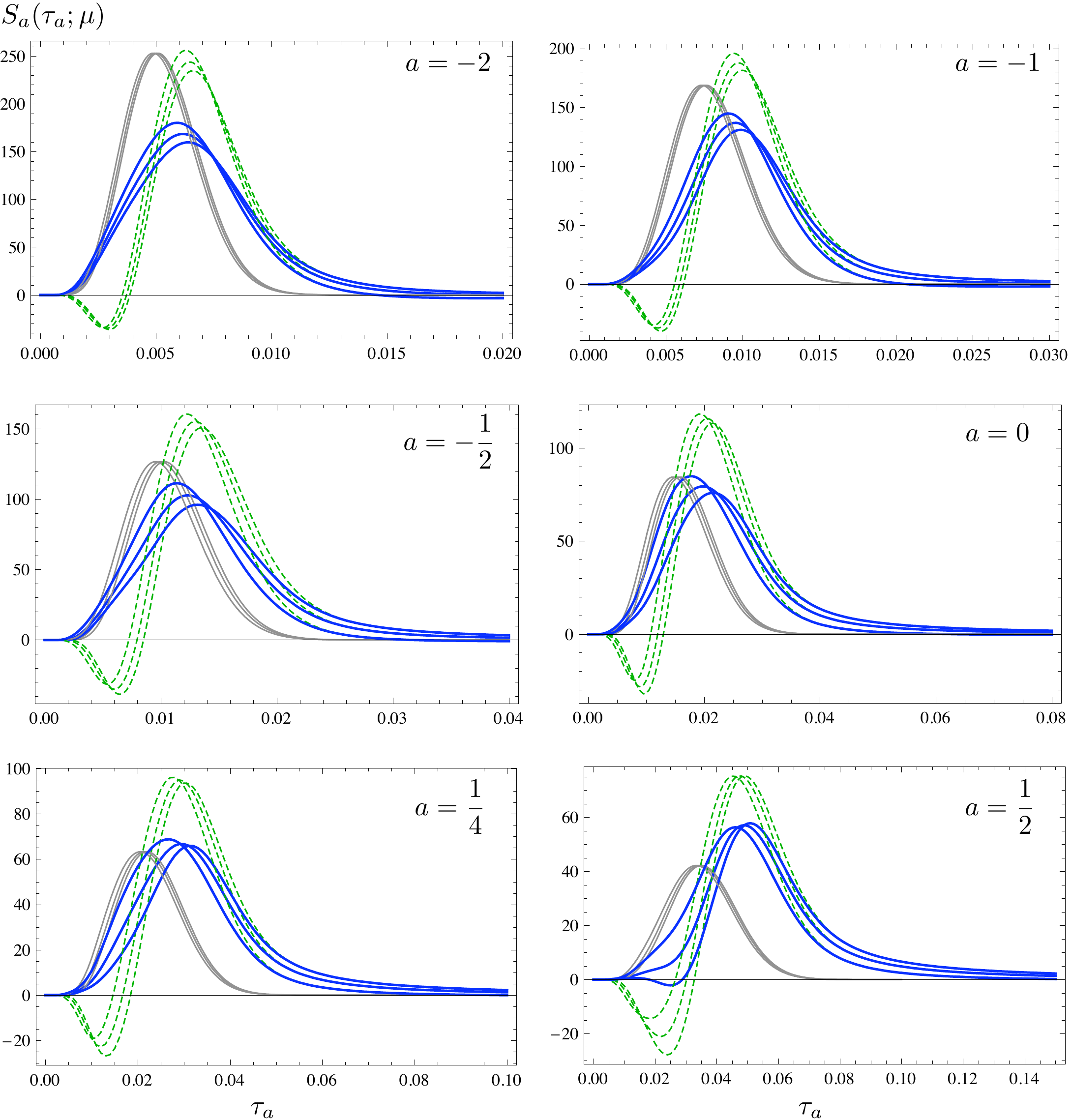}}}
{\caption[1]{Angularity soft functions with a gap parameter, at tree-level (solid gray) and at one-loop with (solid blue) and without (dashed green) renormalon subtraction, for $Q=100\GeV$, for several values of $a$ as labeled on each plot. The variation of the soft functions with the scale $\mu$ is illustrated by first setting $\mu_S^{\text{min}} = 1.0\GeV$ in \eq{musoft} and choosing $\mu$  to be $(0.8,1,1.2)$ times the formula in \eq{musoft}, with the plots for smaller values of $\mu_S$ peaking earlier in $\tau_a$. For the model parameters we take $A=2.5,B=-0.4,\Lambda=0.55\GeV$. In the renormalon subtraction \eq{deltamu}, we have chosen $R=200\MeV$.}
\label{softplot} }
}

In Fig.~\ref{softplot}, we plot $S_a(\tau_a;\mu)$ for six values of $a$ between $-2$ and $1/2$.   In each plot, we show the tree-level (LO) soft function with a gap parameter (solid gray), the one-loop (NLO) soft function with a gap parameter but without renormalon subtraction (dashed green), and the one-loop soft function with a gap and renormalon subtraction (solid blue). For the parameters in the model function \eq{fexpdef} we take $A=2.5,B=-0.4,\Lambda=0.55\GeV$, as extracted from a fit to the jet mass distribution \cite{Korchemsky:2000kp}. For the scale dependence of the gap parameter, we choose $\bar\Delta_0(1\GeV) = 100\MeV$ and use \eq{Deltamu} to evolve to other scales. We choose $R=200\MeV$ in the renormalon subtraction \eq{deltamu} and the minimum value of the scale in \eq{musoft} to be $\mu_S^{\rm min} = 1\GeV$. We illustrate the variation of $S_a(\tau_a;\mu)$ with the scale $\mu$ by varying it between 0.8 and 1.2 times the formula in \eq{musoft}. The tree-level soft functions depend on $\mu$ only through the gap parameter $\bar \Delta_a(\mu)$ and thus artificially have smaller scale variation than the one-loop soft functions, at which order the nontrivial $\mu$ dependence is first probed.

The one-loop soft functions in Fig.~\ref{softplot}  display unphysical behavior near $\tau_a=0$  by taking negative values, due to the renormalon ambiguity in the perturbative series for the partonic soft function. By cancelling the renormalon ambiguity between the partonic soft function and the nonperturbative gap parameter $\Delta_a$ through \eq{SNLO2}, we obtain the renormalon-free one-loop soft functions. One of the plots of the soft function for $a=1/2$ still exhibits a small negative dip after renormalon subtraction, but it is nevertheless much smaller than the original negative dip, and from its size may be expected to an effect of higher-order power corrections. The dip does not appear in the total cross-section calculated below in Sec.~\ref{sec:results}.

\section{Numerical Results for the Full Distribution}
\label{sec:results}

In this section we plot the angularity distributions $\df\sigma/\df\tau_a$ which include LO and NLO perturbative hard, jet, and soft function contributions, resummation of large logarithmic terms to NLL accuracy, matching to QCD at $\cO(\as)$, and the effects of the nonperturbative gapped soft functions.

In Fig.~\ref{PTplot} we plot the angularity distributions given by \eq{totaldist}, plugging in the NLL resummed partonic distribution given by \eq{sigmaNLL} and matched according to \eq{QCDmatch}. We keep the same soft model function parameters as in the previous section. As noted earlier, the logarithms in the hard, jet, and soft functions are minimized by choosing $\mu_H = Q$, $\mu_J\sim Q\tau_a^{1/(2-a)}$, and $\mu_S\sim Q\tau_a$. In order to avoid the Landau pole in $\as$ as $\tau_a\to 0$, we choose the scales as in \eq{musoft}\,,

\begin{subequations}
\label{muSJ}
\begin{align}
\mu_S &= \sqrt{\theta(Q\tau_a-\mu_S^{\rm min})(Q\tau_a-\mu_S^{\rm min})^2 + (\mu_S^{\rm min})^2} \\
\mu_J &= \sqrt{\theta(Q\tau_a^{1/(2-a)}-\mu_J^{\rm min})(Q\tau_a^{1/(2-a)}-\mu_J^{\rm min})^2 + (\mu_J^{\rm min})^2} \,.
\end{align}
\end{subequations}
We may vary $\mu_{S,J}^{\rm min}$ independently, or choose them in a correlated fashion suggested by their natural scaling $\mu_S\sim Q\lambda,\mu_J\sim Q\lambda^{1/(2-a)}$, that is, 
\begin{align}
\label{muJ0}
\mu_J^{\rm min} = Q^{(1-a)/(2-a)}(\mu_S^{\rm min})^{1/(2-a)} 
\,.\end{align} 
In Fig.~\ref{PTplot} we have done the latter. The NLL/NLO distributions exhibit negative values for small $\tau_a$ as a result of the renormalon ambiguity. Performing the renormalon subtraction in the soft function removes this pathology.

\FIGURE{
\centerline{\resizebox{15cm}{!}{\includegraphics{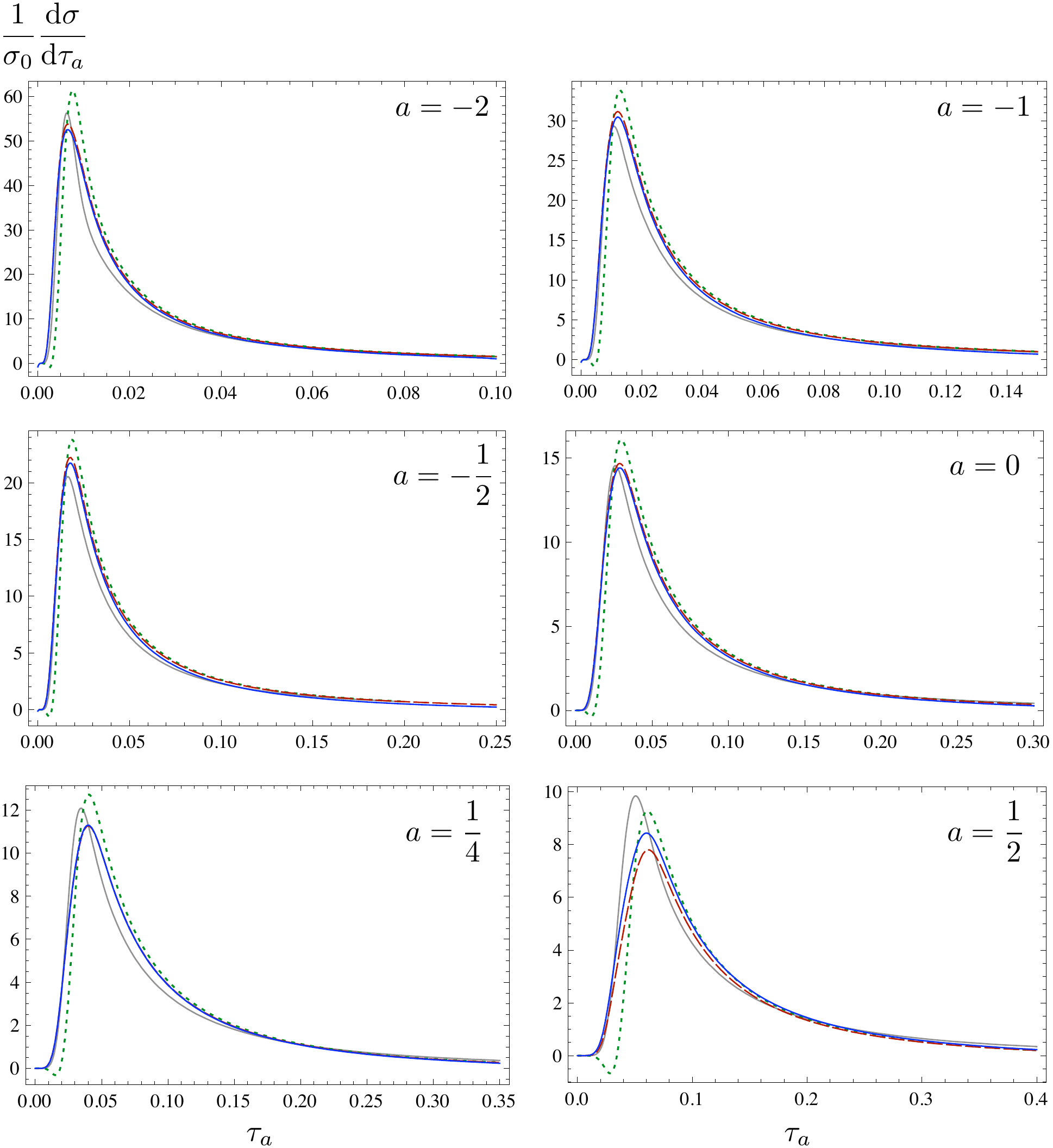}} }
\vspace{-.5cm}
{\caption[1]{Angularity distributions at $Q=100\GeV$ for six values of $a$ between $-2$ and $1/2$. The solid gray curves are the LO partonic distributions resummed to NLL and convoluted with the gapped soft model function. The dotted green curves are NLL/NLO convoluted with the gapped soft function but without renormalon subtraction. The dashed red curves are the same as the green but with renormalon subtraction, and the solid blue curves are the same as the red but matched to fixed-order QCD at $\cO(\as)$. We choose the scales $\mu=Q,\mu_S^{\rm min}=1\GeV$, and $\mu_J^{\rm min}$ given by \eq{muJ0}. For the gap parameter we take $\bar\Delta_0(1\GeV) = 100\MeV$ and in the renormalon subtraction $R=200\MeV$.}
\label{PTplot} }
}

\FIGURE{
\centerline{\resizebox{10cm}{!}{\includegraphics{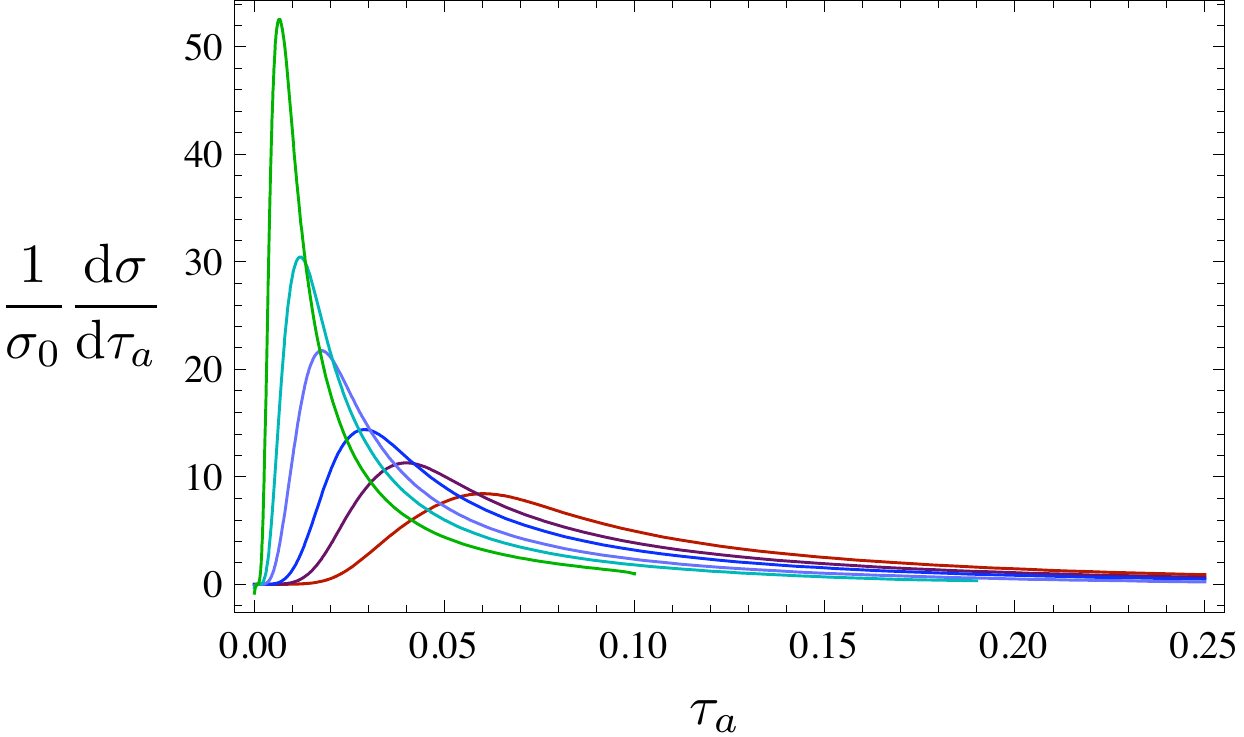}}}
\vspace{-.5cm}
{\caption[1]{Angularity distributions at $Q=100\GeV$. The full, NLL/NLO resummed, renormalon-subtracted distributions in Fig.~\ref{PTplot} are here shown all on the same scale. The parameters are chosen the same as in Fig.~\ref{PTplot}. From highest to lowest peak value, the curves are for $a=-2,-1,-\frac{1}{2}, 0, \frac{1}{4},\frac{1}{2}$.}
\label{allplots} }
}

\FIGURE{
\centerline{\resizebox{15cm}{!}{\includegraphics{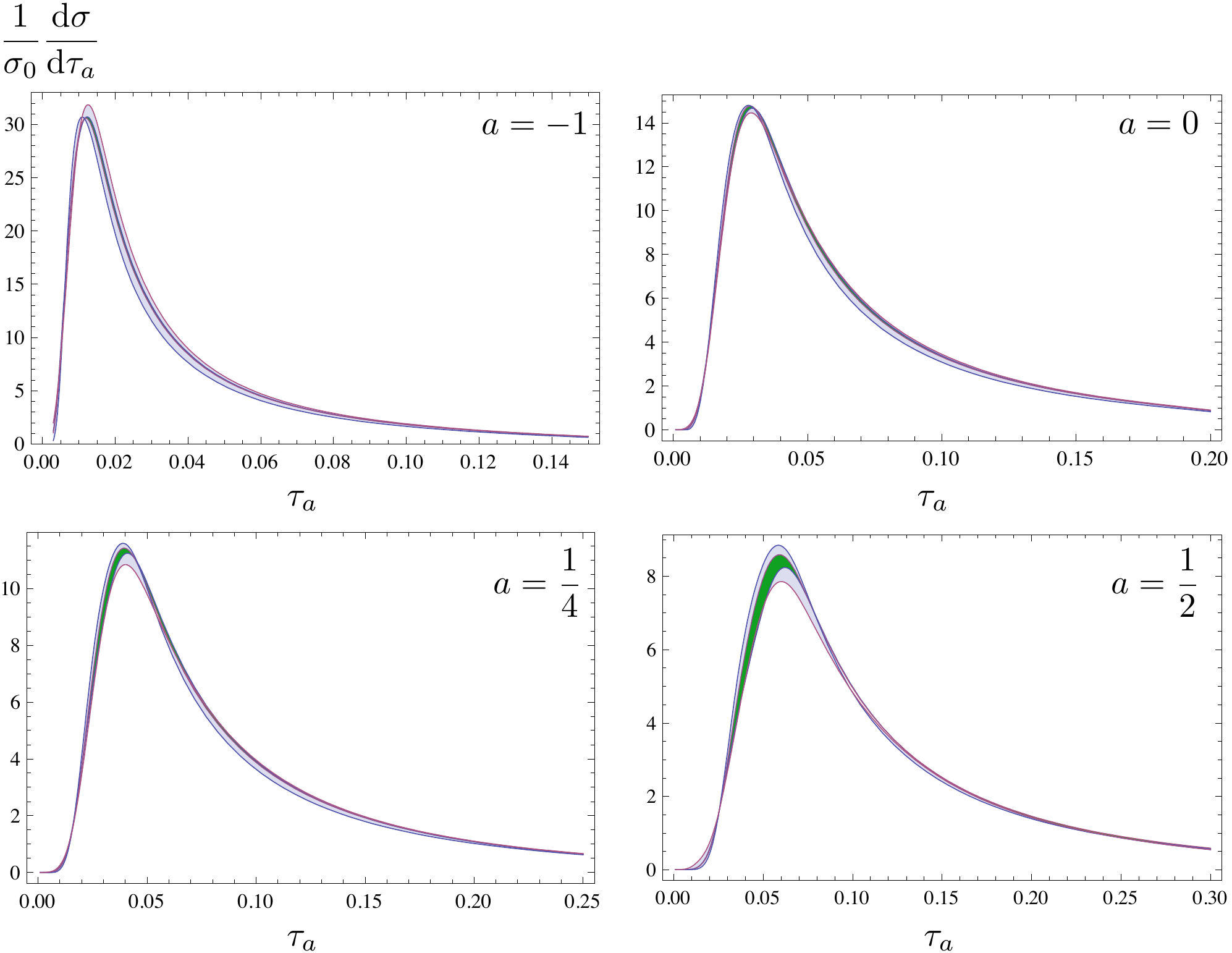}}}
\vspace{-.5cm}
{\caption[1]{Hard scale variation (dark green band) and correlated jet and soft scale variation (light blue band) of the NLL/NLO resummed, renormalon-subtracted angularity distributions at $Q=100\GeV$ for $a=-1$, $a=0$, $a=1/4$, and $a=1/2$. For the hard scale variation, $\mu_H$ varied between $Q/2$ and $2Q$ and for the correlated scale variation, $\mu_J$ and $\mu_S$ are varied between half the values given in \eq{muSJ} and twice these values.}
\label{scalevariation} }
} 

\clearpage

In Fig.~\ref{allplots} we plot angularity distributions for the values of $a$ used in Fig.~\ref{PTplot} on the same figure to illustrate clearly how they change with $a$.  The range of $\tau_a$ populated by two-jet-like events grows with increasing $a$, so that the peak regions are populated by jets of increasing narrowness with increasing $a$. This is reflected in the scales $\mu_{J,S}$ in \eq{muSJ} drawing closer as $a$ grows to 1. 

 In Fig.~\ref{scalevariation} we vary the hard, jet, and soft scales and plot the resulting variation of our final predictions for the distributions. First we vary the hard scale $\mu_H$ between $Q/2$ and $2Q$, plotting the result in the dark green band. Then we vary the collinear and soft scales $\mu_{J,S}$ between half and twice the values we chose in \eq{muSJ} and plot the result in the light blue band. 
 
Although published data on $e^+e^-$ angularity distributions for $a\not =0$ are not yet available, data for the $a=0$ (thrust) distribution are of course plentiful. The remaining difference between our prediction in Fig.~\ref{PTplot} and existing measurements of the $a=0$ distribution can be accounted for by higher-order perturbative corrections (see, for example, Fig.~6 in Ref.~\cite{Becher:2008cf}), which are known but have not been included here, since we calculated the other angularity distributions only to NLL/NLO. For $a$ sufficiently smaller than 1, we expect our predictions of all angularity distributions to agree with data to the same accuracy that the NLL/NLO $a=0$ prediction agrees with the thrust data. 
  
\section{Comparison to Previous Results and Classic Resummation}
\label{sec:relation}

To compare to previous predictions of angularity distributions \cite{Berger:2003iw,Berger:2003pk} and focus more generally on the differences between SCET and alternative approaches to factorization and resummation, in this section we restrict our attention to the perturbative distribution both before matching, \eq{sigmaNLL}, and after matching, \eq{QCDmatch}, leaving out  the nonperturbative model of \sec{sec:model}.

Our result for the unmatched NLL resummed distribution \eq{sigmaNLL} involves an evolution factor $U_a^\sigma$, which resums all leading and next-to-leading logarithms (for example the $(1/\tau_a) \ln \tau_a$ and $1/\tau_a$ terms in the fixed-order $D_a(\tau)$ of \eq{DA}), and a multiplicative NLO prefactor $1+ f_H +2 f_J + f_S = 1+ \mathcal{O}(\as)$. 
Both the evolution factor and the NLO prefactor are sensitive to physics at the three distinct scales $\mu_H$, $\mu_J$, and $\mu_S$. Keeping these scales arbitrary until after solving the RG equations in \sec{sec:NLL} and retaining the freedom to choose them only at the end provides a flexibility which is indispensable in achieving reliable predictions in the SCET approach. This approach has significant advantages  over what we refer to as the classic approach to resummation in QCD \cite{Catani:1992ua}.

\FIGURE[t]{
\centerline{\resizebox{15cm}{!}{\includegraphics{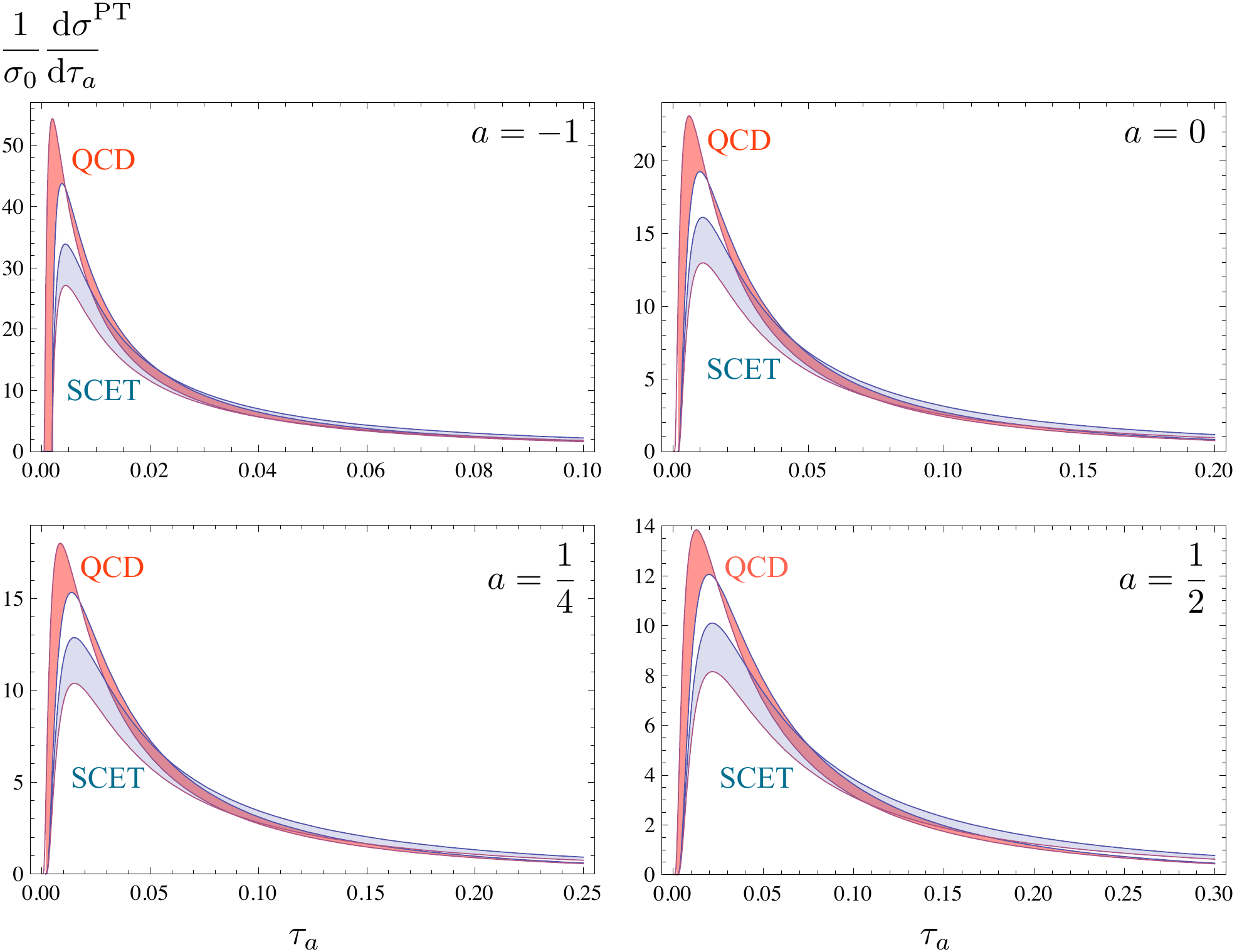}}}
\vspace{-.5cm}
{\caption[1]{Factorization scale $\mu$ variation of the (unmatched, partonic) SCET NLL/LO (light blue band) and the classic QCD NLL/LO  (red band) resummed results for angularity distributions. $\mu$ is varied over the range $\frac{Q}{2}\leq \mu \leq 2Q$ with $Q=100\GeV$ for the cases $a=-1$, $a=0$, $a=1/4$, and $a=1/2$. To make a direct comparison to the QCD results, the scales in the SCET results have been chosen as $\mu = \mu_H = Q$, $\mu_J = Q\tau_a^{1/(2-a)}$, and $\mu_S = Q\tau_a$.}
\label{hardscalevariation} }
}

To illustrate these advantages, we compare our results for angularity distributions to those obtained in full QCD \cite{Berger:2003iw,Berger:2003pk}. The analysis in Ref.~\cite{Berger:2003iw} used a formalism of factorization and resummation of logarithms through renormalization-group evolution paralleling that of SCET, in principle containing all the advantages that we emphasize here, but which were not fully realized. 
Before arriving at the explicit prediction for the NLL resummed distribution $\df \sigma/\df \tau_a$ given in Ref.~\cite{Berger:2003pk}, the factorized result of Ref.~\cite{Berger:2003iw} was first converted into the form of a resummed event shape distribution that would be obtained using the classic approach (and has been for $a=0$). 

One major advantage of the SCET approach over the classic approach is the presence of Landau pole singularities in the results of the classic approach that are not in the results from SCET, as also found in  the cases of DIS and Drell-Yan \cite{Manohar:2003vb,Becher:2006nr,Becher:2006mr}. We can illustrate why SCET avoids this for the case of angularities by returning to our results for the resummed jet and soft functions and for the final resummed distribution. From the expressions for the resummed soft function $S_a(\taus)$, \eq{SoftNLL}, and for the resummed jet function $J_a^n(\taun)$, \eq{JetNLL}, one might be tempted to set $\mu_S = Q\taus$ and $\mu_J = Q (\taun)^{1/(2-a)}$, since the logarithms in \eqs{SoftNLL}{JetNLL} are minimized for these choices. The problem with this choice is that the soft and jet functions still enter the convolution in the factorization theorem \eq{fact-theorem} and thus the scales in $\as(\mu_{J/S})$ run below $\tau_a^{n, s} = \Lqcd / Q$  even for $\tau_a > \Lqcd / Q$ (where $\tau_a = \tau_a^n + \tau_a^\bn + \tau_a^s$) if these 
$\tau_a^{n, s}$-dependent scales are chosen. However, for a $\taus$-independent choice of $\mu_S$ in the case of the soft function, for instance, the full functional dependence of the resummed $S(\taus; \mu)$ on $\taus$ and $\mu_S$ is such that after the integrals over $\taus$, $\taun$, and $\tau_a^\bn$ needed to get to the final resummed distribution, \eq{sigmaNLL}, are performed, the resulting dependence on $\mu_S$ only comes in the combination $\mu_S/Q \tau_a$ in logarithms (and similarly for the jet functions). The proper choice is thus  $\mu_S \sim Q \tau_a$ (and $\mu_J \sim Q \tau_a^{1/(2-a)}$) and not $\mu_S \sim Q \taus$. With this choice, Landau pole singularites never affect our result for $\tau_a  > \Lqcd/Q$. Setting $\mu_S  = Q \taus$ before doing the convolution \eq{fact-theorem} is equivalent to setting $\mu_S = Q /\nu$ in the Laplace transform with respect to $\nu$ of the distribution, which is the scale choice made in Ref.~\cite{Berger:2003iw} needed to reproduce the classic result for $a=0$. Thus, when transforming back to get $\df \sigma/\df \tau_a$, one inevitably runs into spurious Landau pole singularities with this scale choice\footnote{There are also inherent Landau pole singularities in the classic approach before transforming back to $\tau_a$-space and thus not associated with making $\nu$-dependent scale choices for $\mu_{J, S}$. In the classic approach, a prescription to avoid both types of Landau pole singularities is employed, but at the expense of introducing unphysical power corrections \cite{Catani:1992ua, Catani:1996yz}. The results of \cite{Berger:2003pk} plotted in Fig.~\ref{hardscalevariation} used the prescription of \cite{Catani:1992ua}.}, confirming the similar observation of \cite{Becher:2006mr}.

Another difference between the explicit results we give and those given in \cite{Berger:2003iw} is that while both achieved resummation of logarithms to NLL accuracy, the latter does not include a full NLO calculation of the jet and soft functions in the distribution $\df \sigma/\df \tau_a$, that is, effectively does not have the prefactors $f_{H,J,S}$. As with our SCET results, the results of \cite{Berger:2003iw} are not as accurate as fixed-order QCD in the large-$\tau_a$ region and need to be matched. This matching was subsequently performed numerically to at $\cO(\as^2)$ in Ref.~\cite{Berger:2003pk}. We summarize this by saying that we have resummed logarithms of $\tau_a$ to NLL/NLO with $\cO(\as)$ matching and Ref.~\cite{Berger:2003pk} has resummed to NLL/LO with $\cO(\as^2)$ matching. 

The explicit dependence of the NLO prefactor on the separate scales $\mu_{H,J,S}$ makes it distinct from what is obtained by NLO matching to QCD in the large-$\tau_a$ region where the three scales are comparable. Specifically, it improves the accuracy in the smaller-$\tau_a$ region where the distribution depends on physics at the three widely disparate scales separately, as revealed by the factorization theorem.  We emphasize that even though the effects of including this NLO piece are formally of next-to-next-to-leading logarithmic (NNLL) accuracy (using the counting $\as \ln \tau_a \sim \mathcal{O}(1)$), it {\it is} natural to include it in our NLL resummed result since the dependence on the arbitrary scales $\mu_{H, J, S}$ is cancelled to order $\as$ in our NLL/NLO calculation.\footnote{More generally, in an ${\rm N}^n {\rm LL}/ {\rm N}^m {\rm LO}$ calculation, the dependence on $\mu_{H,J,S}$ cancels up to order $\as^{\text{min}\{n,m\}}$, as  the $ \mu_{H, J, S}$ derivative of the logarithm of the distribution receives contributions from the prefactor at order $\as^m$ and from the anomalous dimension at order $\as^n$.}

Finally, we point out that while SCET can incorporate $\cO(\as^2)$ matching with, for example, an $\cO(\as^2)$ QCD calculation or an event generator, the classic approach by itself is less easily generalized to achieve full NLL/NLO accuracy. 
The reason for this difference is that SCET predicts the evolution boundary conditions for the hard, jet, and soft functions, $H(Q; \mu_H)$ and $F(\tau_a; \mu_F)$ ($F = J, S$) in \eq{Fconvol}, for arbitrary scales $\mu_{H, J, S}$ order by order in perturbation theory. On the other hand, as discussed in Ref.~\cite{Schwartz:2007ib}, the classic approach in contrast must effectively use the evolution boundary conditions $F(\tau_a; \mu_0) = \delta(\tau_a)$, which are LO in the SCET point of view. 
An implication of this difference is that, since our NLO prefactor is formally part of the NNLL series, full NNLL resummation is a nontrivial task in the classic approach (e.g. \cite{Bozzi:2003jy, deFlorian:2004mp}) whereas it is straightforward in SCET, using no new techniques additional to the ones described above.

In Fig.~\ref{hardscalevariation}, we compare our result with the classic result obtained in \cite{Berger:2003iw}. To make this comparison, we  truncate our result to NLL/LO accuracy and make the scale choices that are equivalent to those that were made in Ref.~\cite{Berger:2003iw} for the purpose of arriving at the classic resummed form. Namely,
 we run the jet and soft functions from their respective natural scales, $\mu_J = Q \tau_a^{1/(2-a)}$ and $\mu_S  = Q \tau_a$, to the hard scale set to $\mu_H = Q$. In addition, in Ref.~\cite{Berger:2003iw} the factorization scale $\mu$ was also chosen to be $\mu=\mu_H$, effectively turning off running between $\mu_H$ and $\mu$.  Thus, to make a genuine comparison, we vary $\mu$ both in the classic result given in \cite{Berger:2003pk} and in our result  \eq{sigmaNLL} over the range $Q/2$ to $2Q$, fixing $\mu_H = \mu$ in our result. Notice from the plots that the peak position appears to be more stable in the SCET results relative to the classic results and that there is a discrepancy in the overall normalization in the peak region, both of which may be attributed to power corrections arising from the spurious Landau poles present in the classic result.

\section{Conclusions}
\label{sec:conclusions}

We have calculated angularity distributions in $e^+ e^-$ collisions for $a<1$ to $\mathcal{O}(\alpha_s)$ in fixed-order accuracy, resummed leading and next-to-leading large logarithms in the perturbative series, incorporated the effects of a nonperturbative model for the soft function with a gap parameter, and cancelled the leading renormalon ambiguities in the perturbative expansion of the distribution and the gap parameter. Our new results for the one-loop jet and soft functions for all $a<1$ and the NLL resummation of logarithms of $\tau_a$ with explicit analytical dependence on the scales $\mu_{H,J,S}$ made possible what we believe are the most precise predictions of angularity distributions to date.

These predictions, especially after extension to higher orders in perturbation theory and resummation of logarithms, can prove useful in improving extraction of the strong coupling $\alpha_s$ or the parameters of nonperturbative models for the soft function. At the present time, in the absence of a new linear collider, such extractions would require the re-analysis of LEP data to extract the angularity distributions.

We also gain insight into the steps that will be required to predict jet observables in hadronic collisions, a broad range of which have been studied in \cite{Banfi:2004yd, Banfi:2004nk} using the classic approach. An SCET-based framework to factorize jet observables in this environment was developed in \cite{Bauer:2008jx}. Our analysis of angularities suggests that the study of any set of jet observables which vary in their sensitivity to narrower or wider jets or which depend on a jet algorithm picking out narrower or wider jets should be scrutinized in the same way as we did for angularities to determine whether the contributions of collinear and soft modes to each observable can be clearly separated. Also, our calculations of light quark angularity distributions in $e^+e^-$ collisions can be extended to calculating individual jet shapes for jets of various origins to higher accuracy, contributing to strategies to use such jet shapes to distinguish experimentally different types of jets \cite{Almeida:2008tp,Almeida:2008yp}.

While we have used SCET to calculate and explore the behavior of angularity distributions, the variation in behavior of the angularities has in turn shed light on the behavior and applicability of the effective theory. Varying $a$ essentially varies the collinear scale of SCET, in effect interpolating between (and extrapolating beyond) \SCETa and \SCETb, and so angularities provide an ideal testing ground for the behavior of these effective theories.

It is natural and straightforward to consider further improvement of our predictions to higher perturbative accuracy and reduced nonperturbative uncertainty.
We believe by using the cut diagram methods described above to obtain the angularity distributions to $\mathcal{O}(\as)$ we  can extend our results to $\mathcal{O}(\as^2)$ in a straightforward manner. 
Also, all of the ingredients necessary for NNLL resummation at $a=0$ are already known \cite{Becher:2008cf}, and we would only need to calculate those pieces which change with $a$.
The three-loop $\Gamma_{J,S}$ part of the jet and soft anomalous dimensions for arbitrary $a$ can be obtained from the known three-loop $\Gamma_{\cusp}$ \cite{Becher:2006mr} and the all-orders proportionality $ \Gamma_{J, S}\propto\Gamma_{\cusp} $ which we verified in Appendix~\ref{app:cusp}. The only unknown ingredients are the two-loop non-cusp part of the jet and soft anomalous dimensions. These can be obtained solely from the UV divergences of the two-loop graphs, and would immediately extend our results to NNLL accuracy. As for nonperturbative effects in the angularity distributions, we have treated these effects in the soft function in the simplest manner possible, adapting the $a=0$ soft model function to all $a$ by rescaling its first moment. Comparison of these predictions to $e^+e^-$ data can shed light on the reliability of this choice. 

Angularities and other event shapes have proven to be powerful probes of QCD and its effective theories, and promise to play a key role in the new era of collider physics searching for signals of new physics amid a sea of jets and strong interactions.

\acknowledgments
We are particularly grateful to C. Bauer, G. Sterman, and I. Stewart for a detailed review of an early draft and much valuable feedback and insight. We also thank M. Fickinger, S. Fleming, and A. Jain for enlightening discussions. CL and GO are grateful to the Institute for Nuclear Theory at the University of Washington for its hospitality during a portion of this work. AH is supported in part by an LHC Theory Initiative Graduate Fellowship. This work was supported in part by the U.S. Department of Energy under Contract DE-AC02-05CH11231, and in part by the National Science Foundation under grant  PHY-0457315.

\appendix

\section{Relation Among Hard, Jet, Soft, and Cusp Anomalous Dimensions}
\label{app:cusp}

In \eq{kernelparamsNLL} we used that the $\Gamma_F[\as]$ part of the jet or soft function anomalous dimension, defined in \eq{anomgeneral}, is proportional to the cusp anomalous dimension $\Gamma_{\text{cusp}}$ to all orders in $\alpha_s$. This fact is well known for the standard $a=0$ jet function and soft functions. In this section we verify that this relation remains true for all $a$. Our strategy will be to show that $\Gamma_{J,S}[\alpha_s]$ must always remain proportional to $\Gamma_H[\alpha_s]$, which is independent of $a$ and is already known to be proportional to $\Gamma_{\cusp}$.

The consistency of the  factorization theorem \eq{finalformula} requires a relation among the hard, jet, and soft function renormalization counterterms, and, thus, among the anomalous dimensions (see, e.g., \cite{Berger:2003iw,Fleming:2007xt}). This relation can be derived by requiring that \eq{finalformula} remain true when written in terms of either the bare or renormalized hard, jet, and soft functions on the right-hand side. This requires that
\begin{equation}
\label{Zrelation}
Z_H^{-1}(\mu) \delta(\tau_J - \tau_S) = \int \df\tau'\int \df\tau'' Z_J(\tau_J-\tau';\mu)Z_J(\tau'-\tau'';\mu)Z_S(\tau''-\tau_S;\mu)\,,
\end{equation}
to all orders in $\alpha_s$. To $\mathcal{O}(\alpha_s)$, we can easily verify this relation using \eqss{ZO}{Zsoft}{Zjet} with $Z_H(\mu) = \vert Z_{\mathcal{O}} (\mu)\vert^{-2}$. This relation amongst the counterterms requires in turn that the anomalous dimensions satisfy
\begin{equation}
\label{gammarelation}
-\gamma_H(\mu)\delta(\tau) = 2\gamma_J(\tau;\mu) + \gamma_S(\tau;\mu)\,.
\end{equation}

To all orders in $\alpha_s$ the hard anomalous dimension takes the form of \eq{anomhardallorders} and the jet and soft anomalous dimensions take the general form of \eq{anomgeneral} \cite{Grozin:1994ni} ,
where the constant $j_F$ is $j_J = 1/(2-a)$ for the jet function and $j_S = 1$ for the soft function. The constraint \eq{gammarelation} then requires the three independent relations
\begin{align}
0 &= \frac{4}{j_J} \Gamma_J[\alpha_s] + \frac{2}{j_S}\Gamma_S[\alpha_s] \,, \label{relation1} \\
-\Gamma_H[\alpha_s]  &= 2\Gamma_J[\alpha_s] + \Gamma_S[\alpha_s] \,, \label{relation2} \\
-\gamma_H[\alpha_s] &= 2\gamma_J[\alpha_s] + \gamma_S[\alpha_s] \,, \label{relation3}
\end{align}
to all orders in $\alpha_s$. These relations can be verified to $\mathcal{O}(\alpha_s)$ from \eq{anomhard} and Table~\ref{tab:jvalue}. The first two relations \eqs{relation1}{relation2} taken together imply that
\begin{equation}
\Gamma_S[\alpha_s] = \frac{1}{1-a}\Gamma_H[\alpha_s]\,,\qquad \Gamma_J[\alpha_s] = -\frac{1-a/2}{1-a}\Gamma_H[\alpha_s]\,,
\end{equation}
to all orders in $\alpha_s$ and for all $a<1$. Since $\Gamma_H[\alpha_s] \propto \Gamma_{\text{cusp}}$ and is independent of $a$, both $\Gamma_{S,J}[\alpha_s] \propto \Gamma_{\text{cusp}}$ as well. 

\section{Evaluation of Resummed Jet and Soft Functions and Full Distribution}
\label{app:RGE}

To evaluate the resummed jet and soft functions, we used the following method. First, note that from the expressions for the evolution equation, \eq{Fconvol}, the form of the evolution kernel, \eq{kernelF}, and the generic form of the NLO jet and soft functions,
\begin{equation}
    F(\tau;\mu_0)=c_1 \delta(\tau)+c_2\left(\frac{1}{\tau}\right)_++c_3\left(\frac{\ln{\tau}}{\tau}\right)_+
    \label{Fren}
\,,\end{equation}
the resummed jet and soft functions are proportional to
\begin{align}
F(\tau; \mu) \propto  &\int \!\df\tau' \left[\frac{\theta(\tau-\tau')}{(\tau-\tau')^{1+\omega}}\right]_+ F(\tau';\mu_0)=c_1 W_1+c_2 W_2+c_3 W_3
 \,,\end{align}
 where
 \begin{align}
 &W_1=\int \!\df\tau' \left[\frac{\theta(\tau-\tau')}{(\tau-\tau')^{1+\omega}}\right]_+ \delta(\tau') \,,
\nn\\
  &W_2=\int \!\df\tau' \left[\frac{\theta(\tau-\tau')}{(\tau-\tau')^{1+\omega}}\right]_+\left [\frac{\theta(\tau')}{\tau'}\right]_+ \,, \nn\\ 
 &W_3=\int \!\df\tau' \left[\frac{\theta(\tau-\tau')}{(\tau-\tau')^{1+\omega}}\right]_+  \left[\frac{\theta(\tau')\ln(\tau')}{\tau'}\right]_+
 \label{W3}
\,.\end{align}
Next, note that from the definitions of the plus functions, \eqs{lognplusdef}{omegaplusdef}, we can find $W_i$ as the coefficient of $\delta^i$ in the Taylor series of $ W(\delta)$, where $W(\delta)$ is defined as
\begin{align}
  W(\delta)\equiv \int \!\df\tau' \left[\frac{\theta(\tau-\tau')}{(\tau-\tau')^{1+\omega}}\right]_+\left [\frac{\theta(\tau')}{\tau'^{1+\delta}}\right]_+=\frac{\Gamma(-\omega)\Gamma(-\delta)}{\Gamma(-\omega-\delta)} \left[\frac{\theta(\tau)}{\tau^{1+\omega+\delta}}\right]_+
    \label{Wtilde}
\,.\end{align}
\eq{Wtilde} follows from the fact that
\begin{equation}
    \int \!\df\tau''   \left[\frac{\theta(\tau-\tau'')}{(\tau-\tau'')^{1+\omega_1}}\right]_+ \left[\frac{\theta(\tau''-\tau')}{(\tau''-\tau')^{1+\omega_2}}\right]_+=\frac{\Gamma(-\omega_1)\Gamma(-\omega_2)}{\Gamma(-\omega_1-\omega_2)} \left[\frac{\theta(\tau-\tau')}{(\tau-\tau')^{1+\omega_1+\omega_2}}\right]_+
    \label{plusidentity}
    \,.\end{equation}

By expanding both sides of \eq{Wtilde} in $\delta$ and comparing like powers of $\delta$, we find that
\begin{align}
   W_1 &= \left [\frac{\theta(\tau)}{\tau^{1+\omega}}\right]_+ \,, \qquad 
    W_2 =\bigg[\bigg(\ln(\tau)-H(-1-\omega) \bigg) \bigg(\frac{\theta(\tau)}{\tau^{1+\omega}}\bigg)\bigg]_+ \,,\nn\\
    W_3&= \bigg[\bigg(\frac{1}{2}\ln^2(\tau)-\ln(\tau)H(-1-\omega)+\frac{\pi^2}{12} \nn\\
    &\qquad \qquad \quad +\frac{1}{2}H(-1-\omega)^2-\frac{1}{2}\psi^{(1)}(-\omega)\bigg)\bigg(\frac{\theta(\tau)}{\tau^{1+\omega}}\bigg)\bigg]_+
\,.\end{align}
Here, $H(z)$ is the harmonic number function and $\psi^{(\nu)}(z)$ is the polygamma function.

The same technique can be used to analytically calculate the fully resummed cross-section, \eq{finalformula}, directly from the unresummed jet and soft functions. The resummed cross-section is of the form
\begin{align}
\frac{1}{\sigma_0}\frac{\df \sigma}{\df\tau}^{\PT} \propto \prod_{i=1}^3 \left( \int \! \df\tau_i \, \df\tau'_i \, F_i(\tau'_i; \mu_i) \left[\frac{\theta(\tau_i  - \tau'_i)}{(\tau_i-\tau'_i)^{1+\omega_i}} \right]_+  \right) \delta(\tau-\tau_1-\tau_2-\tau_3)
\label{sigmatot}
\,.\end{align}
where the jet and soft functions $F_i(\tau_i, \mu_i)$ are all of the form given in \eq{Fren}. These integrals can be done most easily by replacing the $F_i(\tau_i; \mu_i)$ on the right-hand side of \eq{sigmatot} with $\big[\theta(\tau)/\tau^{1+\delta_i} \big]_+$, expanding in $\delta_i$ before and after combining all the plus distributions using \eq{plusidentity}, and comparing like powers of the $\delta_i$. The result for the resummed cross-section \eq{sigmaNLL} then follows.

\section{Angularity Distribution in QCD to $\cO(\as)$}
\label{app:QCD}

In Sec.~\ref{sec:results} we matched the NLL resummed two-jet angularity distributions in SCET onto the $\mathcal{O}(\as)$ fixed-order distributions in full QCD using the remainder function $r_a(\tau_a)$, defined in \eq{rtau}. In this section we provide some details of how we calculate the QCD contribution to $r_a(\tau_a)$ away from $\tau_a = 0$, $A_a(\tau_a)$. In the process, we show that for $a \lesssim -1.9$ the angularities of events with more two-jet like kinematics become degenerate with those of more three-jet like events and contribute to the same $\tau_a$, and that for $a \lesssim -2.6$ the maximally symmetric three-jet event contributes to a smaller $\tau_a$ then some more two-jet like events. Thus, for small enough $a$, angularities fail to separate two-jet and three-jet like events. 

Both the one loop $q \bar q$ and tree-level $q\bar q g$ final states contribute to $\df\sigma/\df\tau_a$ at $\mathcal{O}(\alpha_s)$. However, the $q \bar q$ final states' contribution is proportional to $\delta(\tau_a)$ and hence only contributes to $A^\delta_a$. Thus to find $A_a(\tau_a)$ we only need to consider the tree-level $q \bar q g$ final states. Their contribution can be writtten as
\begin{equation}
\label{QCDNLOdist}
\frac{1}{\sigma_0}\frac{\df\sigma}{\df\tau_a} ^{q \bar q g} = \left( \frac{\as}{2 \pi}\right)A_a(\tau_a) \,,
\end{equation}
where
\begin{equation}
\label{Aa-dist}
A_a(\tau_a) = \CF \int \df x_1\,\df x_2\, \frac{x_1^2 + x_2^2}{(1-x_1)(1-x_2)}\delta\bigl(\tau_a - \tau_a(x_{1}, x_2)\bigr)\,,
\end{equation}
and where $x_{1,2} \equiv 2E_{1,2}/Q$ are the energy fractions of any two of the three final-state partons. By momentum conservation, $x_1 + x_2 + x_3 = 2$. For a three-particle final state, the thrust axis is given by the direction of the particle with the largest energy. The $x_{1,2}$ phase space can be divided into three regions, as illustrated in Fig.~\ref{fig:phasespace}A, according to which parton has the largest energy. In the region in which $x_i$ is larger than $x_{j,k}$, the angularity $\tau_a(x_1,x_2)$ is given by
\begin{equation}
\label{tauax1x2}
\tau_a(x_1,x_2)\Bigr\rvert_{x_i>x_{j,k}} = \frac{1}{x_i}(1-x_i)^{1-a/2} \left[(1- x_j)^{1-a/2}(1-x_k)^{a/2} + (1-x_j)^{a/2}(1-x_k)^{1-a/2}\right]\,.
\end{equation}
At each fixed value of $\tau_a = c$ in the distribution \eq{QCDNLOdist}, the delta function restricts the integral over $x_{1,2}$ to a linear contour determined by the equation $\tau_a(x_1,x_2) = c$, where $\tau_a(x_1,x_2)$ is given by \eq{tauax1x2}. Examples of these integration contours are shown in Fig.~\ref{fig:phasespace}B.  
\FIGURE[t]{
\centerline{\resizebox{15cm}{!}{\includegraphics{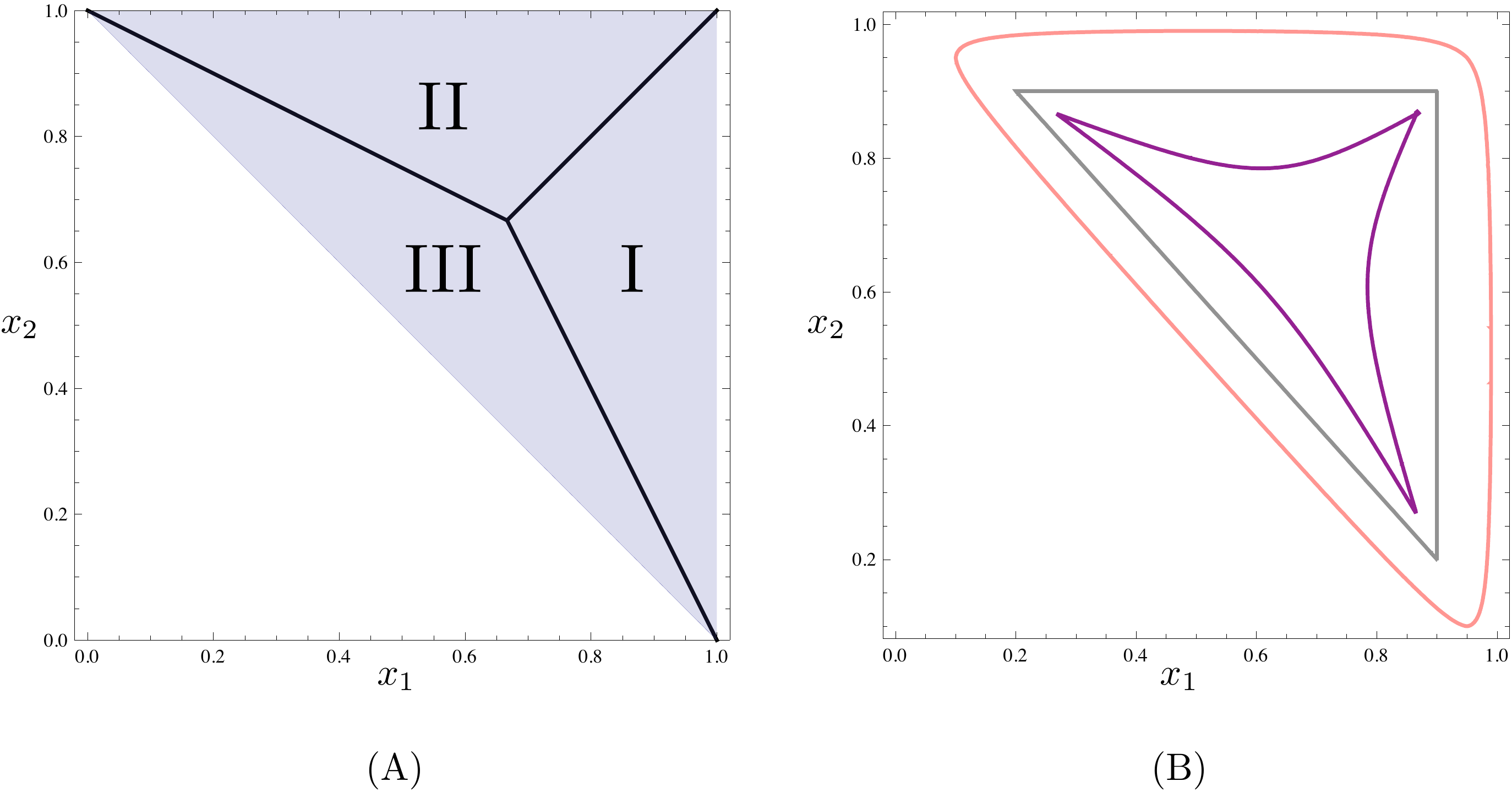}}}
\vspace{-.2in}
{ \caption[1]{(A) Phase space for three-particle $q\bar q g$ final state. The energy fractions $x_i = 2E_i/Q$ of the three particles satisfy $x_1+x_2+x_3=2$. In region I, $x_1>x_{2,3}$, in region II, $x_2>x_{1,3}$, and in region III, $x_3>x_{1,2}$. The thrust axis is in the direction of the particle with the largest energy. (B) Contours of constant $\tau_a = 1/10$ for $a=-1$ (purple), $a=0$ (gray), and $a=1$ (pink). The differential cross-section $\df\sigma/\df\tau_a$ is given by integrals over these contours in the $x_{1,2}$ phase space.}
\label{fig:phasespace} }
}

It is
sufficient to consider the part of the phase space corresponding to region
III shown in Fig.~\ref{fig:phasespace}, where $x_3>x_{1,2}$. Integration over the remaining two regions can be related to the
integration over region III by a  trivial shift of variables of integration.
Thus we need to solve
\begin{equation}
\label{contoureq}
  c=\frac{1}{2-x_1-x_2}
(x_1+x_2-1)^{1-a/2}\left[(1-x_1)^{1-a/2}(1-x_2)^{a/2}+(1-x_1)^{a/2}(1-x_2)^{1-a/2}\right]\,,
\end{equation}
where $x_{1,2}$ lie in region III. To find an explicit one-variable parameterization for $x_{1, 2}(w)$ which satisfies
\eq{contoureq}, we first absorb the factor
$1/(2-x_1-x_2)$ inside the brackets and define
\begin{align}
 w\equiv\frac{1-x_1}{2-x_1-x_2} 
 \label{wdef}\,.
\end{align}
In terms of $w$, \eq{contoureq} can be written as
\begin{equation}
\label{contour}
  c=(x_1+x_2-1)^{1-a/2}\left[w^{1-a/2}(1-w)^{a/2}+w^{a/2}(1-w)^{1-a/2}\right] \,.
\end{equation}
Solving Eqs. (\ref{wdef}, \ref{contour}) for $x_1, x_2$ gives:
\begin{align}
 &x_1(w)=1-w+w\left(\frac{c}{w^{1-a/2}(1-w)^{a/2}+w^{a/2}(1-w)^{1-a/2}}\right)^{\frac{1}{1-a/2}} \,,\nn\\
  &x_2(w)=x_1(1-w).
\end{align}

\FIGURE[t]{
\centerline{\mbox{ \hbox{\epsfysize=4.5truecm\epsfbox{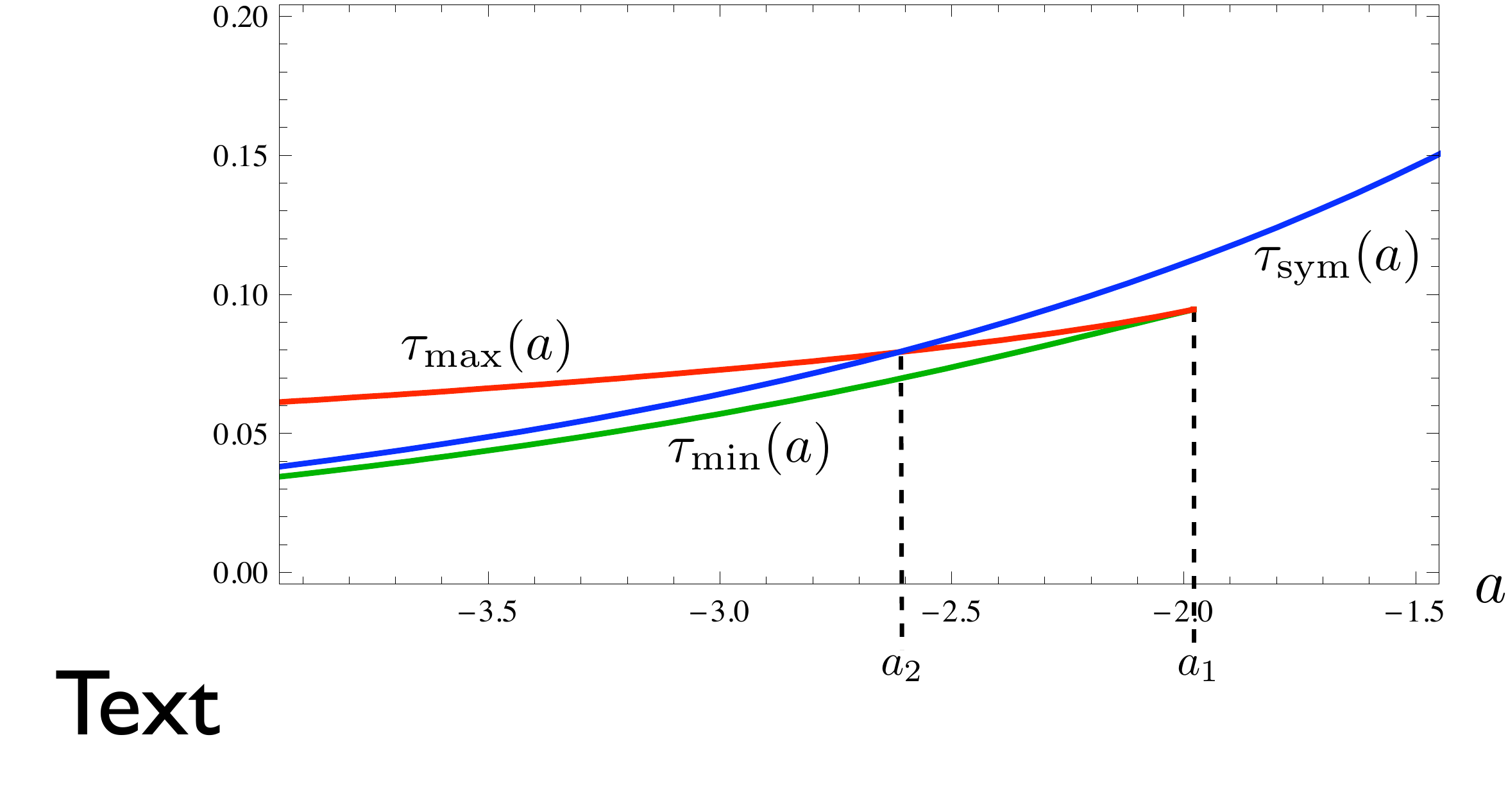}}  }}
\vspace{-.25in}
{\caption[1]{The local minimum (green line) and maximum (red line) of the function $F_a(w)$ over the range $ 0 < w < 1/2$ coincide at the point $a \equiv a_1\approx -1.978$.
At $a \equiv a_2\approx-2.618$, the value of angularity for the maximally symmetric three-jet case, $\tau_{\text{sym}}(a)=1/3^{1-a/2}$ (blue line), intersects the local maximum and so for $a < a_2$, the value of
maximum angularity for such $a$ corresponds not to the maximally symmetric case but to a more two-jet like event.  }
\label{tauaplots} }
}

\FIGURE[h]{
\includegraphics[totalheight = .45\textheight]{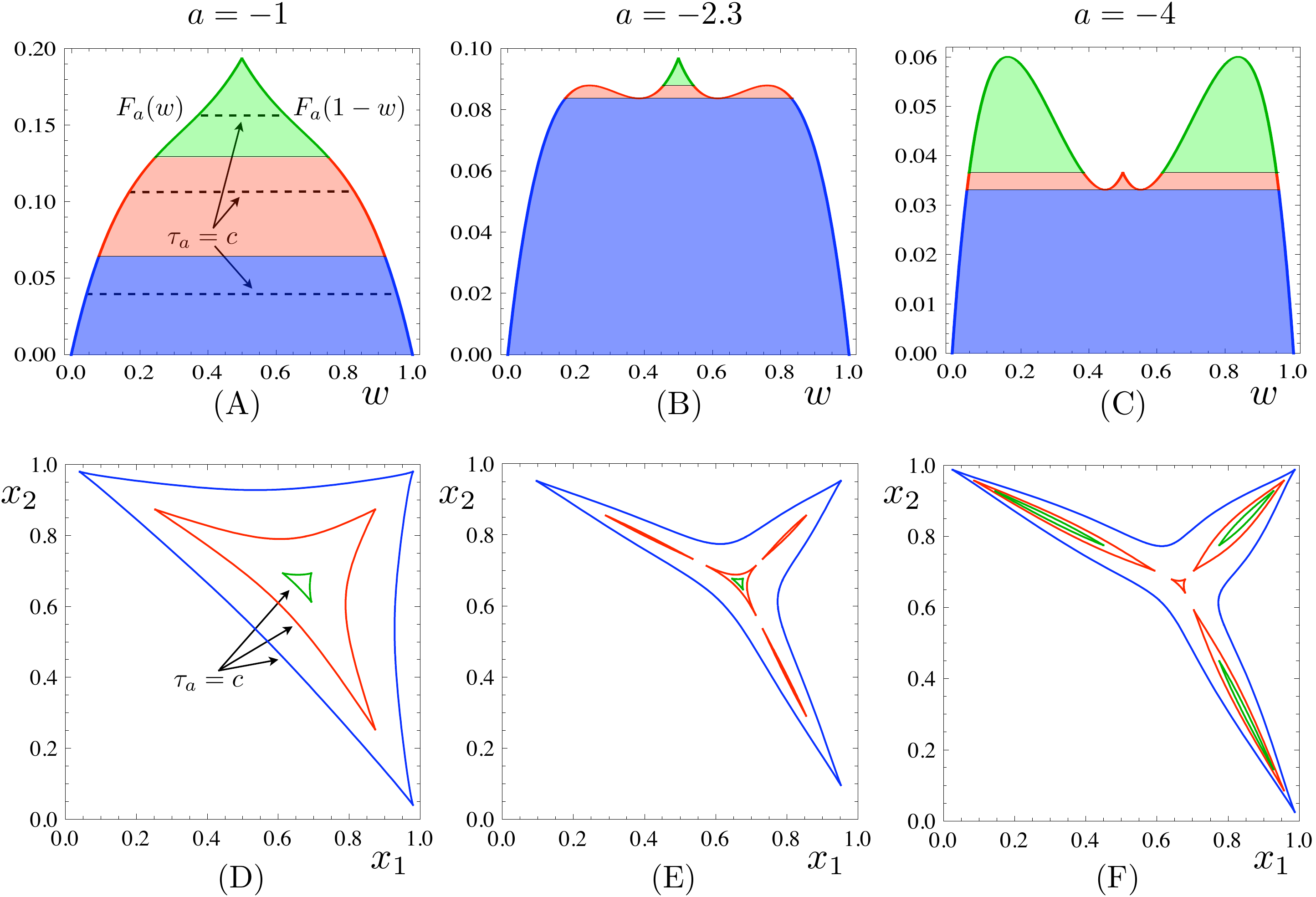}
\vspace{-.2in}
{\caption[1]{(A), (B), (C) Allowed regions for the parameter $w$ as a function of fixed $\tau_a = c$ are bounded by the curves $F_a(w)$ and $F_a(1-w)$. For (A), (D) $a=-1$, the integration is over a single, continuous domain for all fixed $\tau_a = c$ but for (B), (E) $a= -2.3$ and (C), (F) $a= -4$, there are multiple disjoint regions of integration for large enough values of $c$. In (D), (E), and (F), the blue, red, and green curves represent contours of integration for fixed $\tau_a = c$, in order of increasing $c$, and correspond to integration over a range of $w$ given by the lines of constant $\tau_a = c$ in the regions of the same color in (A), (B) and (C), respectively.}
\label{omegaregions} }
} 

Clearly from \eq{wdef},
$w$ lies in the interval $0\leq w\leq 1$. The precise range of values for $w$ is
determined from the conditions $x_1(w)\le 2-x_1(w)-x_2(w)$ and $ x_2(w)\le
2-x_1(w)-x_2(w)$. These inequalities can be simplified to

\begin{align}
   &c\le \text{min} \left \{ F_a(w), F_a(1-w) \right \} =
   \begin{cases} 
   F_a(w) &\text{for} \quad 0\le w\le 1/2\\
   F_a(1-w) &\text{for} \quad 1/2\le w \le 1
   \end{cases}  
\,, \end{align}
where
\begin{align}
   F_a(w)\equiv\frac{w(1-w)^{a/2}}{(1+w)^{1-a/2}}(w^{1-a}+(1-w)^{1-a})\,.
   \label{inequal}
\end{align}

The function $F_a(w)$ is monotonically increasing over the range $0 < w < 1/2$ only for $2 > a \ge
a_1\approx-1.978$, but for $a < a_1$  turns out to have exactly one
local maximum, $\tau_{\rm max}(a)$, and one local minimum, $\tau_{\rm min}(a)$. At $a=a_2\approx -2.618$, $\tau_{\rm max}(a)$ is equal to the
angularity of the symmetric three-jet configuration $x_1=x_2=x_3$ (where $w = 1/2$), 
$\tau_{\text{sym}}(a)=1/3^{1-a/2}$. Thus, the global maximum of $\tau_a$ over the whole range $0 \le w \le 1$, defined as $\tau_a^{\rm max}$, is $\tau_{\rm max}(a)$ for $a  \le a_2$ and is $\tau_{\rm sym}(a)$ for $a \ge a_2$.

In Fig.~\ref{tauaplots}, we show how the maximum and minimum of the
function $F_a(w)$ depend on $a$, along with the $a$ dependence of the symmetric three-jet configuration, and plot the special points $a_1$ and $a_2$.

In Fig.~\ref{omegaregions} we plot the boundary of $\tau_a$ ($F_a(w)$ for $0\le w\le 1/2$ and $F_a(1 - w)$ for $1/2\le w\le 1$) together with the contours of constant
$\tau_a(x_1,x_2)=c$ for different values of $c$ in the full $x_1$-$x_2$ plane for the cases $a = -1$, $a = -2.3$, and $a = -4$, which qualitatively represent the three cases $a> a_1$, $a_1> a > a_2$, and $a_2 >a$, respectively. 
From this analysis we conclude that for
$a<a_1$ and especially $a<a_2$ angularities fail to separate two-jet like and three-jet like events.

To obtain $A_a(\tau_a)$, we evaluate the integral in \eq{QCDNLOdist} over the appropriate contours in the $x_{1,2}$ phase space numerically, except for $a=0$, for which the integral can be evaluated analytically, giving (cf. \cite{DeRujula:1978yh})
\begin{equation}
\label{thrustdist}
A_0(\tau_0) = \CF\left[\frac{2(2-3\tau_0 + 3\tau_0^2)}{\tau_0(1-\tau_0)}\ln\left(\frac{1-2\tau_0}{\tau_0}\right) - \frac{3(1-3\tau_0)(1+\tau_0)}{\tau_0}\right]\,.
\end{equation}


\bibliography{NLL}

\end{document}